\pdfoutput=1
\documentclass[11pt,a4paper]{article}

\usepackage[a4paper,top=30mm,bottom=25mm,left=25mm,right=25mm]{geometry}
\usepackage[T1]{fontenc}
\usepackage[utf8]{inputenc}
\usepackage{newtxtext,newtxmath} 
\usepackage{microtype}
\usepackage{graphicx}
\usepackage{booktabs}
\usepackage{float}

\usepackage{siunitx}
\usepackage{xspace}

\usepackage[numbers,sort&compress]{natbib}
\usepackage[hidelinks]{hyperref}

\makeatletter
\renewcommand{\maketitle}{%
  \begin{flushleft}
    {\bfseries\LARGE\@title\par}
    \vspace{0.8em}
    {\normalsize \@author \par}
    \vspace{1.0em}
  \end{flushleft}
}
\makeatother

\makeatletter
\newlength\absleft \newlength\absright
\renewenvironment{abstract}{%
  \vspace{0.25\baselineskip}%
  \begin{list}{}{\leftmargin=\absleft \rightmargin=\absright}%
  \item[]\noindent\textbf{Abstract. }\setlength{\parindent}{0pt}\normalsize
}{%
  \end{list}%
  \vspace{0.25\baselineskip}%
}
\makeatother

\newenvironment{keywords}{%
  \vspace{-1.3em}
  \begin{list}{}{\leftmargin=\absleft \rightmargin=\absright}%
  \item[]\noindent\textbf{Keywords: }\normalsize
}{%
  \end{list}%
  \vspace{0.3em}
}

\usepackage{tocloft}
\cftsetindents{section}{0em}{2.2em}      
\cftsetindents{subsection}{2.2em}{3.0em} 
\cftsetpnumwidth{2.0em}                  

\newcommand{\met}{\ensuremath{p_T^{\text{miss}}}\xspace}
\newcommand{\none}{\ensuremath{\tilde{\chi}^0_1}\xspace}
\newcommand{\ntwo}{\ensuremath{\tilde{\chi}^0_2}\xspace}
\newcommand{\nthree}{\ensuremath{\tilde{\chi}^0_3}\xspace}
\newcommand{\cone}{\ensuremath{\tilde{\chi}^\pm_1}\xspace}

\title{Supersymmetry searches in CMS Run~2: A complete review}

\author{%
  {\large Sezen Sekmen}\\[3pt]
  {\small\itshape The Center for High Energy Physics, Kyungpook National University, Daegu, South Korea}\\
  {\small \href{mailto:ssekmen@cern.ch}{\texttt{ssekmen@cern.ch}}}%
}
\date{} 

\hypersetup{
  pdftitle={Supersymmetry searches in CMS Run 2: A complete review},
  pdfauthor={Sezen Sekmen},
  pdfsubject={High Energy Physics - Experiment},
  pdfkeywords={Beyond the standard model, supersymmetry, LHC, CMS, Run 2, analysis}
}

\begin{document}

\maketitle

\begin{abstract}
The Run 2 data-taking period of the CERN Large Hadron Collider during years 2015--2018 provided about 140 fb$^{-1}$ of proton-proton collisions at 13 TeV, offering an unprecedented opportunity to explore supersymmetry (SUSY) across a wide range of experimental signatures.  CMS responded with a broad and diverse search program, carrying out dozens of analyses that probed a multitude of final states and systematically explored different regions of the SUSY parameter space.  No significant deviations from standard model predictions were observed, and the results were used for constraining the SUSY landscape.  In this review, I provide a comprehensive account of the CMS Run 2 SUSY program, covering its strategy, targeted models, and analysis methods. I then present the full set of searches and conclude with their combined impact through simplified model and phenomenological MSSM interpretations.
\end{abstract}

\begin{keywords}
Beyond the standard model, supersymmetry, LHC, CMS, Run~2, analysis
\end{keywords}

\vspace{-0.5cm}

\begin{center}
{\it Published in Highlights in High-Energy Physics.}
\end{center}

\vspace{0.7cm}

\tableofcontents

\newpage

\section{Introduction }

Supersymmetry (SUSY) remains one of the most compelling frameworks for physics beyond the standard model and has long been a central focus of searches at the CERN Large Hadron Collider (LHC).  Its appeal lies in its ability to address various open questions in particle physics through an inherently elegant theoretical structure.  SUSY can manifest itself in many different ways, predicting a wide spectrum of signatures, hence providing a rich landscape for exploration at the LHC.

During Run 1, the Compact Muon Solenoid (CMS) Collaboration carried out an extensive program of searches with about 25 fb$^{-1}$ of proton-proton data at 7 and 8 TeV, collected between 2010 and 2012. In Run 2, the center-of-mass energy increased to 13 TeV and about 140 fb$^{-1}$ of data were collected between 2015 and 2018. This substantial gain in energy and luminosity opened a new level of sensitivity, enabling CMS to probe regions of SUSY parameter space that had previously been out of reach.

The Run 2 SUSY program was broad and ambitious.  It ranged from inclusive searches, sensitive to many different models at once, to targeted analyses focusing on scenarios predicting final states with compressed mass spectra, low momentum objects, signatures strongly resembling those of the standard model, small production cross sections (such as direct slepton pair production), and new types of long-lived particles. Using the Run 2 dataset, the collaboration carried out dozens of dedicated searches across a wide array of signatures. These efforts were made possible by advances in reconstruction and analysis techniques, supported by developments in machine learning. While no significant excess has been observed, the combined program has substantially reshaped our understanding of the viable SUSY parameter space, excluding broad classes of models and highlighting the regions where sensitivity remains limited.

In this review, I present a complete account of the CMS Run 2 SUSY program. I begin with the overall strategy, the models investigated, and the tools and techniques developed. I then summarize the full set of analyses, organized by their targeted final states, and conclude with a global view of the impact of CMS searches on the SUSY landscape, through collective interpretations and combinations in simplified models and in the phenomenological minimal supersymmetric standard model (pMSSM).

\section{The general strategy and evolution}

As the LHC began operations, we greeted the first data with high expectations for SUSY.  We set out to build a broad CMS SUSY program encompassing a diverse range of viable and experimentally accessible SUSY signatures.  The earliest CMS searches focused on so-called ``vanilla" SUSY signatures, with clear characteristics that made them easy to distinguish from the SM backgrounds. These arose in scenarios with low sparticle masses, leading to high production cross sections, and sizeable mass differences that produced object-rich decays.  Studies prioritized prominent gluino, squark, and electroweakino decays, yielding final states with multiple objects, high transverse activity, and, above all, large missing transverse momentum $\met$ (the negative vector sum of the transverse momenta of visible objects in the event).
The early analyses were categorized by the presence of objects such as jets, b-tagged jets, leptons, or photons, with $\met$ as a common hallmark.  These general purpose analyses also served as a testing ground for a range of discriminating variables and techniques. As the center-of-mass energy increased from 7 TeV to 8 TeV and the dataset grew, more targeted searches emerged, including those for third-generation squarks, staus, boosted objects, and long-lived particles.  By the end of Run 1, CMS searches using 5 fb$^{-1}$ at 7 TeV and 19.6 fb$^{-1}$ at 8 TeV set meaningful constraints on gluinos, squarks, the lightest charginos and neutralinos, and staus.

The increase from 8 TeV to 13 TeV in center-of-mass energy brought a substantial boost to the production cross sections of many SUSY processes. For instance, the cross section for top squark pair production with a mass of 800 GeV increases by roughly a factor of four, whereas the cross section for standard model (SM) $t\bar{t}$ production grows by only about a factor of two.  This relative gain gave SUSY searches a welcome advantage. Moreover, the integrated luminosity collected during Run 2 reached nearly 140 fb$^{-1}$, almost seven times that of Run 1. Together, these improvements allowed us to probe new regions of parameter space, including higher mass regimes that had previously been out of reach. We could also revisit already covered areas with enhanced sensitivity, and extend the program to more complex or subtle signatures that had earlier seemed prohibitively challenging.

The Run 2 SUSY search program was designed, for the most part, as a coordinated effort to explore SUSY across a broad and complementary landscape of signatures.  At its core, the program remained signature-driven: we organized our searches according to the observable features of the final states rather than committing to any specific theoretical model. This strategy preserved our sensitivity to a wide variety of models while ensuring that results remained interpretable and re-usable in different theoretical contexts. 

Building on the inclusive searches of Run 1, the Run 2 program broadened its scope to include both general-purpose and highly targeted analyses. Inclusive searches in 0-, 1-, 2- leptons plus multijet plus $\met$ final states formed a core foundation, offering broad coverage for strong production of gluinos and squarks. In parallel, more specialized analyses addressed final states featuring taus, photons, same-sign dileptons, soft objects, or low hadronic activity.  In parallel, searches with long-lived particles, which had already begun in Run 1 became more refined and versatile.  This diversity expanded sensitivity from canonical decay chains, to more elusive signatures arising from compressed mass spectra, electroweak production of charginos, neutralinos, and sleptons, R-parity violation, and similar.  

A guiding principle throughout was complementarity.  Searches were intentionally designed to be orthogonal where possible, and overlapping where necessary, to allow for eventual combinations.  We aimed to ensure that the most motivated regions of parameter space were covered in multiple ways, such that the strength of one analysis could compensate for the limitations of another.  The program was also structured with reinterpretation in mind: the use of simplified models and standardized categorization helped make results broadly applicable across theoretical scenarios.

For many SUSY searches, analyses were first performed using the 2016 dataset, corresponding to an integrated luminosity of 35.9 fb$^{-1}$. As additional data became available, these early 13 TeV searches were revisited using the full Run 2 dataset. In most cases, this update involved refining the methodology, incorporating improved techniques, expanding the scope of the final states, or redefining the search regions to enhance sensitivity or explore new territory. In some instances, multiple earlier analyses were merged into a single, more comprehensive search.  In this review, I will cover and refer to these more mature, legacy searches performed on the full Run 2 data, with a few exceptions that used partial data because of technical restrictions. 

Over the course of Run 2, the CMS SUSY program evolved into a coordinated and versatile framework, able to address the most compelling scenarios, as I will describe in the coming sections, while remaining flexible enough to capture the unexpected. 

\section{The SUSY model landscape}

The CMS Run 2 SUSY search program was designed to address a broad range of theoretical scenarios. While the searches were often inspired by complete SUSY frameworks, they were formulated and interpreted primarily in terms of simplified models. These are defined by a minimal set of new particles with specified masses, production modes, and decay channels, while all other new particles are assumed to be heavy and decoupled. A simplified model typically is described by just a few parameters, such as the relevant masses and, for long-lived cases, the proper lifetime, and typically fixes the branching fractions to 1 (unless stated otherwise, for example, when alternative decay modes of a particle is considered in the same model).  This approach captures the essential kinematics and topology of a process without the complexity of a full SUSY spectrum, enabling a systematic exploration of distinct signatures and allowing results to be reinterpreted in a wide variety of more complete scenarios.

The majority of searches assumed R-parity conserving SUSY, in which superpartners are pair-produced and decay to a stable lightest supersymmetric particle (LSP), typically taken to be the lightest neutralino. In these models, the presence of undetected LSPs leads to signatures with $\met$, which is a key requirement in many CMS SUSY searches.  Particular emphasis was placed on scenarios motivated by naturalness, where light Higgsinos, top squarks, and gluinos are expected to play a central role in stabilizing the electroweak scale. These considerations guided a significant portion of the program, including searches targeting final states with top and bottom quarks, W, Z, or Higgs bosons, and large $\met$. Notably, Higgsino scenarios can also give rise to long-lived particle (LLP) signatures when the mass splittings within the electroweakino sector are small, leading to displaced decays or disappearing tracks.

In parallel, other CMS searches explored a number of nonminimal SUSY scenarios with distinctive experimental signatures. In R-parity violating (RPV) models, the LSP decays promptly into SM particles, leading to final states with high lepton or jet multiplicities and typically low (or no) $\met$. Stealth SUSY scenarios, in which the SUSY decay chain ends in nearly mass-degenerate particles, suppress visible kinematic signatures and require alternative approaches, such as the use of initial-state radiation (ISR) or soft object reconstruction. In gauge-mediated SUSY breaking (GMSB) models, the LSP is assumed to be a light gravitino $\tilde{G}$, and the next-to-lightest SUSY particle (NLSP) can decay into a photon, Z, or Higgs boson, producing final states with photons and $\met$, and in some cases delayed or displaced signatures depending on the NLSP lifetime. A subset of searches also targeted dark matter-motivated scenarios, including those involving coannihilation mechanisms, where the LSP and a nearly degenerate partner (e.g., a stau) freeze out together in the early universe. These models often predict compressed spectra and lead to challenging final states with soft visible objects and moderate $\met$, motivating the development of dedicated search strategies.  The program also covered extensions of the MSSM gauge structure, with a search targeting a U(1)'-extended MSSM model derived from embedding SUSY in an $E_6$ grand unified theory.  This construct features a heavy, neutral Z' resonance, with a leptophobic nature, decaying into charginos, and leading to final states with leptons and $\met$. 

While simplified models formed the backbone of the individual search interpretations, a subset of results was also interpreted in the context of the phenomenological MSSM (pMSSM), a phenomenologically feasible realization of MSSM in 19 free parameters defined at the SUSY scale. This approach provides a more global view of the viable SUSY parameter space. A dedicated combination of several Run 2 searches was performed in this framework, allowing the impact of the full CMS SUSY program to be assessed in a more comprehensive and correlated way.

\section{Analysis techniques and tools}

Although CMS SUSY searches cover a broad range of final states, most analyses follow a common structure. We start by reconstructing, identifying, and selecting physics objects (such as electrons, muons, jets, photons, etc.), then apply preselection criteria. From there, we categorize events by defining signal, control, and validation regions to isolate candidate signals, constrain the dominant backgrounds, and validate the background estimation strategy.  A substantial part of the work involves evaluating systematic uncertainties associated with detector effects, background modeling, and theoretical inputs. Final results are obtained through a statistical analysis, most often performed using the CMS Combine framework, to extract potential signals or set exclusion limits, where systematic effects are incorporated as nuisance parameters to the fit.

Analyses in Run 2 broadly followed this common structure but also benefited from several key innovations. Object reconstruction and identification improved significantly, achieving better performance for standard objects and coverage for a wider variety of long-lived particle signatures. Trigger capabilities also expanded, allowing lower thresholds and dedicated designs targeting challenging scenarios. New kinematic variables were developed to improve signal-background separation. Machine learning (ML) became central, enhancing both object tagging and event level discrimination. These advances, together with more mature statistical treatments and software tools, formed the technical foundation for the progress in Run 2 SUSY searches.

A major development during Run 2 was the substantial improvement in physics object reconstruction and identification, aided by the rapid adoption of ML techniques. CMS deployed deep neural network (DNN)-based taggers for b jets (DeepCSV~\cite{Guest:2016iqz, CMS:2017wtu}, DeepJet~\cite{Bols:2020bkb,CMS-DP-2018-058}) and for boosted objects, including W, Z, Higgs bosons, and top quarks (DeepAK8~\cite{CMS:2020poo}, DeepDoubleB~\cite{CMS-DP-2018-046}, ParticleNet~\cite{Qu:2019gqs}). These taggers were essential in searches for heavy SUSY particles whose decays produce highly Lorentz-boosted SM or SUSY states, with subsequent decays leading to spatially collimated particles that merge into single reconstructed objects. We also introduced boosted leptonic jet tagging to identify objects decaying to leptons and quarks, which added sensitivity to RPV LSPs decaying in this way and, more generally, to final states with leptonic decays of boosted top quarks.  Direct searches for top squarks also made use of ML-based resolved top taggers, such as DeepResolved, to maintain efficiency in nonboosted kinematic regimes.  Over the course of Run 2, lepton reconstruction improved steadily, allowing lower transverse momentum selection thresholds and better isolation allowing the inclusion of soft leptons in compressed spectrum searches. Tau identification advanced with the DeepTau algorithm~\cite{CMS:2022prd}, which improved both efficiency and background rejection, while reducing misidentification rates. Together, these developments expanded the range of final states we could address with confidence.

About a third of the Run 2 SUSY analyses explored models predicting long-lived particles. These rely on specialized reconstruction techniques. CMS developed or improved algorithms for identifying heavy charged stable particles, disappearing tracks, delayed photons, delayed jets, displaced vertices, displaced tracks, displaced jets, trackless delayed jets.  ML techniques were used for efficiently identifying disappearing tracks, trackless delayed jets, displaced tracks, and displaced vertices.  These searches have additional backgrounds from mismeasurements, and noncollision sources such as cosmic rays, beam halo, noise or spurious sources. Dedicated data-driven background estimation methods were developed to reduce and estimate such backgrounds.

Trigger developments also played an important role in improving our searches. CMS operated with lower thresholds on lepton triggers compared to Run 1, for example, having 3 GeV muon thresholds in a dimuon trigger, which expanded the reach of searches involving soft leptons targeting compressed spectra.  Dedicated triggers were developed for LLP searches, including a trigger for displaced jets using track impact parameters, and delayed photon triggers using calorimeter timing.

Event selection and signal extraction rely on variables whose distributions differ between SUSY signals and SM  backgrounds. Many of these observables were first developed and studied during Run 1, and their use has been refined in Run 2.  Widely used examples include inclusive quantities such as object multiplicities, $\met$, hadronic transverse momentum $H_T$ (the scalar sum of jet transverse momenta), $S_T$ (the scalar sum of photon and jet transverse momenta), $L_T$ (the sum of lepton transverse momenta and $\met$), and $M_J$ (the sum of large-radius jet masses). These variables emphasize the central activity characteristic of many SUSY signals. Angular separations between objects, or between objects and $\met$, are also used to enhance the signal-to-background ratio by suppressing backgrounds with different event topologies. Other variables are designed to exploit the mass structure of the SUSY decay chain, such as invariant mass, transverse mass $m_T$, stransverse mass $m_{T2}$~\cite{Lester:1999tx, Barr:2003rg}, and the razor kinematic variables $M_R$ and $R^2$~\cite{Rogan:2010kb}.  In Run 2, additional specialized observables, including recursive jigsaw reconstruction variables~\cite{Buckley:2013kua, Jackson:2016mfb, Jackson:2017gcy} and the ``topness'' discriminant~\cite{Graesser:2012qy}, have been employed in analyses targeting compressed or otherwise challenging scenarios.

ML techniques were also widely adopted for event level signal discrimination. Boosted decision trees (BDTs) and DNNs were trained to separate SUSY signals from SM backgrounds using multiple low or high-level input variables.  In an increasing number of analyses, the output of these discriminants was used as the primary search variable, binned and fitted within signal (and sometimes control) regions. A notable advance was the use of parameterized neural networks (pNNs), in which the SUSY particle masses (or other model parameters) are included as additional inputs to the network. This allows a single network to learn the full dependence of the event kinematics on these parameters, and to interpolate smoothly between simulated mass points. As a result, the trained model can provide predictions for intermediate mass values without the need for separate training, enabling more efficient signal modeling over large kinematic ranges and improving the reinterpretation potential of the analysis.  The use of ML was not limited to signal discrimination. In selected analyses, it was also integrated into background modeling strategies. One example is ABCDiscoTeC~\cite{CMS:2025cvw}, an NN-based method for learning transfer functions between control and signal regions. 

With the increase in integrated luminosity, the granularity of signal region definitions also grew. Many analyses defined several hundred exclusive bins, and in at least one case the number exceeded a thousand. This high level of resolution helped maintain sensitivity across a wide range of decay topologies, mass configurations, and signal kinematics. It also increased the challenges on background estimation and the modelling of systematic uncertainties, which adapted and matured over the course of Run 2.

A notable methodological shift during Run 2 was the transition to simultaneous fits across signal and control regions, using a shared statistical model.  Earlier analyses often treated control region background estimates and signal region fits as distinct stages, esimating backgrounds first, then performing the statistical analysis in the signal regions. Though this practice still continued in Run 2, most Run 2 searches transitioned into integrating all regions into a unified statistical model. This approach allowed for consistent propagation of systematic uncertainties, better constraints on background normalization and shapes, and improved statistical power.

Finally, the Run 2 SUSY program benefitted from increasingly mature software infrastructure, common analysis tools, and analysis preservation practices. The widespread use of the CMS Combine statistical framework~\cite{CMS:2024onh} enabled consistent statistical treatment of uncertainties and facilitated combination studies. 

These developments in reconstruction, triggering, background estimation, statistical treatment, and software infrastructure provided a robust and flexible toolkit for Run 2 SUSY searches. They allowed each analysis to be tailored to its targeted signature while still benefiting from a common foundation of techniques and tools. In the following, we will go through the individual analyses and show how these methods were applied and adapted to address the wide variety of SUSY scenarios explored during Run 2.

\section{Signatures and analyses}

The rich variety of Run 2 CMS SUSY searches can be grouped according to the primary experimental signature they target: i) inclusive searches, ii) top squark searches, iii) electroweakino and slepton searches, iv) searches for compressed mass spectra, v) RPV and stealth searches, and iv) searches for long-lived particles. We will go through each category in turn, listing the individual analyses it contains and summarizing their target topology, notable experimental techniques, and main interpretations. Some analyses naturally span more than one category. When this occurs, they are placed where they most clearly belong, with additional mentions in other relevant contexts.

\subsection{Inclusive searches}
\label{sec:inclusive}

\begin{figure}[H]
\centering
\includegraphics[width=0.32\textwidth]{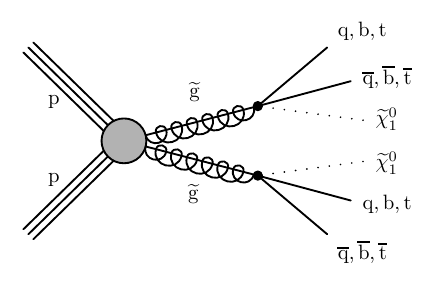} 
\includegraphics[width=0.32\textwidth]{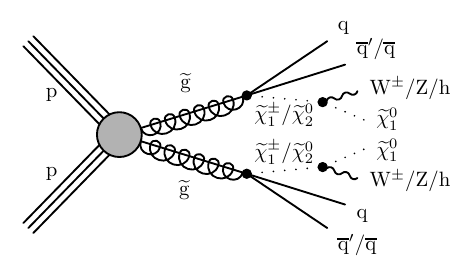} 
\includegraphics[width=0.32\textwidth]{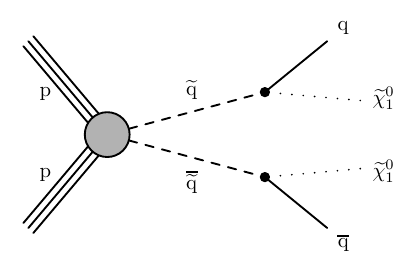} 
\caption{Diagrams for a nonexhaustive example set of simplified models covered by the inclusive searches.}
\label{fig:diag_inclusive}
\end{figure}

Inclusive searches are characterized by broad selection criteria targeting general SUSY event topologies, with minimal assumptions about the specific decay chains involved. These analyses typically categorize events based on object multiplicities, $\met$, or other global features, partitioning the selected sample into hundreds of exclusive search regions. This structure allows for sensitivity to a wide range of models in a largely signature-agnostic manner. Inclusive searches are particularly effective at probing strong production modes involving gluinos or squarks, but they are also interpreted in scenarios featuring electroweakinos or R-parity violation. Their generality makes them well suited to uncovering unexpected signals and highly valuable as benchmarks in reinterpretation studies.  Figure~\ref{fig:diag_inclusive} shows a nonexhaustive example set of simplified models interpreted by the inclusive searches.

A flagship inclusive search, one that also served as a training ground for generations of CMS SUSY analysts, is the fully hadronic analysis historically known as ``RA2b''~\cite{CMS:2019zmd}. This search targets events with at least two jets and significant missing transverse momentum. Events were partitioned into 174 exclusive search regions defined in four dimensions: jet multiplicity, number of b-tagged jets, the scalar sum of jet transverse momenta ($H_T$), and the magnitude of the vectorial sum of jet transverse momenta ($H_T^{miss}$).  The analysis also played a key role in validating and refining inclusive background estimation methods, which remain widely used across CMS SUSY searches.
The results were interpreted in simplified models of gluino and squark pair production. Gluino decays considered included $\tilde{g}\to t\bar{t}\none$, $\tilde{g}\to b\bar{b}\none$, $\tilde{g}\to q\bar{q}\none$, and cascade decays via electroweakinos, $\tilde{g}\to q\bar{q}\ntwo/\cone \to q\bar{q}\,(Z/W)\,\none$. Squarks were assumed to decay directly to a quark and a neutralino, $\tilde{q}\to q\none$. Under these assumptions and for 100\% branching fractions, the analysis excluded gluino masses up to 2000--2310~GeV and squark masses up to 1190--1630~GeV, depending on the production and decay scenario.

Another fully hadronic inclusive analysis made use of the stransverse mass variable $M_{T2}$ to enhance signal discrimination~\cite{CMS:2019ybf}. It extended the RA2b phase space by including monojet topologies. For events with two or more jets, signal regions are defined based on jet and b-tag multiplicities and $M_{T2}$. For monojet events, the jet $p_T$ serves as the signal discriminant. The results were interpreted in simplified models of gluino and squark pair production, excluding gluino masses up to 2250~GeV and $\none$ masses up to 1525~GeV. Squark mass limits reached up to 1710~GeV (light flavor), 1240~GeV (bottom), and 1200~GeV (top), with corresponding $\none$ mass exclusions up to 870, 700, and 580~GeV.  The analysis also has a disappearing track category, which will be mentioned in Section~\ref{sec:LLP}. 

Fully hadronic searches were complemented by inclusive analyses targeting final states with a single isolated lepton, multiple jets, and significant $\met$. One such search explicitly requires the presence of b jets and used $M_J$, the sum of masses of large-radius jets with $R = 1.4$, to define signal regions across the $n_{\text{jet}}$, $n_{\text{b}}$, $\met$, and $M_J$ dimensions~\cite{CMS:2020cur}. It targeted gluino pair production with decays to $t\bar{t}\none$, excluding gluino masses up to about 2150~GeV.  

A second inclusive single-lepton search took a more general approach, defining categories with zero or at least one b-tagged jet~\cite{CMS:2022idi}. It employs the variable $\Delta\phi$, the azimuthal angle between the lepton and its reconstructed W boson candidate, to suppress backgrounds, and uses both boosted and resolved top quark and W npspm tagging. Signal regions are defined in terms of $n_{\text{jet}}$, $L_T$, $H_T$, $\Delta\phi$, and the number of tagged boosted W bosons or top quarks. This analysis excluded gluino masses up to 2130~GeV for $t\bar{t}\none$ decays and up to 2280~GeV for $q\bar{q}W\none$ decays.

Inclusive searches also extended to final states with two or more leptons, offering complementary sensitivity to scenarios with cascade decays. One such search targeted events with two oppositely charged, same-flavor leptons and large $p_T^{\text{miss}}$~\cite{CMS:2020bfa}. It probed three distinct features: an excess of dilepton events near the Z boson mass peak, a kinematic edge in the dilepton invariant mass, and an enhancement in the nonresonant dilepton production. The results were interpreted in simplified models involving gluino and squark pair production with decays via \cone, \ntwo, and $\tilde{\ell}$, as well as direct production of \cone, \ntwo, and $\tilde{\ell}$. The search excluded gluino masses up to 1870~GeV, light-flavor and bottom squark masses up to 1800 and 1600~GeV, respectively, chargino and neutralino masses up to 750 and 800~GeV, and slepton masses up to 700~GeV.

Another analysis studied final states with at least two jets and either two same-charge leptons or three or more charged leptons~\cite{CMS:2020cpy}. The search explored a wide range of gluino, squark, and electroweakino processes, including those with RPV decays, through 168 search regions defined by $n_{\text{jets}}$, $n_b$, $H_T$, $m_T$, and $p_T^{\text{miss}}$. It excluded gluino masses up to 2.1~TeV, expanding the reach of previous searches by about 200~GeV, and top and bottom squark masses up to 0.9~TeV. It also provided model-independent limits as a function of $p_T^{\text{miss}}$ and $H_T$, together with background predictions and data yields in a set of simplified signal regions.

To increase sensitivity at high masses and large mass splittings, several analyses focused on final states with boosted objects. One early search explored gluino cascades via neutralinos involving Z bosons through a boosted Z boson and \met\ signature, using the mass of the boosted Z boson reconstructed with the anti-$k_\text{T}$ algorithm, with size 0.8 (i.e., an AK8 jet), and excluded gluino masses up to 1920~GeV~\cite{CMS:2020fia}. Another analysis targeted events with two Higgs bosons decaying as $H \to b\bar{b}$, accompanied by \met, identifying both boosted and resolved Higgs bosons to maintain coverage across the full kinematic range~\cite{CMS:2022vpy}. The results were interpreted in models of gluino-mediated or direct production of heavy neutralinos decaying to Higgs bosons plus lighter neutralinos or goldstinos, excluding gluino and neutralino masses up to 2330 and 1025~GeV, respectively.

\begin{figure}[H]
\centering
\includegraphics[width=0.49\textwidth]{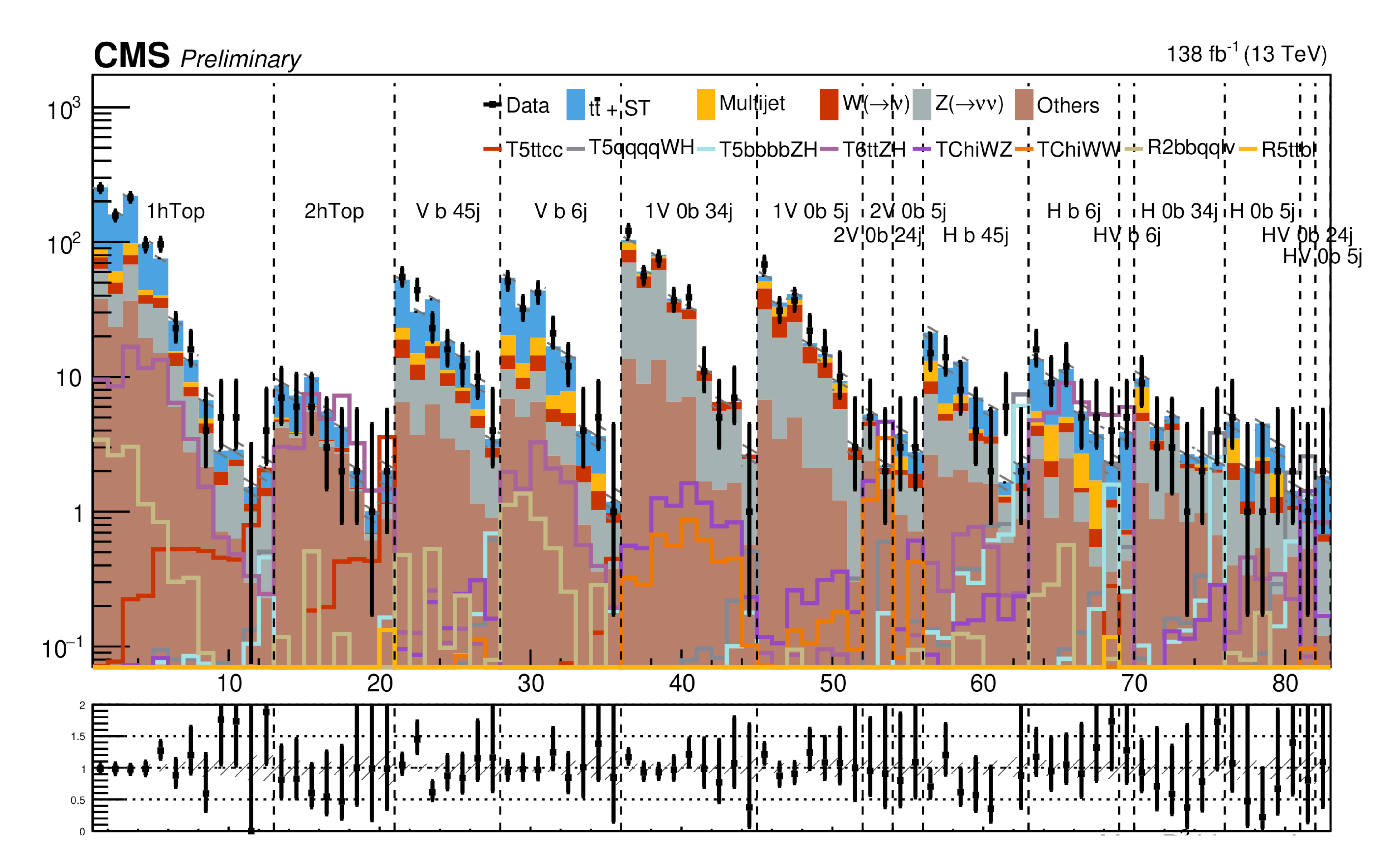} 
\includegraphics[width=0.49\textwidth]{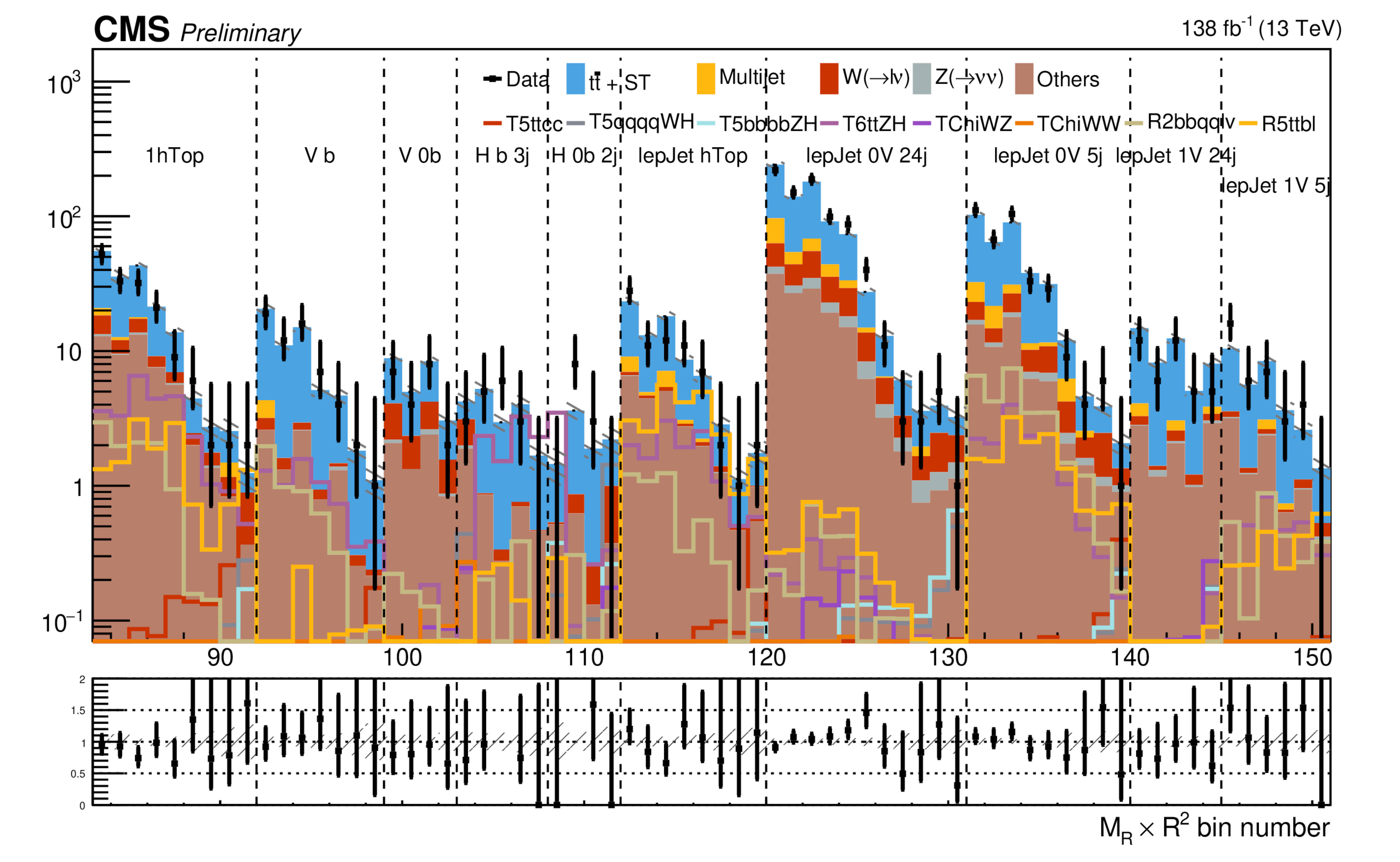} 
\caption{Highlights from inclusive analyses: Hadronic (left) and isolated and nonisolated leptonic (right) search regions and bins from the inclusive boosted analysis with razor variables. }
\label{fig:highlights_inclusive}
\end{figure}

The most recent and most inclusive boosted search targeted final states containing hadronically decaying boosted W, Z, and Higgs bosons and top quarks~\cite{CMS-PAS-SUS-23-014}. It also included, for the first time in CMS SUSY searches, leptonic decays of boosted SM or SUSY particles, reconstructed as boosted leptonic jets. The analysis used the razor kinematic variables, which characterize events with massive particles decaying to visible and invisible states as a peak above a smoothly falling background. Events are separated into channels with zero leptons, an isolated lepton, or a nonisolated lepton, and organized into 25 disjoint signal regions defined by object multiplicities, further divided into 150 bins of razor variables, shown in Figure~\ref{fig:highlights_inclusive}. In representative R-parity-conserving models, the search excluded gluino masses up to 2.35~TeV and top squark masses up to 1.45~TeV. In RPV scenarios, it excluded bottom squark masses up to 0.97~TeV and gluino masses up to 1.82~TeV, while electroweak production of nearly mass-degenerate charginos and neutralinos was excluded up to 1.05~TeV.

An additional inclusive search targeted final states with a photon, jets, and \met, where the photon provides a characteristic signature of neutralino decays into a gravitino LSP. The analysis used $S_T$ to suppress backgrounds and divided events into 45 search bins defined by $n_{\text{jets}}$, $n_b$, \met, and the presence of hadronically decaying W, Z, or Higgs bosons identified via the mass of AK8 jets. In a range of production processes, the search excluded gluino masses up to 2.35~TeV, squark masses up to 1.43~TeV, and electroweakino masses up to 1.23~TeV.

\subsection{Top squark searches}
\label{sec:stop}

\begin{figure}[H]
\centering
\includegraphics[width=0.32\textwidth]{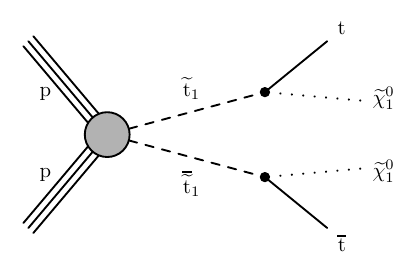} 
\includegraphics[width=0.32\textwidth]{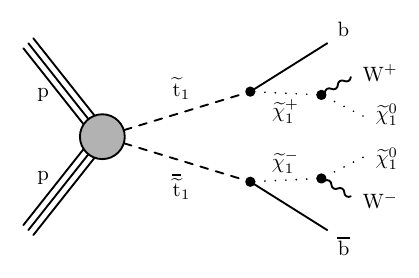} 
\includegraphics[width=0.32\textwidth]{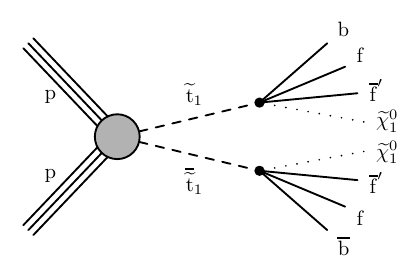} 
\caption{Diagrams for a nonexhaustive example set of simplified models covered by the top squark searches.}
\label{fig:diag_stop}
\end{figure}

Top squarks play a central role in many SUSY models, particularly in naturalness-motivated scenarios where relatively light top squarks can cancel the dominant top quark loop corrections to the Higgs boson mass. During Run 2, CMS carried out a coherent set of searches covering a wide range of decay modes and kinematic regimes, from large mass splittings between the top squark $\tilde{t}$ and $\none$ to compressed spectra where the decay products have low momenta.
The analyses are designed to be complementary, targeting different final states with zero, one, or two leptons, as well as tau leptons, and making use of both resolved and boosted top quark reconstruction. Together, they provide robust coverage of the parameter space, with significant overlap in sensitivity to enhance discovery potential and improve exclusion reach.  Figure~\ref{fig:diag_stop} shows a nonexhaustive example set of simplified models covered by the top squark searches.

\begin{figure}[H]
\centering
\raisebox{0.25\height}{\includegraphics[width=0.34\textwidth]{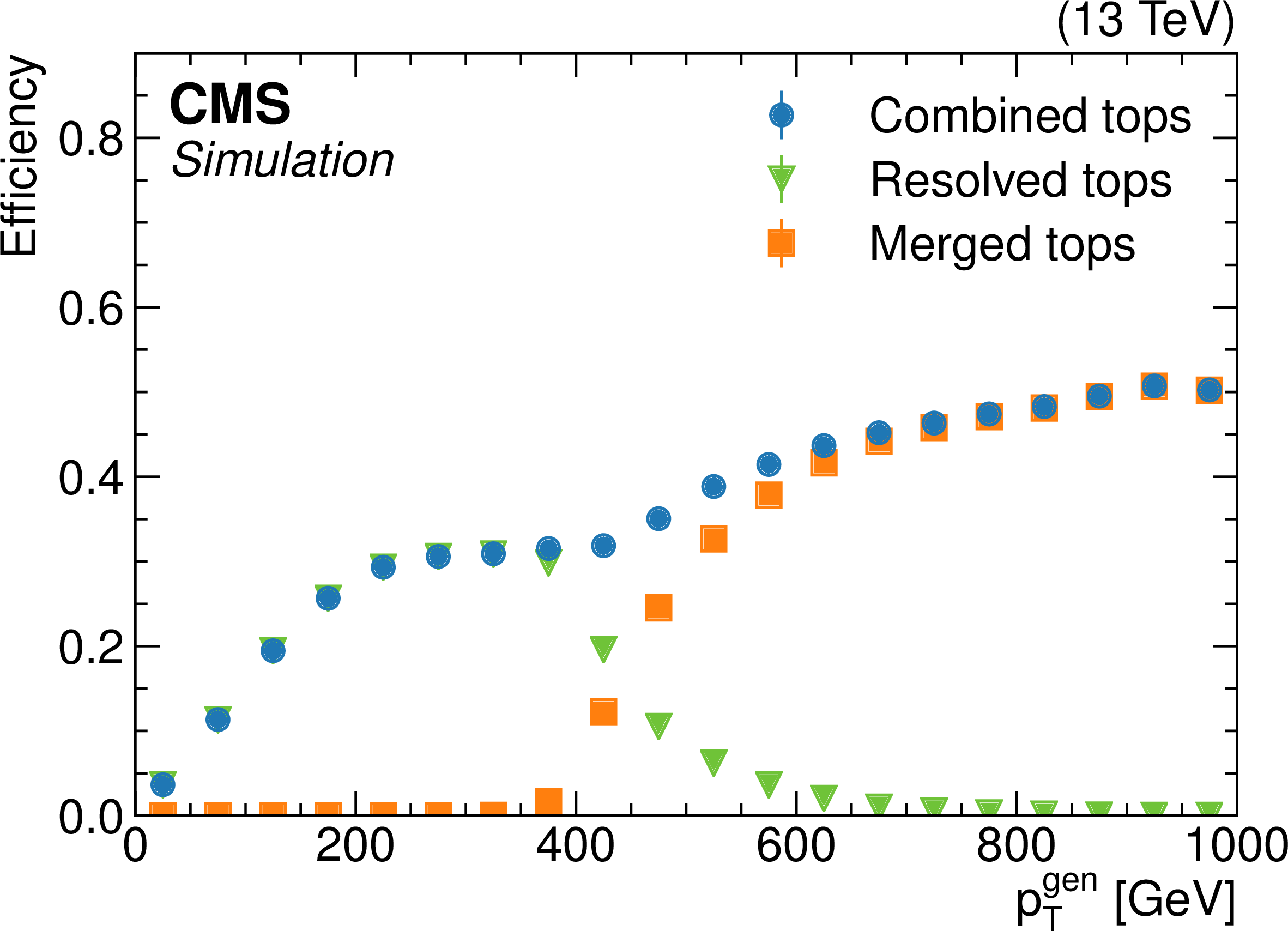}} \quad
\raisebox{0.2\height}{\includegraphics[width=0.33\textwidth]{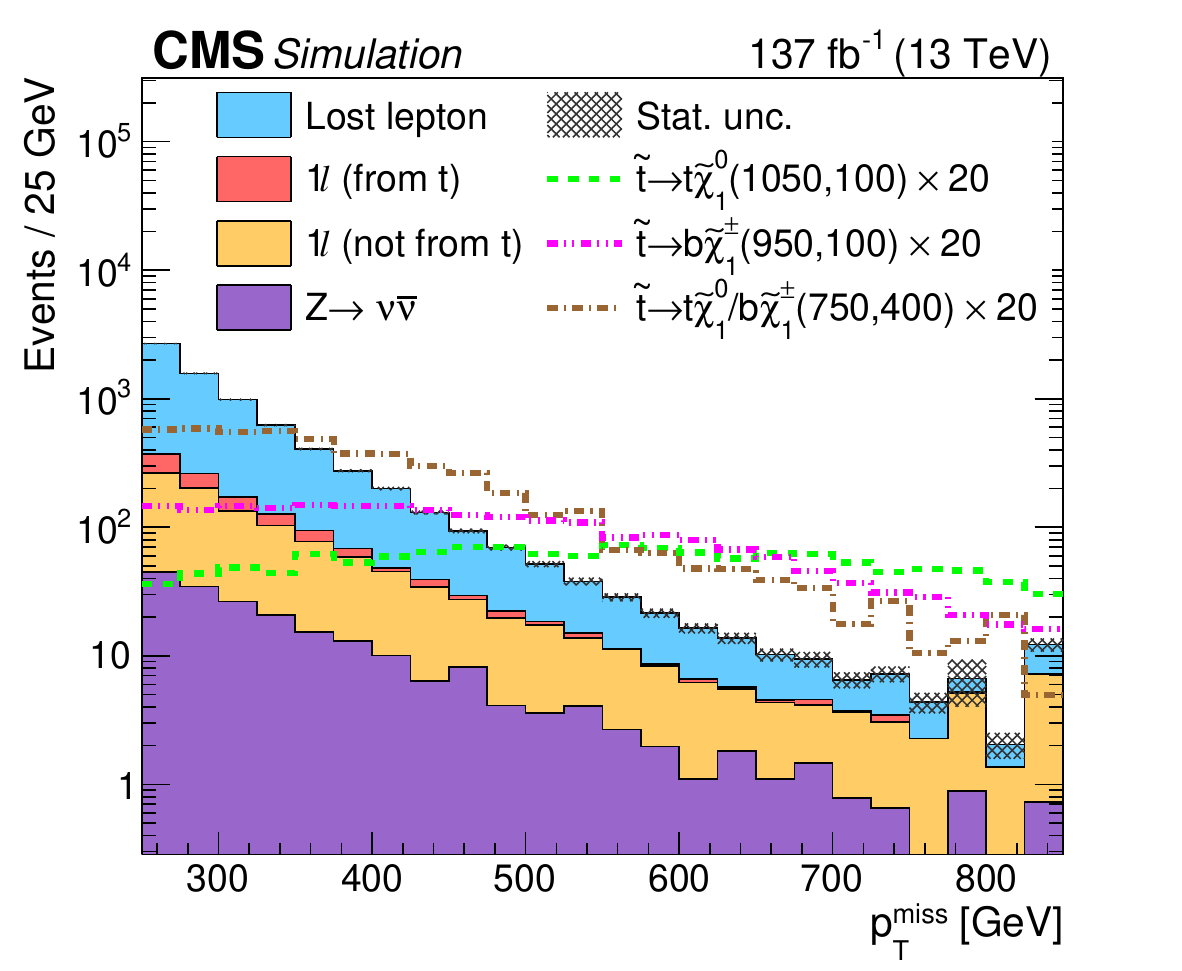}} 
\includegraphics[width=0.26\textwidth]{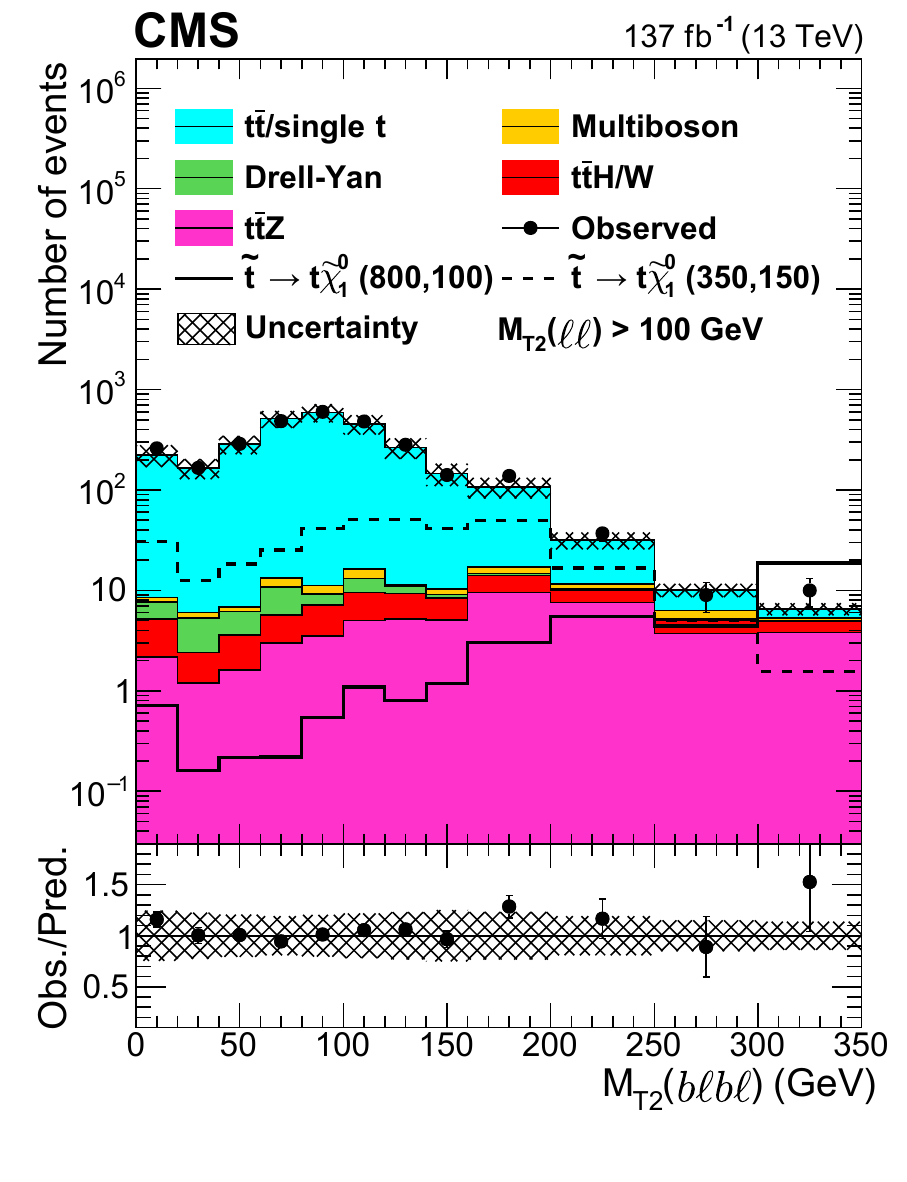} 
\caption{Highlights from top squark searches: Complementarity of boosted and resolved top tagging efficiencies at function of generator level top quark $p_T$ from the fully hadronic top squark search (left); $\met$ distribution after preselection in the single lepton top squark search (center); and $M_{\text{T2}}(b\ell b\ell)$ distribution for the preselection from the dilepton top squark search (right).}
\label{fig:highlights_stop}
\end{figure}

The fully hadronic search for direct and gluino mediated top squark production targeted final states with at least two jets and $\met$ for scenarios with both low and high mass splitting ($\Delta m$) between the top squark and $\none$~\cite{CMS:2021beq}. It employed the DNN-based DeepResolved and DeepAK8 taggers to identify hadronically decaying boosted W bosons and resolved and boosted top quarks, whose complementary efficiency performance is shown as function of generator level top quark $p_T$ in Figure~\ref{fig:highlights_stop} (left).  In the low $\Delta m$ case, sensitivity is enhanced through the use of an ISR jet and secondary vertices for low-$p_T$ b-quark tagging.  Events are split into 53 bins for the low-$\Delta m$ case, defined by $n_\text{jets}$, $n_b$, $n_\text{SV}$, $m_T^b$, $p_T^\text{ISR}$, $p_T^b$, and \met, and 130 bins for the high-$\Delta m$ case, defined by $m_T^b$, $n_\text{jets}$, $n_b$, $n_\text{top}$, $n_W$, $n_\text{resolved top}$, $H_T$, and \met.  The search excluded top squarks up to 1310 GeV for direct production, and gluinos up to 2260 GeV for gluino-mediated top squark production.  

The single lepton top squark search was performed in a similar spirit, exploring both noncompressed and compressed mass splitting cases distinctly, and employing the DNN-based taggers~\cite{CMS:2019ysk}.  It also features a modified version of the topness variable, $t_{\text{mod}}$, a $\chi^2$-like discriminator constructed from reconstructed top quark and W boson  masses, their resolutions, and the momenta of their decay products, to improve separation from the dominant $t\bar{t}$ background. Figure~\ref{fig:highlights_stop} (center) shows the \met distribution after preselection. The noncompressed selection defines 39 signal regions based on $n_{\text{jets}}$, $t_{\text{mod}}$, $M_{\ell b}$, top tagging category, and $\met$. The compressed selection targets both $\Delta m(\tilde{t},\none) \approx m_{\text{top}}$ and $\Delta m(\tilde{t},\none) \approx m_W$ in 10 $\met$-binned regions. The analysis excluded top squark masses up to 1.2 TeV for direct production.  

Complementing the single lepton search, the dilepton search employed $\met$ significance variable together with the stransverse mass variables $m_{T2}(\ell,\ell)$ and $m_{T2}(\ell b,\ell b)$, shown in Figure~\ref{fig:highlights_stop} (right) to define 12 signal regions~\cite{CMS:2020pyk}. It probed direct production in direct, cascade, and double-cascade decay modes, excluding top squark masses up to 925 GeV for $\tilde{t} \to t\none$, 850 GeV for $\tilde{t} \to b\cone \to bW^\pm\none$, and 1.4 TeV for $\tilde{t} \to b\cone \to b\nu\tilde{\ell} \to b\nu\ell\none$, assuming the chargino mass to be the mean of the top squark and neutralino masses.  

The fully hadronic. single lepton, and dilepton searches were combined in~\cite{CMS:2021eha}. That study also added a dedicated search targeting the so-called "corridor" region where $\Delta m_\text{cor} \equiv |\Delta m(\tilde{t} - \none| - m_t| = 30$~GeV, and $m_{\tilde{t}} < 175$~GeV, based on a dilepton selection and a DNN for signal extraction.  

There were also searches focusing on challenging final states of top squark pair production. One targeted the compressed regime, where $\Delta m(\tilde{t}, \none) < m_W$, with prompt four-body decays $\tilde{t} \to b f \bar{f}' \none$, where $f$ and $\bar{f}'$ denote either a lepton and a neutrino or a quark-antiquark pair~\cite{CMS:2023ktc}. It used the single lepton, jets, and $\met$ channel, and employed a BDT trained to exploit the small $\Delta m$ kinematics for signal extraction. The search excluded top squark masses up to 480 GeV and 700 GeV for $\Delta m$ of 10 GeV and 80 GeV, respectively. 

Another search targeted double cascade decays via charginos and staus or tau sneutrinos, $\tilde{t} \to b \cone \to b \nu \tilde{\tau}_1 \to b \nu \tau \none$ or $\tilde{t} \to b \cone \to b \tau \tilde{\nu} \to b \nu \tau \none$ in final states with at least one hadronically decaying $\tau$ lepton identified with the DNN-based DeepTau algorithm~\cite{CMS:2023yzg}. The analysis covers the $e\tau_h$, $\mu\tau_h$, and $\tau_h\tau_h$ channels, each divided into 15 search bins defined by $m_{T2}$, $\met$, and $H_T$.  It probed top squark masses up to 1150 GeV for a nearly massless neutralino.

\subsection{Searches for electroweakinos and sleptons}
\label{sec:ewkslep}

\begin{figure}[htbp]
\centering
\includegraphics[width=0.32\textwidth]{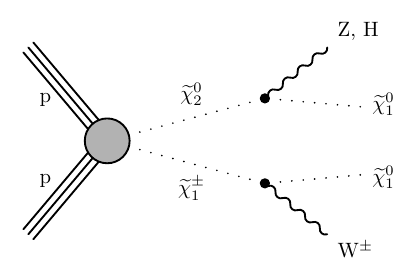} 
\includegraphics[width=0.32\textwidth]{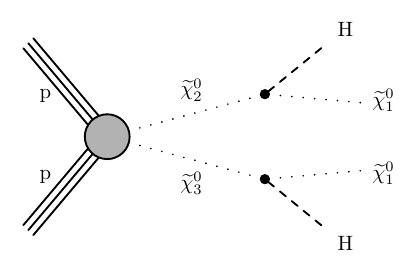} 
\includegraphics[width=0.32\textwidth]{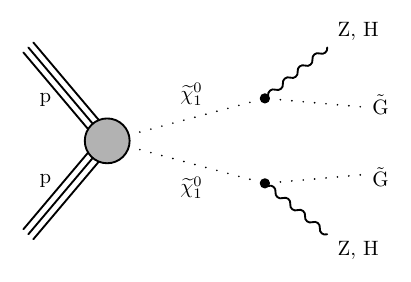} 
\includegraphics[width=0.32\textwidth]{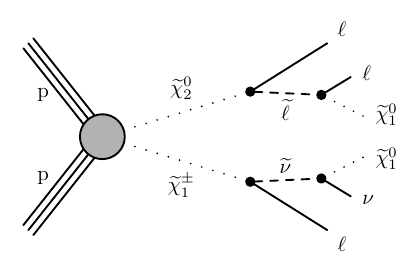} \includegraphics[width=0.32\textwidth]{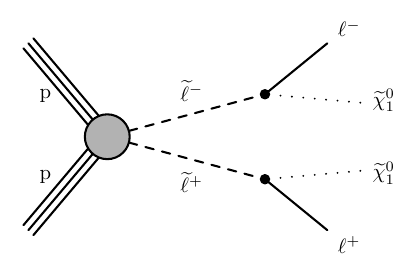} \includegraphics[width=0.32\textwidth]{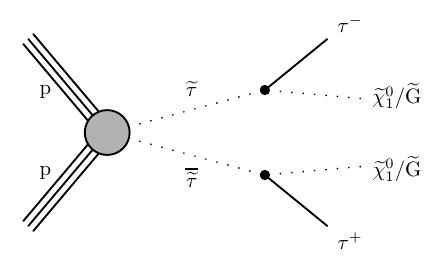} 
\caption{Diagrams for a nonexhaustive example set of simplified models covered by the electroweakino and slepton searches.}
\label{fig:diag_ewk}
\end{figure}

Searches targeting electroweak production of charginos, neutralinos, and sleptons complement strong-production searches, particularly in scenarios where colored SUSY particles are heavy. These processes often lead to final states with leptons, $\met$, and sometimes Higgs or gauge bosons, and require dedicated strategies to overcome their much smaller production cross sections.  Run 2 searches explored a wide range of electroweak signatures, from multilepton channels to final states with hadronic boson decays, as well as searches targeting staus and their distinctive $\tau$-rich final states.  Figure~\ref{fig:diag_ewk} shows a nonexhaustive example set of simplified models covered by the electroweakino and slepton searches.

\begin{figure}[H]
\centering
\raisebox{0.1\height}{\includegraphics[width=0.33\textwidth]{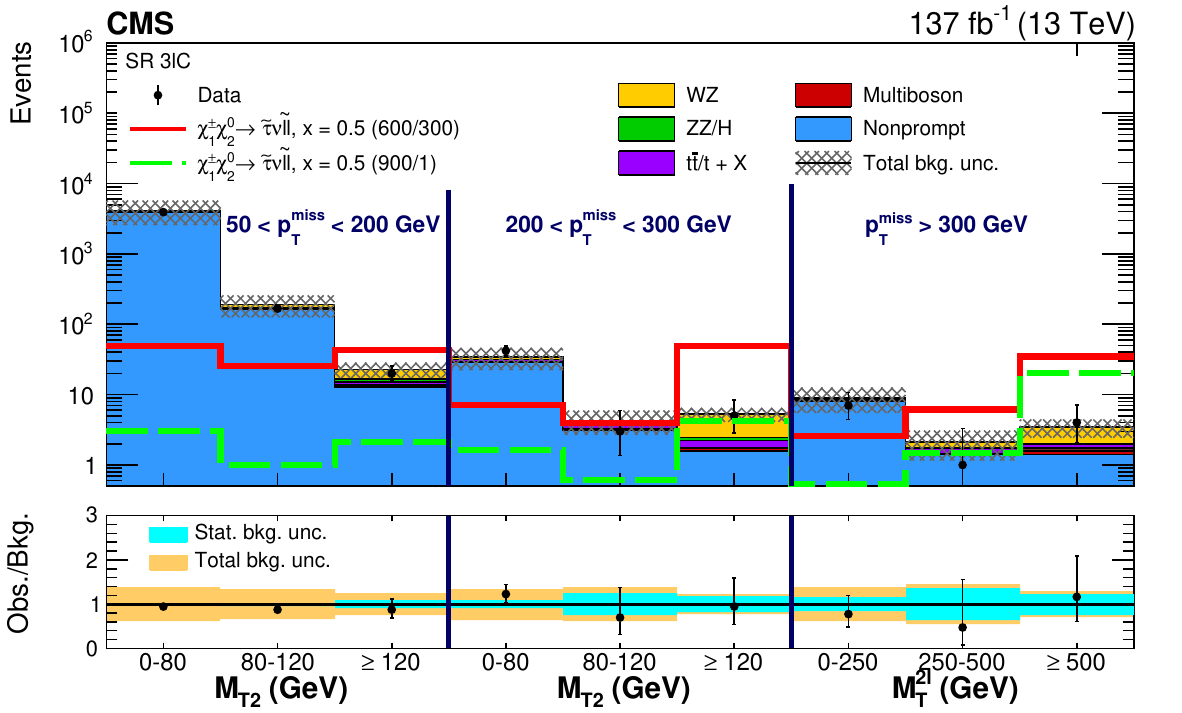}} 
\includegraphics[width=0.32\textwidth]{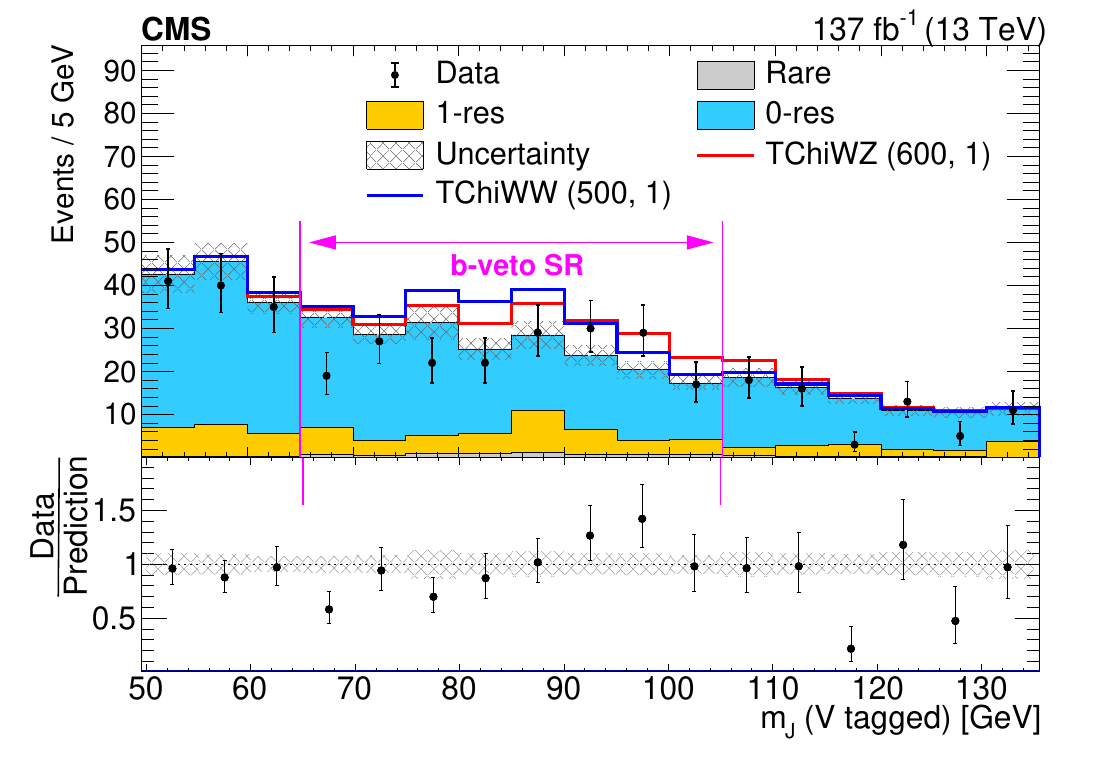} 
\includegraphics[width=0.28\textwidth]{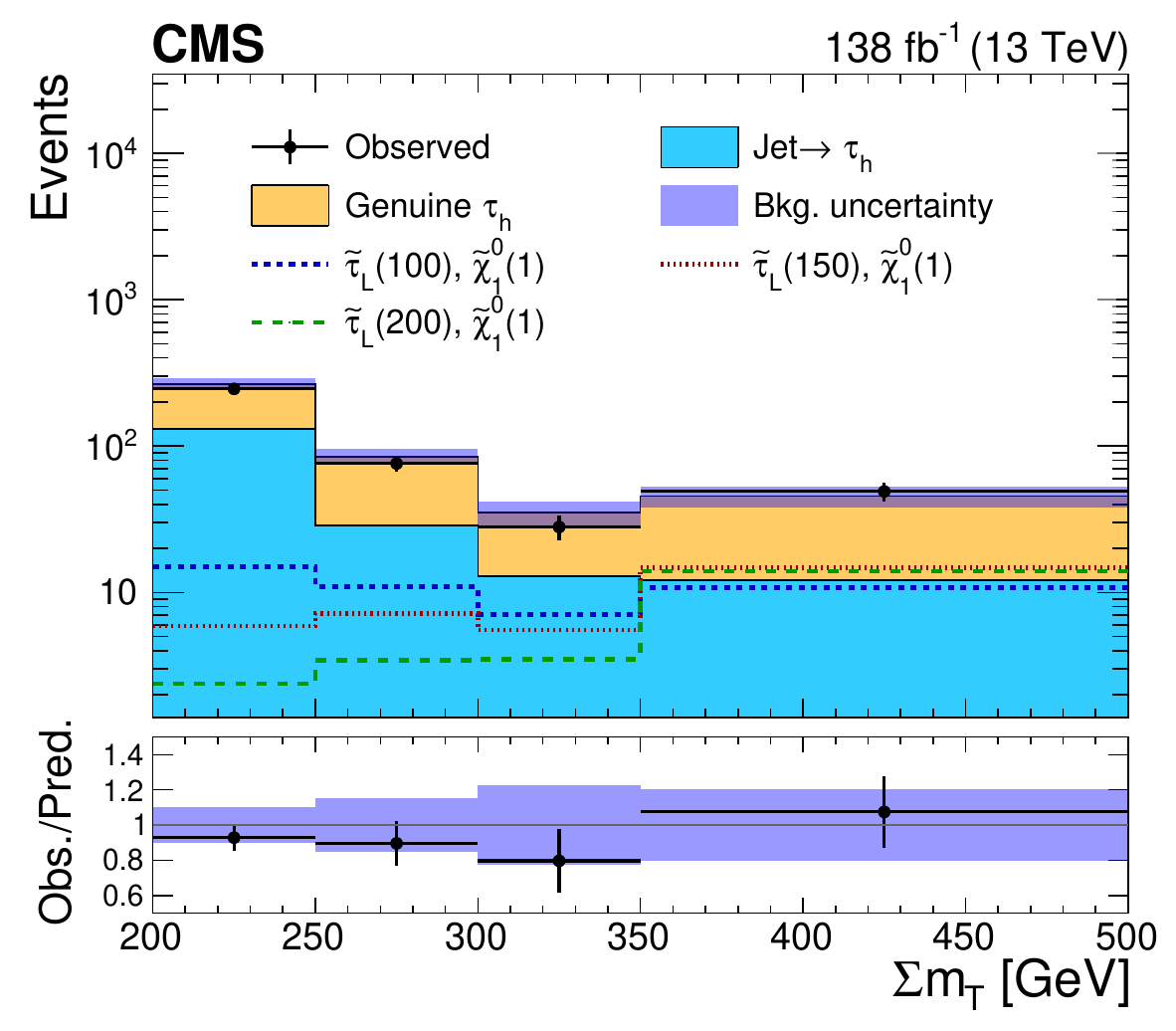} 
\caption{Highlights from the electroweakino and slepton searches: Observed and expected event yields across the signal regions in events with a $\mu^+\mu^-$ or $e^+e^-$ pair and an additional $\tau_h$ candidate from the multilepton electroweakino search (left); distribution of the jet mass for W/Z-tagged AK8 jets in the b-veto signal region from the hadronic electroweaking search (center), and $\sum m_J$ distribution for events passing the preselection for prompt signal regions in the direct stau search (right)}
\label{fig:highlights_ewk}
\end{figure}

An inclusive multilepton search targeted direct $\cone \ntwo$ production with moderate $\Delta m(\cone / \ntwo, \none)$, covering direct decays to $\none$, cascade decays via sleptons or sneutrinos, and direct $\none \none$ production with decays to a Higgs or Z boson and a gravitino~\cite{CMS:2021cox}. Events are categorized into channels with three or four leptons, allowing up to two hadronically decaying $\tau$ leptons, or into two same-sign light lepton channels. Each category is further divided into bins using a variety of invariant mass, $m_T$, $m_{T2}$, and momentum variables. Figure~\ref{fig:highlights_ewk} (left) shows the observed and expected events yields across the signal regions in events with a $\mu^+\mu^-$ or $e^+e^-$ pair and an additional $\tau_h$ candidate. In the three-lepton category with an opposite-sign lepton pair, a parametric neural network, trained as a function of $\Delta m$, was used to suppress backgrounds. In addition, orthogonal search regions are defined to provide model-independent results, facilitating reinterpretation. Depending on the model assumptions, the search excluded charginos and neutralinos with masses up to between 300 and 1450 GeV. 

A more recent search explored the complementary oppositely charged dilepton plus $\met$ final state, targeting chargino pair production with $\cone \to W\none$ and $\cone \to \nu\tilde{\ell} / \tilde{\nu} \ell \to \nu \ell \none$, as well as slepton pair production with $\tilde{\ell} \to \ell \none$~\cite{CMS-PAS-SUS-23-002}. Signal regions are defined by the number of jets, b jets, $\met$, and $m_{T2}(\ell\ell)$. The analysis excluded charginos and neutralinos up to 1100 GeV and 480 GeV, respectively, for the cascade decays, while observed sensitivity to the direct decay remains limited.  Sleptons are excluded up to 700 GeV and neutralinos up to 360 GeV.

A complementary search targeted $\tilde{\chi}^\pm_1 \tilde{\chi}^\mp_1$, $\cone \ntwo$, and $\ntwo \tilde{\chi}^0_3$ production with direct decays to $W\none$, $Z\none$, and $H\none$ in fully hadronic final states with WW, WZ, WH and large $\met$~\cite{CMS:2022sfi}. Hadronically decaying W, Z, and Higgs bosons are identified using the AK8 jet soft-drop mass together with the DeepAK8 DNN-based tagger.  Events are categorized according to the number of b jets and the presence of W, Z, or Higgs bosons, and further divided into $\met$ bins. Figure~\ref{fig:highlights_ewk} (center) shows distribution of the jet mass for W/Z-tagged AK8 jets in the b-veto signal region. The search excluded wino-like mass-degenerate $\ntwo$ and $\cone$ up to 870 GeV and 960 GeV for the $\ntwo \to Z\none$ and $\ntwo \to H\none$ decay modes, respectively. Higgsino-like mass-degenerate $\none$, $\ntwo$, and $\cone$ are excluded for masses between 300 and 650 GeV.  

A more targeted search focused on $\ntwo\cone \to H\none, W\none \to \ell\nu\none, b\bar{b}\none$ in the exclusive final state with one lepton, two b jets from a Higgs boson decay, and $\met$~\cite{CMS:2021few}. Boosted Higgs bosons are identified with DeepAK8, while resolved candidates are reconstructed via the $m_{b\bar{b}}$ requirement. Backgrounds were suppressed using variables such as $m_T$ and the cotransverse mass $m_{CT}$. Signal regions are defined according to the Higgs boson topology (boosted or resolved), jet multiplicity, and $\met$. The search excluded charginos with masses below 820 GeV for a $\none$ lighter than 200 GeV, and $\none$ masses up to about 350 GeV for a $\cone$ mass near 700 GeV, extending Run 1 limits by up to 350 GeV in chargino mass and 250 GeV in neutralino mass.  The above searches (except the recent opposite charge dilepton search) became a part of a grand electroweakino combination study, as we will see in Section~\ref{sec:interp_sms}. 

A recent search, the first of its kind at the LHC, probed a heavy Z' boson decaying to a pair of charginos, each subsequently decaying to a W boson and a neutralino, in the oppositely charged dilepton plus $\met$ channel~\cite{CMS-PAS-SUS-23-006}. To capture the high boost from the heavy Z' decay, the analysis imposes higher lepton $p_T$ thresholds than in other searches. A parametric DNN, taking as input a wide range of mass, momentum, and angular variables constructed from the two leptons and $\met$, is used for signal discrimination. The DNN output bins in the $ee$, $\mu\mu$, and $e\mu$ channels serve as the final signal regions. For $m_{\cone} = 2m_{\none}$, Z' masses are excluded up to 3.5 TeV, while for a Z' mass of 2.9 TeV, $\cone$ masses are excluded from 0.4 to 1.4 TeV.

A dedicated search targeted stau pair production in decays to $\tilde{\tau} \to \tau \none$ or $\tilde{\tau} \to \tau \tilde{G}$, the latter producing long-lived signatures~\cite{CMS:2022syk}. Final states with two hadronically decaying $\tau$ leptons and $\met$ were considered, with $\tau_h$ candidates identified using the DNN-based multiclass DeepTau algorithm. The analysis includes both prompt and displaced categories, the latter designed for the long-lived scenario, and defines 31 disjoint signal regions. In the prompt case, regions are binned in $\sum m_T = m_T(\tau_h^1) + m_T(\tau_h^2)$, shown in Figure~\ref{fig:highlights_ewk} (right), $m_{T2}$, $n_{\text{jets}}$, and the $p_T$ of the leading $\tau_h$. The displaced case is defined by criteria on the significance of the $\tau$ impact parameter relative to the primary vertex in the transverse plane ($d_{xy}$), required to be above 5, and the absolute value of its three-dimensional impact parameter (IP3D) required to be above 100 mm, and is binned in the trailing $\tau$ $p_T$.  For left-handed stau production, $\tilde{\tau}$ masses between 115 and 340 GeV are excluded. For a lifetime corresponding to $c\tau_0 = 0.1$ mm, masses between 150 and 220 GeV are excluded.

\subsection{Searches for compressed mass spectra}
\label{sec:compressed}

\begin{figure}[htbp]
\centering
\includegraphics[width=0.32\textwidth]{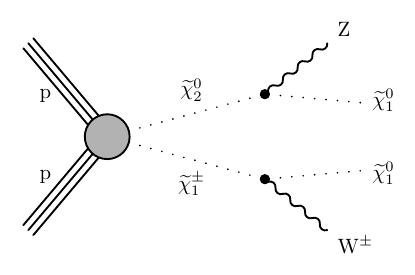} 
\includegraphics[width=0.32\textwidth]{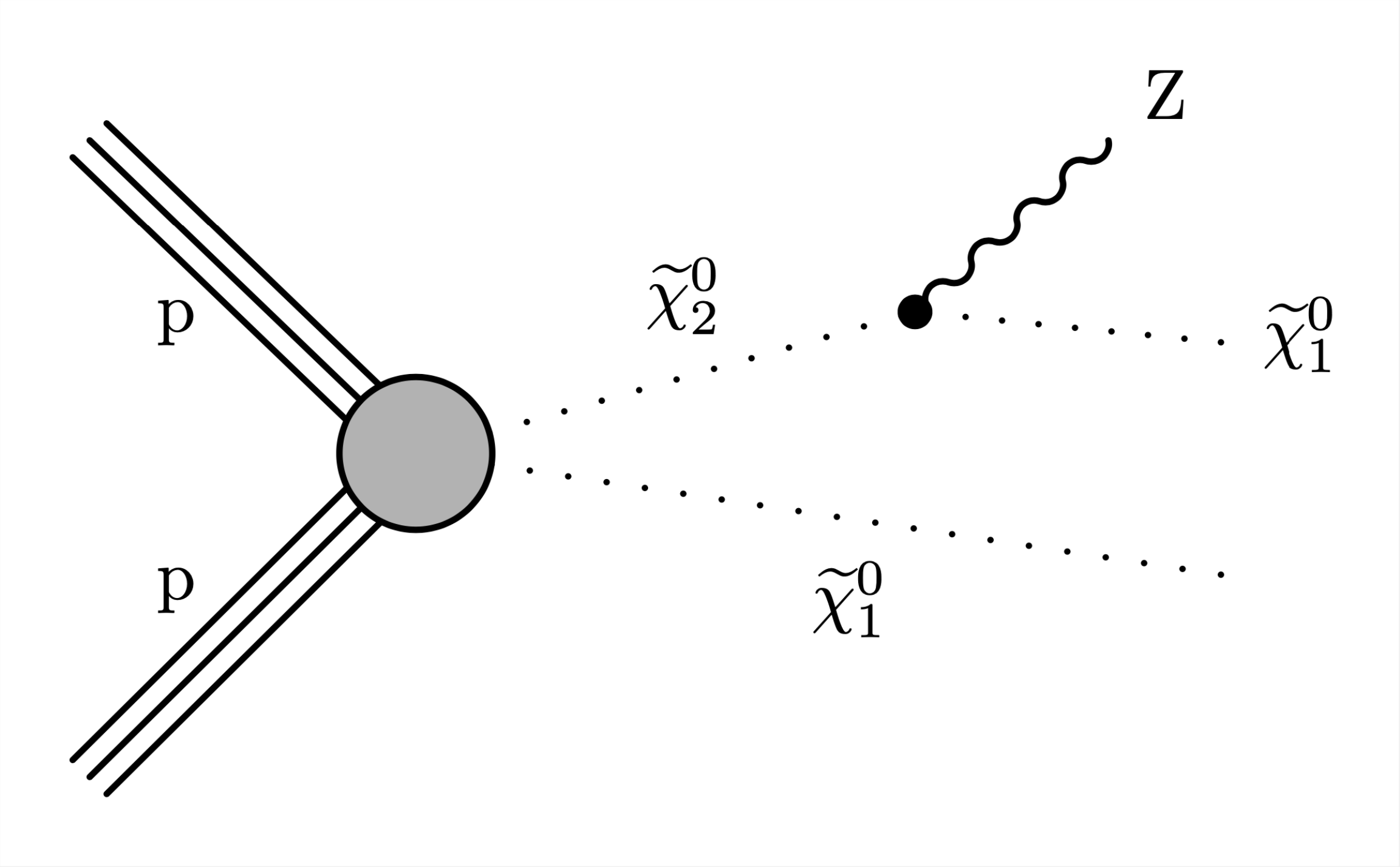} 
\includegraphics[width=0.32\textwidth]{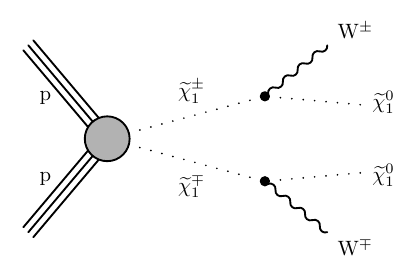} 
\includegraphics[width=0.32\textwidth]{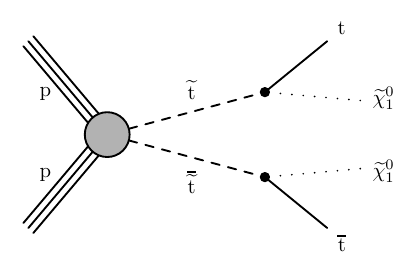} \includegraphics[width=0.32\textwidth]{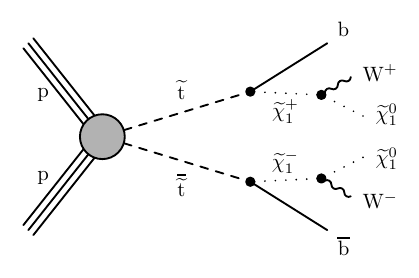} \includegraphics[width=0.32\textwidth]{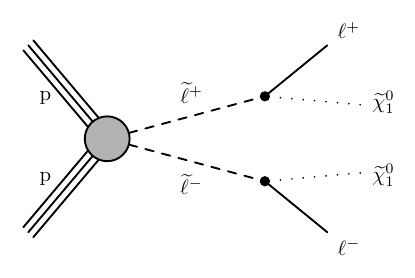} 
\caption{Diagrams for a nonexhaustive example set of simplified models covered by searches targeting compressed mass spectra.}
\label{fig:diag_compressed}
\end{figure}

Compressed SUSY scenarios, where the mass difference between the next-to-lightest and lightest SUSY particle is small, pose a particular challenge for analysis. They produce low-momentum visible objects and low $\met$, making signal reconstruction and background suppression difficult. Depending on the mass difference, signatures can be prompt or long-lived. These scenarios are nevertheless well-motivated theoretically, so CMS has placed special emphasis on them, developing strategies that range from using ISR jets to incorporating dedicated kinematic variables and reconstructing low-momentum particles or long-lived particles.  Figure~\ref{fig:diag_compressed} shows a nonexhaustive example set of simplified models covered by the searches targeting compressed mass spectra.

\begin{figure}[H]
\centering
\raisebox{0.2\height}{\includegraphics[width=0.65\textwidth]{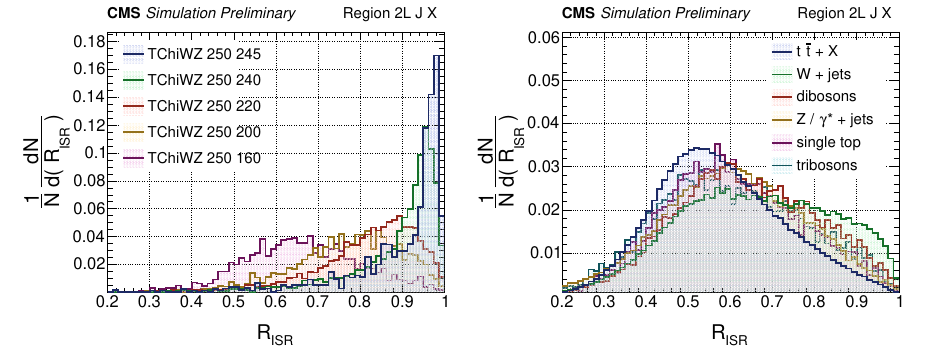}} \quad
\includegraphics[width=0.26\textwidth]{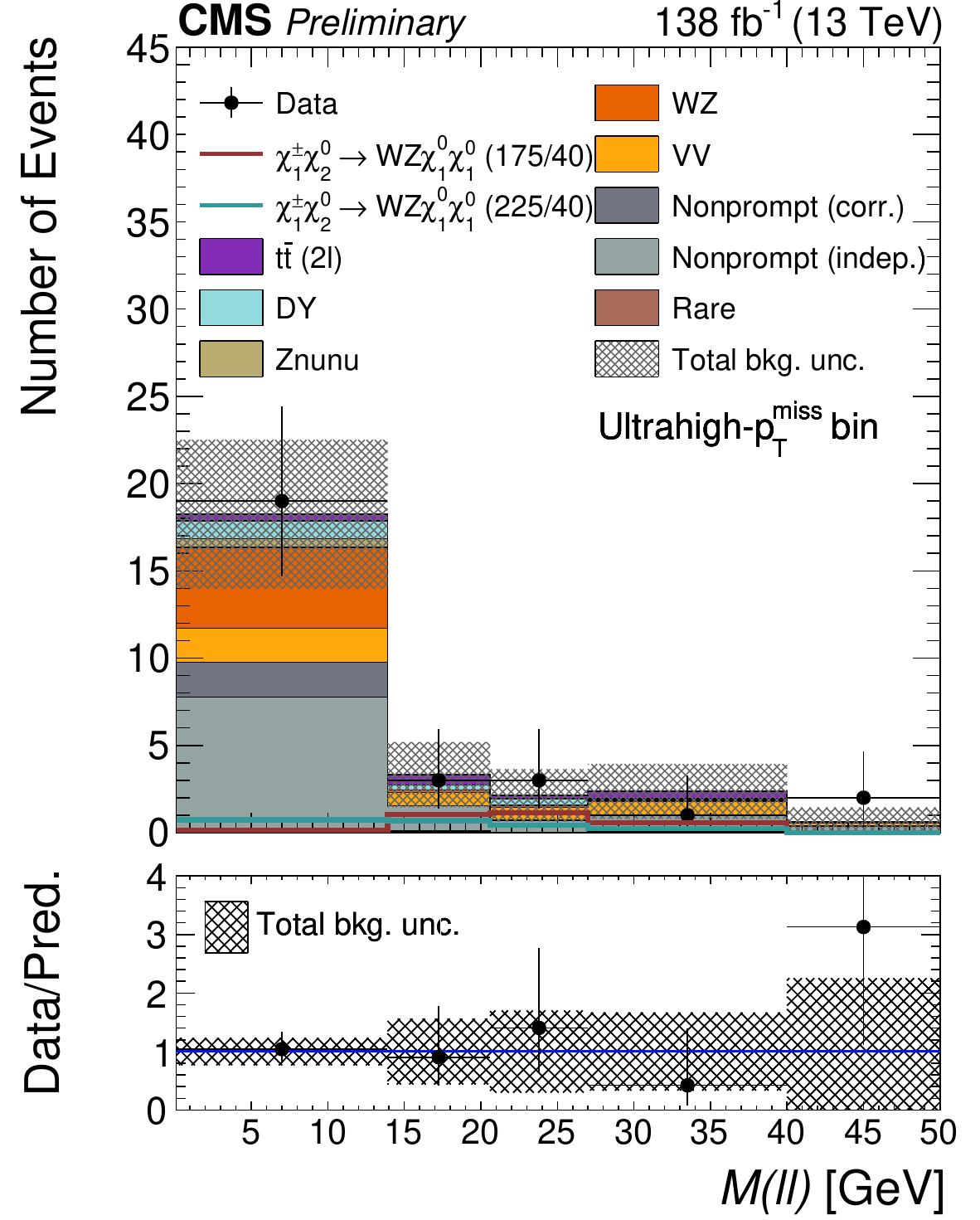} 
\caption{Highlights from searches for compressed spectra: Distributions of the RJR variable $R_{\text{ISR}}$ for simulated signal and background events from the RJR analysis (left and center); postfit $m(\ell \ell)$ distribution in the dimuon signal region for the ultrahigh \met selection from the latest soft opposite charge dilepton and trilepton analysis (right). }
\label{fig:highlights_compressed}
\end{figure}

We start with a generic search that focused on events with a high $p_T$ system from ISR jets recoiling against a potential sparticle system with significant $\met$~\cite{CMS-PAS-SUS-23-003}. The analysis makes use of the recursive jigsaw reconstruction (RJR) algorithm.  Distributions of the RJR variable $R_{\text{ISR}}$ for simulated signal and background events are shown in Figure~\ref{fig:highlights_compressed} (left and center).  The analysis implements a complex categorization based in 0-, 1-, 2-, and 3-lepton final states, based on number of leptons, jets, b tags.  Events in these categories are further subdivided based on RJR and other kinematic variables sensitive to the sparticle masses and mass splittings. The analysis performs an exceptionally involved likelihood fit over 2443 bins in 392 categories.  As a result, it  probed top squark masses up to 780, 620, and 660 GeV, and excluded 750, 550, and 520 GeV for decays via $t\none$, $b\cone$, and $c\none$, respectively, covering mass differences from about 60--175 GeV, 35--140 GeV, and  10--60 GeV. For electroweak production, the anaysis excluded chargino--neutralino masses up to 325 GeV (wino) and 175 GeV (higgsino) over $\Delta m$ ranges of roughly 8--65 GeV and 3--50 GeV, and chargino pair production up to 490 GeV for $\Delta m \sim 55$ GeV in slepton-mediated decays. Slepton masses are constrained up to 270 GeV, with sensitivity maintained across $\Delta m$ from a few GeV to about 80 GeV.

Next we move to two searches targeting electroweakinos with higher compression, down to mass splittings of about 1 GeV, in final states with either two oppositely charged same-flavor soft leptons or three soft leptons, each accompanied by $\met$ \cite{CMS:2021edw, CMS-PAS-EXO-23-017}.
The latter analysis~\cite{CMS-PAS-EXO-23-017} updated~\cite{CMS:2021edw} with improved reconstruction, extending sensitivity by reconstructing electrons down to 1 GeV with a dedicated algorithm. In both analyses muons with $p_T$ as low as 3.5 GeV are used, along with a special dimuon + $\met$ trigger with 3 GeV muon $p_T$ thresholds.
Events are categorized into $\mu\mu$ low-$\met$, $ee/\mu\mu$ high-$\met$, $3\ell$ low-$\met$, and $3\ell$ high-$\met$ regions, each further binned in $m_{\ell\ell}$ with binning optimized for different signals and mass splittings. 
Figure~\ref{fig:highlights_compressed} (right) shows postfit $m(\ell \ell)$ distribution in the dimuon signal region for the ultrahigh \met selection from~\cite{CMS-PAS-EXO-23-017}.
The interpretation was done in two simplified SUSY scenarios. In the higgsino model, where \cone, \ntwo, and \none are nearly mass-degenerate, the more sensitive Ref.~\cite{CMS-PAS-EXO-23-017} excludes next-to-LSP masses up to 225 GeV for $\Delta m(\ntwo, \none) = 10$ GeV, and reaches $m_{\ntwo} = 100$ GeV for $\Delta m = 1$ GeV thanks to the dedicated low-$p_T$ electron reconstruction. In the wino-bino model, where the wino-like \cone and \ntwo are mass-degenerate and decay to a bino-like \none,  \cone masses are excluded up to 310 GeV for $\Delta m = 10$~GeV, and up to 100 GeV for $\Delta m = 0.7$~GeV. The earlier analysis~\cite{CMS:2021edw} reported a local $2.4\sigma$ excess in the wino-bino model around $\ntwo$ mass 125 GeV and mass splitting 40 GeV, arising from data excesses over the background prediction in four bins across different signal regions. The latest analysis~\cite{CMS-PAS-EXO-23-017} has so far presented only preliminary results, which are currently being refined. 

A more targeted search looked at the same final states above, but focusing on Higgsinos with mass splitting of 0.5--5 GeV, and probing them in final states with either two muons or a reconstructed lepton (muon or electron) and an isolated track, large $\met$, and an ISR jet~\cite{CMS-PAS-SUS-24-003}.  The selection focuses on cases where the lepton $p_T$ or the opening angle between the leptons is particularly small, and uses multivariate discriminants to suppress SM backgrounds.  The dimuon channel is designed to be orthogonal to those in~\cite{CMS-PAS-EXO-23-017} by requiring the dimuon angular separation to be smaller than 0.3.  Tracks in the isolated (exclusive) track plus lepton category are selected by a track picking BDT.  Another BDT trained based on variables composed of leptons, tracks, and $\met$ is used to define signal regions.  The search excluded higgsino masses up to 145 GeV for $\Delta m = 4$ GeV, and probed splittings down to 1.5 GeV for $m_{\ntwo} = 100$ GeV, with a modest $\sim 2.2 \sigma$ local excess observed in the most sensitive signal regions.

Another targeted search probed even smaller chargino--neutralino mass splittings, down to about 0.3 GeV, by selecting events with a soft, slightly displaced track accompanied by large $\met$~\cite{CMS-PAS-SUS-24-012}. In the higgsino scenario, where \cone, \ntwo, and \none are nearly mass-degenerate, the analysis uses a parameterized neural network to separate signal from background over a range of track kinematics and topologies. It excludes charginos up to 185 GeV for $\Delta m = 0.55$ GeV, and probes mass splittings of 0.33--1.2 GeV for a 100 GeV chargino, currently setting the most stringent limits in this regime, placing direct pressure on natural SUSY dark matter hypothesis.

Finally, a dedicated search targeted compressed $\tilde{\tau}$ scenarios, motivated by models where coannihilation between the $\tilde{\tau}$ and \none accounts for the observed dark matter relic density~\cite{CMS:2019zmn}. The analysis selects events with exactly one soft, hadronically decaying $\tau$ lepton, large $\met$, and a high-$p_T$ ISR jet to boost the system. It is optimized for scenarios with $\Delta m(\tilde{tau}, \none) \leq 50$~GeV, where the stau is produced either directly or via decays of $\cone$ and $\ntwo$. For $\Delta m(\cone, \none) = 50$~GeV with a 100\% branching fraction to $\tilde{\tau}\nu_\tau \to \tau \none \nu_\tau$, \cone masses up to 290 GeV are excluded, surpassing the previous limits from LEP.

\subsection{Searches for RPV / stealth SUSY}

\begin{figure}[htbp]
\centering
\includegraphics[width=0.32\textwidth]{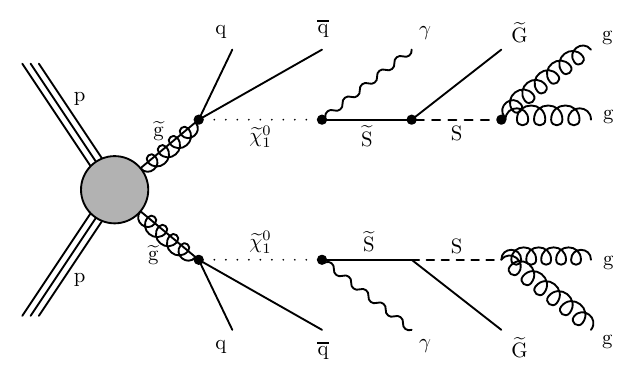} 
\includegraphics[width=0.32\textwidth]{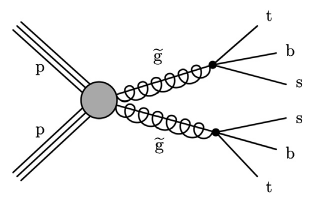} 
\includegraphics[width=0.32\textwidth]{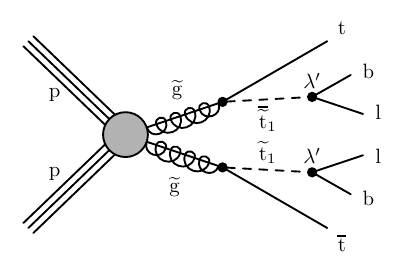} 
\includegraphics[width=0.32\textwidth]{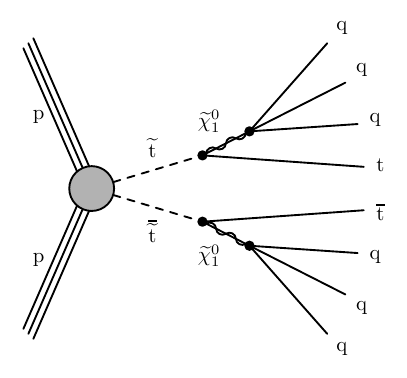} \includegraphics[width=0.32\textwidth]{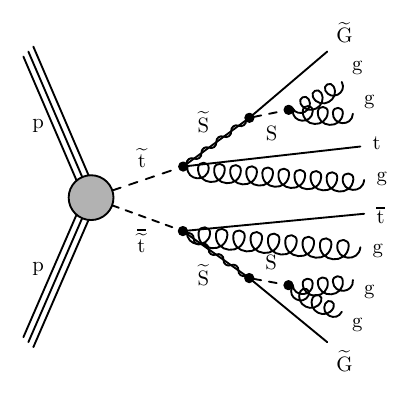} 
\raisebox{0.3\height}{\includegraphics[width=0.32\textwidth]{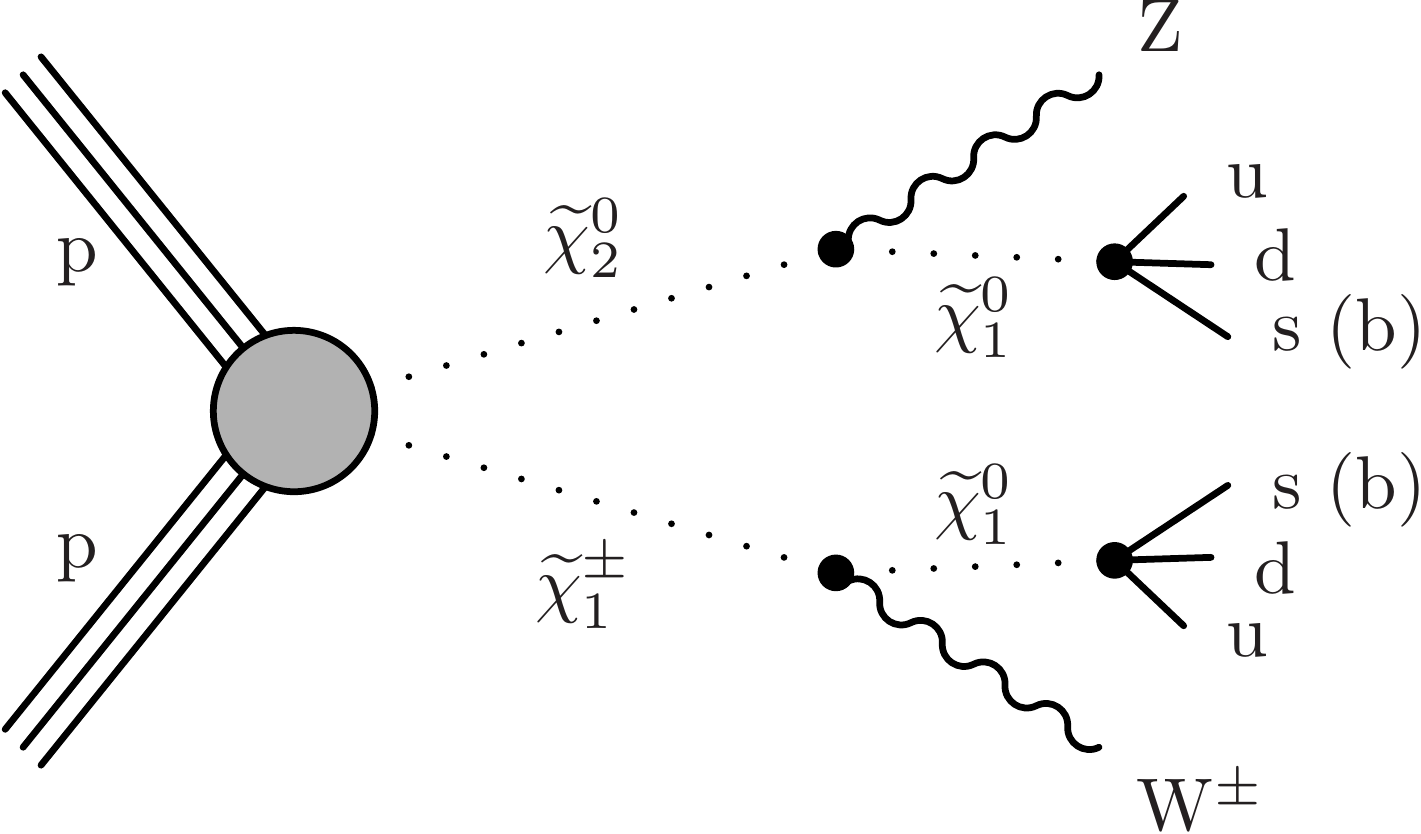}} 
\caption{Diagrams for a nonexhaustive example set of simplified models covered by searches for RPV and stealth SUSY.}
\label{fig:diag_rpvstealth}
\end{figure}

The RPV/stealth SUSY analyses target scenarios with little or no genuine $\met$, where the LSP decays to SM particles or to nearly mass-degenerate states. These models typically lead to high-multiplicity hadronic or multi-lepton final states, sometimes with photons, requiring dedicated background rejection and estimation methods suitable to $\met$  selections.  They make extensive use of jet substructure, event level mass variables, and data-driven background backtround estimation techniques.  CMS Run 2 searches targeted RPV/stealth gluinos, top squarks and electroweakinos, and share the common feature of no direct threshold requirement on $\met$.  Figure~\ref{fig:diag_rpvstealth} shows a nonexhaustive example set of simplified models covered by searches for RPV and stealth SUSY.

\begin{figure}[H]
\centering
\includegraphics[width=0.29\textwidth]{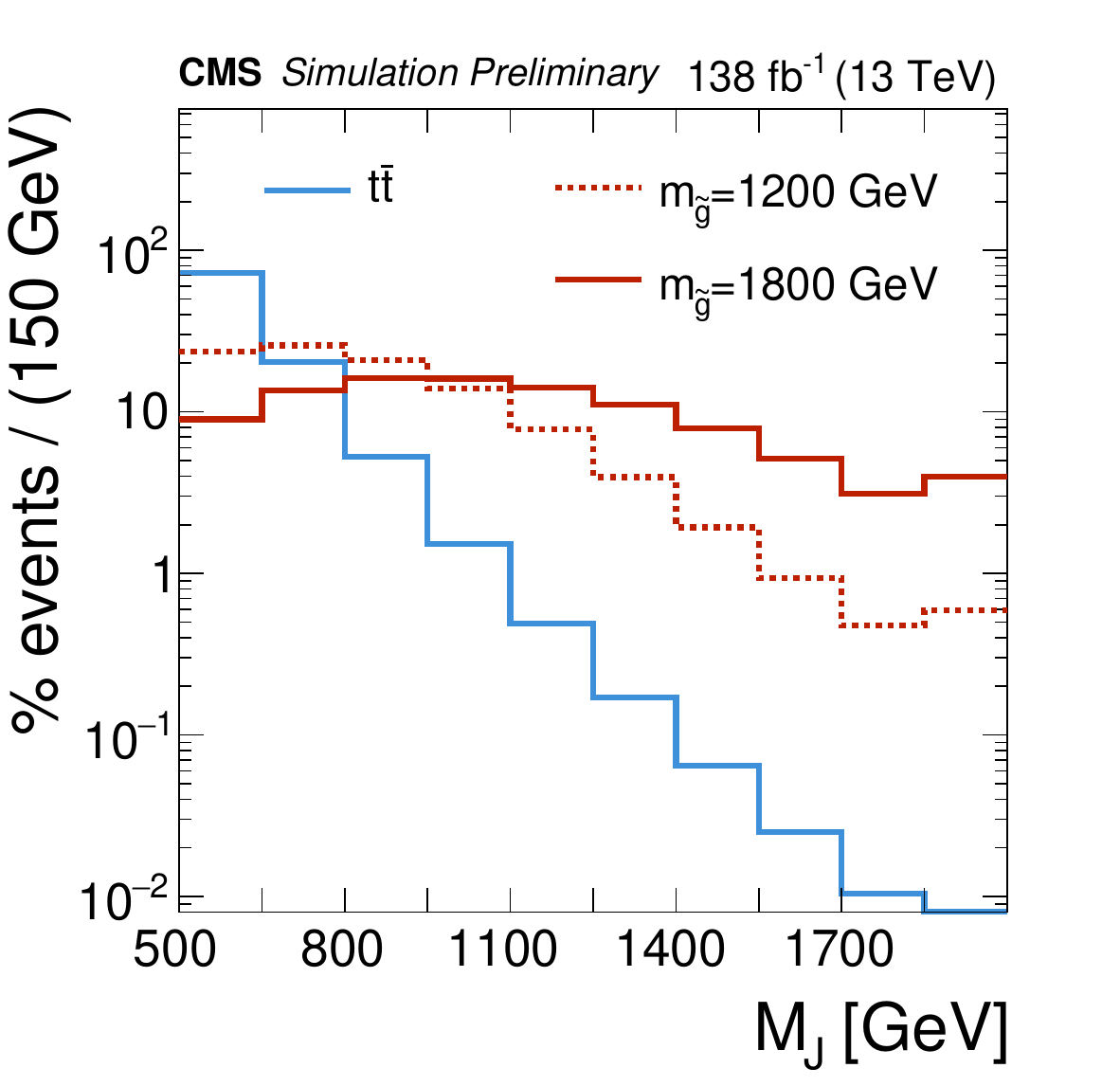}
\raisebox{0.1\height}{\includegraphics[width=0.36\textwidth]{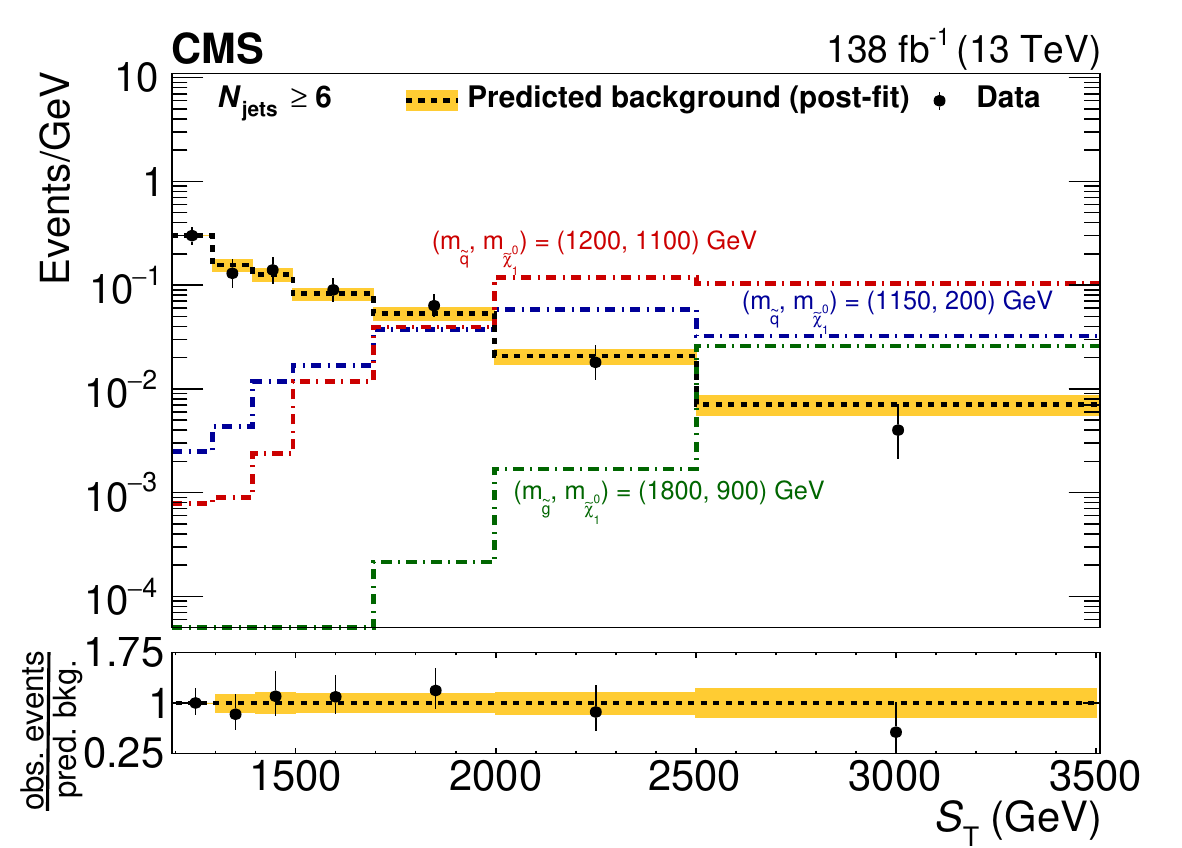}} 
\raisebox{0.05\height}{\includegraphics[width=0.33\textwidth]{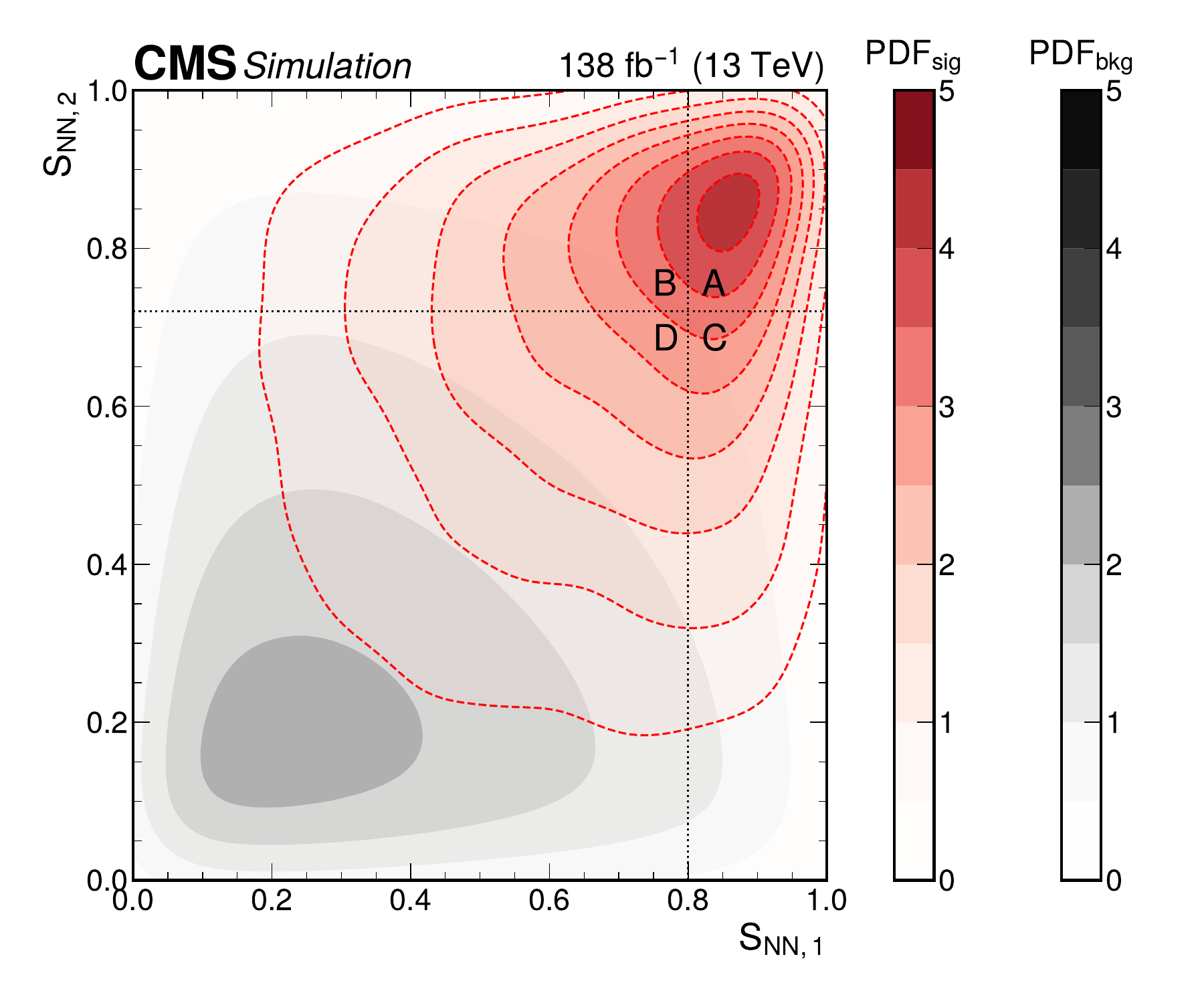}} 
\caption{Highlights from RPV and stealth SUSY searches: Normalized distributions of $M_J$ for simulated signal and background events from the one lepton plus multijets search (left); distributions of $S_T$ comparing data and postfit background predictions for $n_{\text{jets}} > 6$ from the 2 photons plus four jets stealth SUSY search (center); two-dimensional probability distributions of the two ABCDiscoTEC neural network outputs for simulated RPV signal and $t\bar{t}$ background events from the latest RPV top squark search (right).}
\label{fig:highlights_rpvstealth}
\end{figure}

The first in our list is a search for RPV gluino pair production in single-lepton events with high jet and b-jet multiplicities, targeting $\tilde{g} \to tbs$ decays via baryon number violating couplings~\cite{CMS-PAS-SUS-21-005}. Signal separation is achieved using large-radius ($R = 1.2$) jets clustered from AK4 jets and the lepton, with $M_J$, the scalar sum of their masses, serving as the main discriminant. Figure~\ref{fig:highlights_rpvstealth} (left) compares normalized $M_J$ distributions for simulated signal and background. Preselection requires at least four jets, $H_T > 1200$ GeV, and $M_J > 500$ GeV. Events are categorized in $n_{\text{jets}}$ and $n_b$ bins into signal and control regions, and further binned in $M_J$. A simultaneous fit was performed across all bins using $M_J$ templates derived from simulation and corrected through data-simulation comparisons in dedicated control regions. The analysis excluded gluinos up to 1890 GeV.

The next search targets stealth SUSY in events with two photons and at least four jets~\cite{CMS:2023zuu}. Events are 
required to have high $S_T$, the scalar sum of the transverse momenta of all reconstructed objects in the event (including $\met$), above 1200 GeV.  Selected events are partitioned into those with 4, 5, and $\ge 6$ jets, and further binned in $S_T$. Figure~\ref{fig:highlights_rpvstealth} (center) shows distributions of $S_T$ comparing data and postfit background predictions for $n_{\text{jets}} > 6$ signal region. The SM background was modeled from data using control samples, ensuring reliable predictions in the low-$\met$ regime characteristic of stealth SUSY scenarios. The results were interpreted in simplified models with gluino or squark pair production, excluding gluino masses up to 2150 GeV and squark masses up to 1850 GeV, the most stringent limits to date for these models.

The next pair of analyses in this category target top squark pair production in RPV and stealth SUSY models, with decays to two top quarks and multiple light-flavor quarks or gluons.  In the RPV scenario, each top squark decays to a top quark and an RPV neutralino, which subsequently decays into three light-flavor quarks. In the stealth scenario, each top squark decays to a top quark and a singlino $\tilde{S}$, followed by $\tilde{S} \to S\tilde{G}$ and singlet $S \to gg$. The first analysis focused on the single lepton channel, requiring at least seven jets, of which at least one is b tagged~\cite{CMS:2021knz}. Events are categorized with a neural-network-based discriminant, and the dominant $t\bar{t}$ background is constrained from data via a fit to the jet multiplicity spectrum across four bins of the NN score. This search excluded top squark masses up to 700 GeV in the RPV model and 930 GeV in the stealth SUSY model. It also observed a local excess with a significance of 2.8 standard deviations for an RPV top squark mass of 400 GeV, motivating an updated analysis. The follow-up search added 0- and 2-lepton channels and introduced the ABCDisCoTEC method, which uses two uncorrelated neural network outputs in an ABCD-style background estimation framework~\cite{CMS:2025bxo}.  Figure~\ref{fig:highlights_rpvstealth} (right) shows the two-dimensional probability distributions of the two ABCDiscoTEC neural network outputs for simulated RPV signal and $t\bar{t}$ background events. This reduced the dependence on jet multiplicity modeling and improved sensitivity, especially for low top squark masses. In the expanded dataset and channel coverage, the earlier excess was not confirmed, and the analysis excluded top squark masses up to 700 GeV in the RPV model and 930 GeV in the stealth SUSY model.

The final analysis in this category targets RPV neutralinos in multilepton events~\cite{CMS:2025wfw}. The scenario considered features $\cone\ntwo$ pair production, with the $\cone$ and $\ntwo$ decaying via $W$ and $Z$ bosons to $\none$, which subsequently decays promptly to $uds$ or $udb$ through an RPV coupling. The search exploited the distinctive jet scaling patterns expected in the signal by comparing the observed jet multiplicity distributions in events with one, two, and four leptons to those in three-lepton events. Dedicated categories are defined according to the number of $b$-tagged jets to enhance sensitivity to different flavor compositions of the RPV decay. For the considered process, the analysis excluded RPV $\none$ masses up to 275 and 180 GeV for the $uds$ and $udb$ decays.

Apart from these dedicated analyses, the inclusive search with $\ge 2$~jets, 2 same charge leptons or 3 leptons~\cite{CMS:2020cpy} and the inclusive search with boosted objects in the 0- and 1-lepton final state~\cite{CMS-PAS-SUS-23-014} described in Section~\ref{sec:inclusive}, as well as several searches with long-lived particles to be described in the next section had RPV interpretations. 

\subsection{Searches with long-lived particles}
\label{sec:LLP}

\begin{figure}[htbp]
\centering
\includegraphics[width=0.32\textwidth]{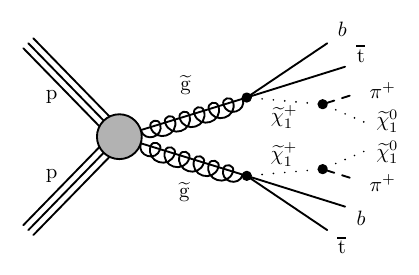} 
\includegraphics[width=0.32\textwidth]{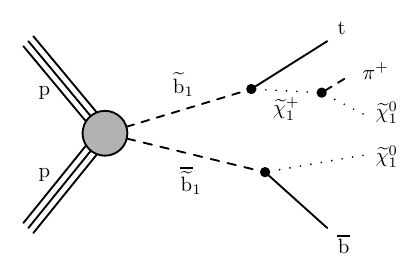} 
\includegraphics[width=0.32\textwidth]{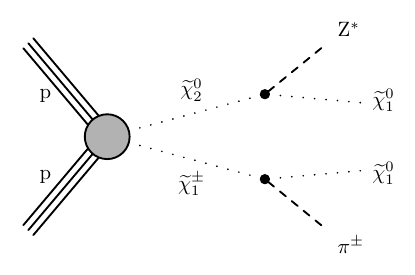} 
\includegraphics[width=0.32\textwidth]{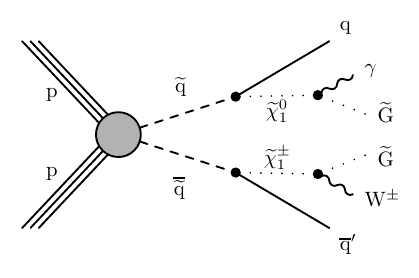} 
\includegraphics[width=0.32\textwidth]{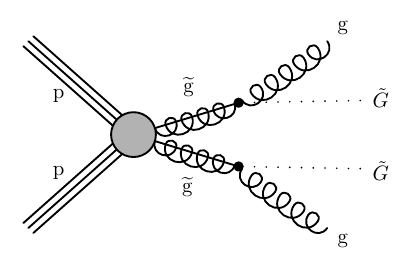} 
\includegraphics[width=0.32\textwidth]{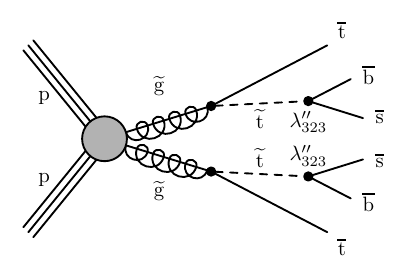}
\includegraphics[width=0.32\textwidth]{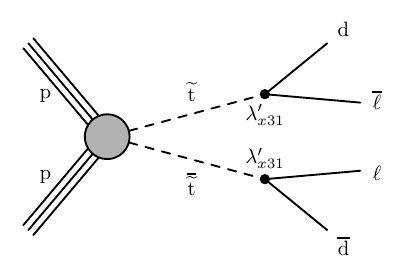} 
\includegraphics[width=0.32\textwidth]{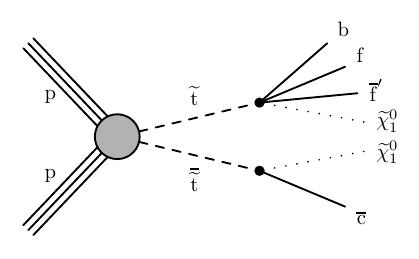} 
\includegraphics[width=0.32\textwidth]{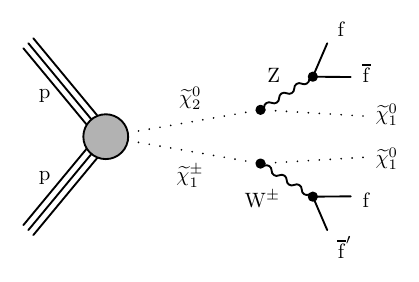} 
\caption{Diagrams for a nonexhaustive example set of simplified models covered by the LLP searches.}
\label{fig:diag_llp}
\end{figure}

Searches with long-lived particles (LLPs) add a new dimension to the search for SUSY. They target scenarios where new states have macroscopic lifetimes, leading to unconventional signatures that can be significantly displaced from the primary interaction point, delayed in time, or appear as anomalous ionization patterns. In SUSY frameworks, LLPs can arise from weakly coupled lightest superpartners, small mass splittings, or suppressed decay modes, and are realized in a wide range of models such as split SUSY, gauge-mediated SUSY breaking, and coannihilation scenarios. CMS Run 2 analyses probed this space through a rich set of techniques, covering signatures from highly ionizing tracks and disappearing tracks to displaced or delayed jets, photons, and vertices.  Figure~\ref{fig:diag_llp} shows a nonexhaustive example set of simplified models covered by the LLP searches.

\begin{figure}[H]
\centering
\raisebox{0.1\height}{\includegraphics[width=0.34\textwidth]{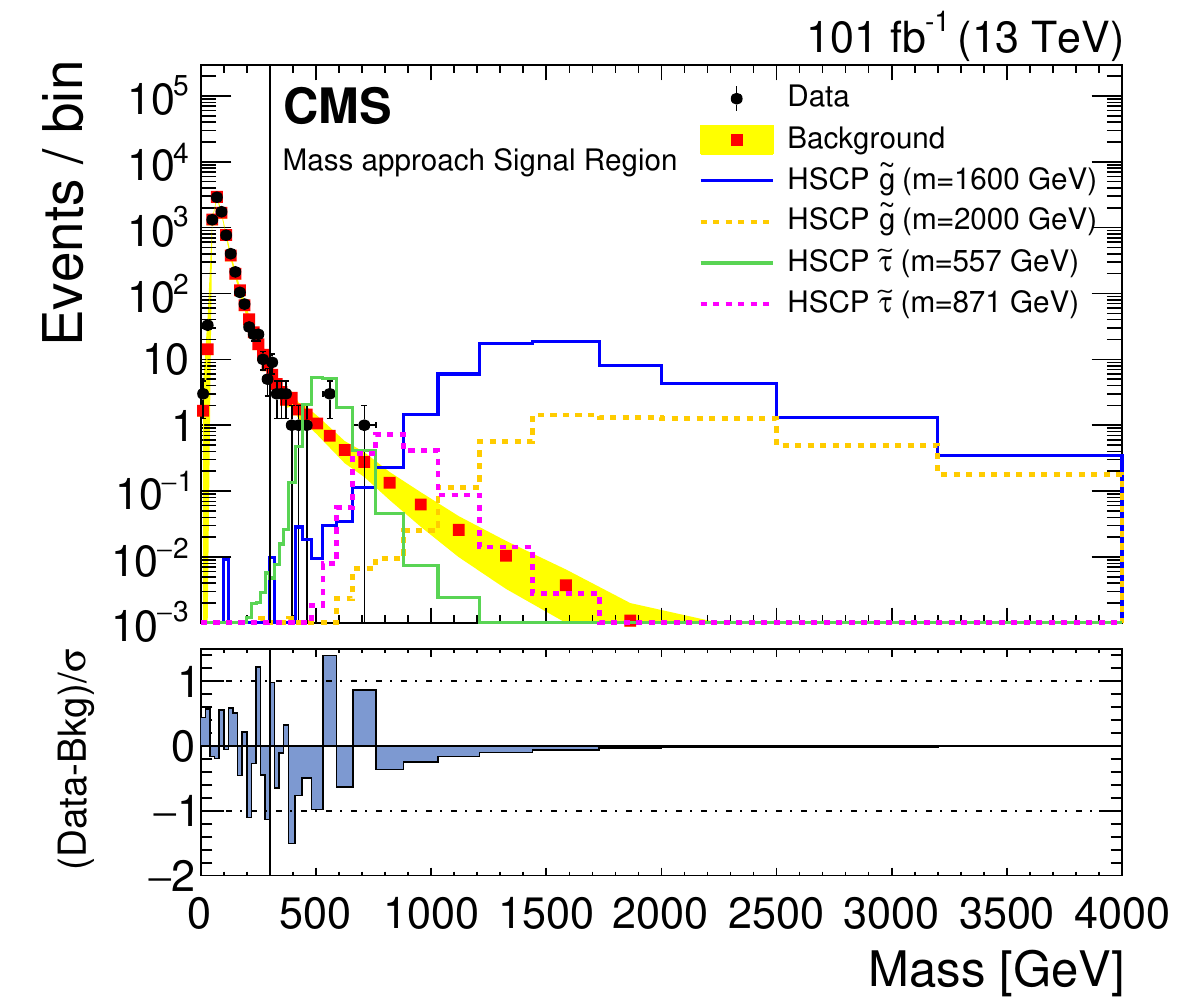}} \quad
\includegraphics[width=0.38\textwidth]{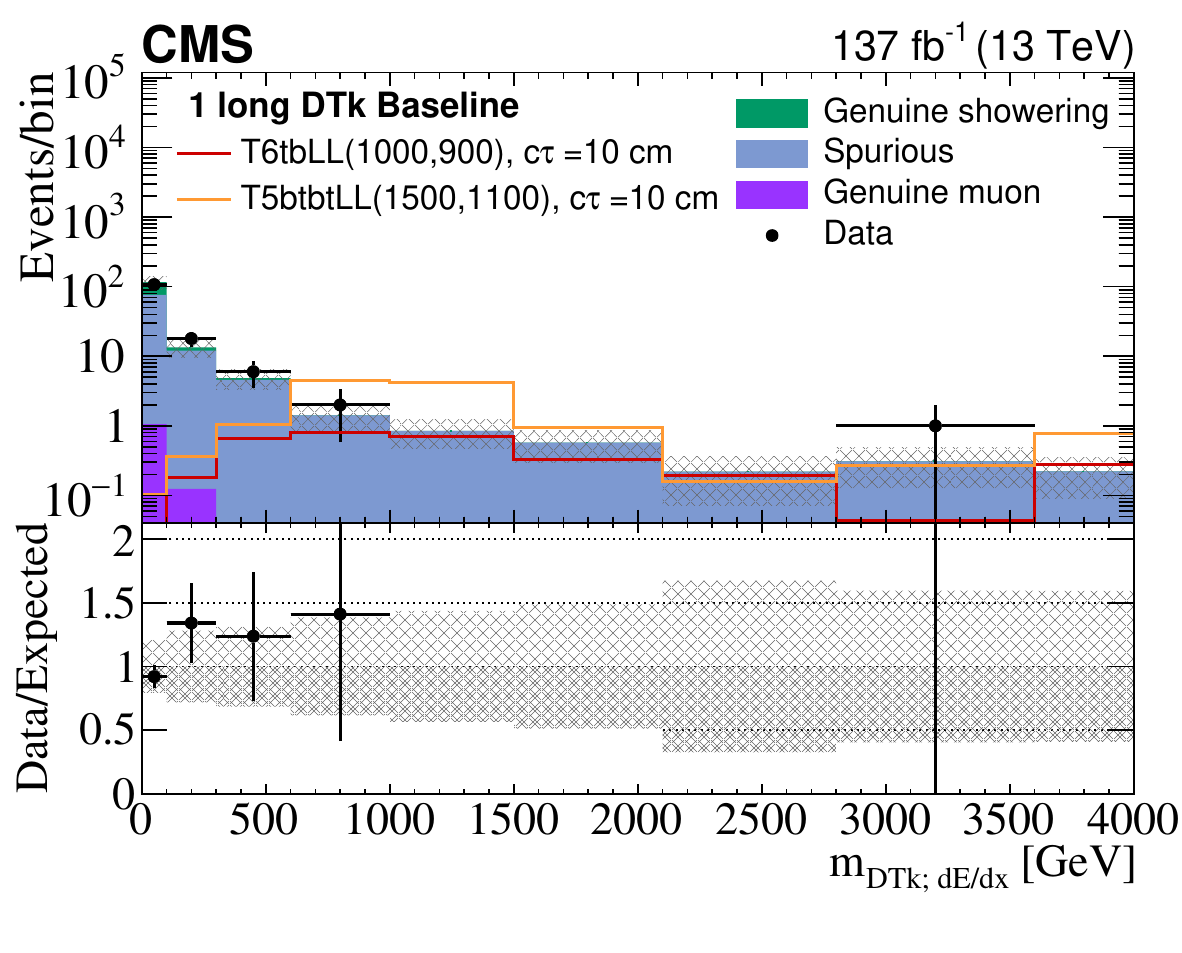}
\includegraphics[width=0.34\textwidth]{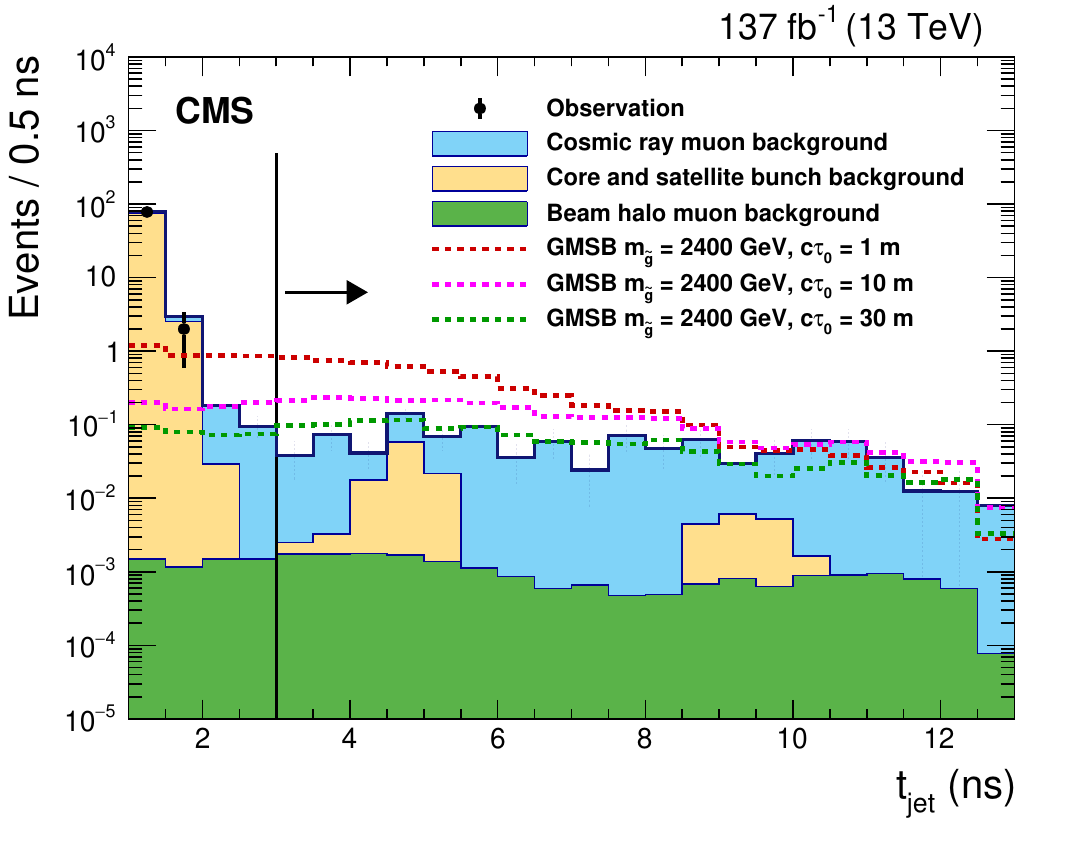} \quad
\raisebox{0.1\height}{\includegraphics[width=0.27\textwidth]{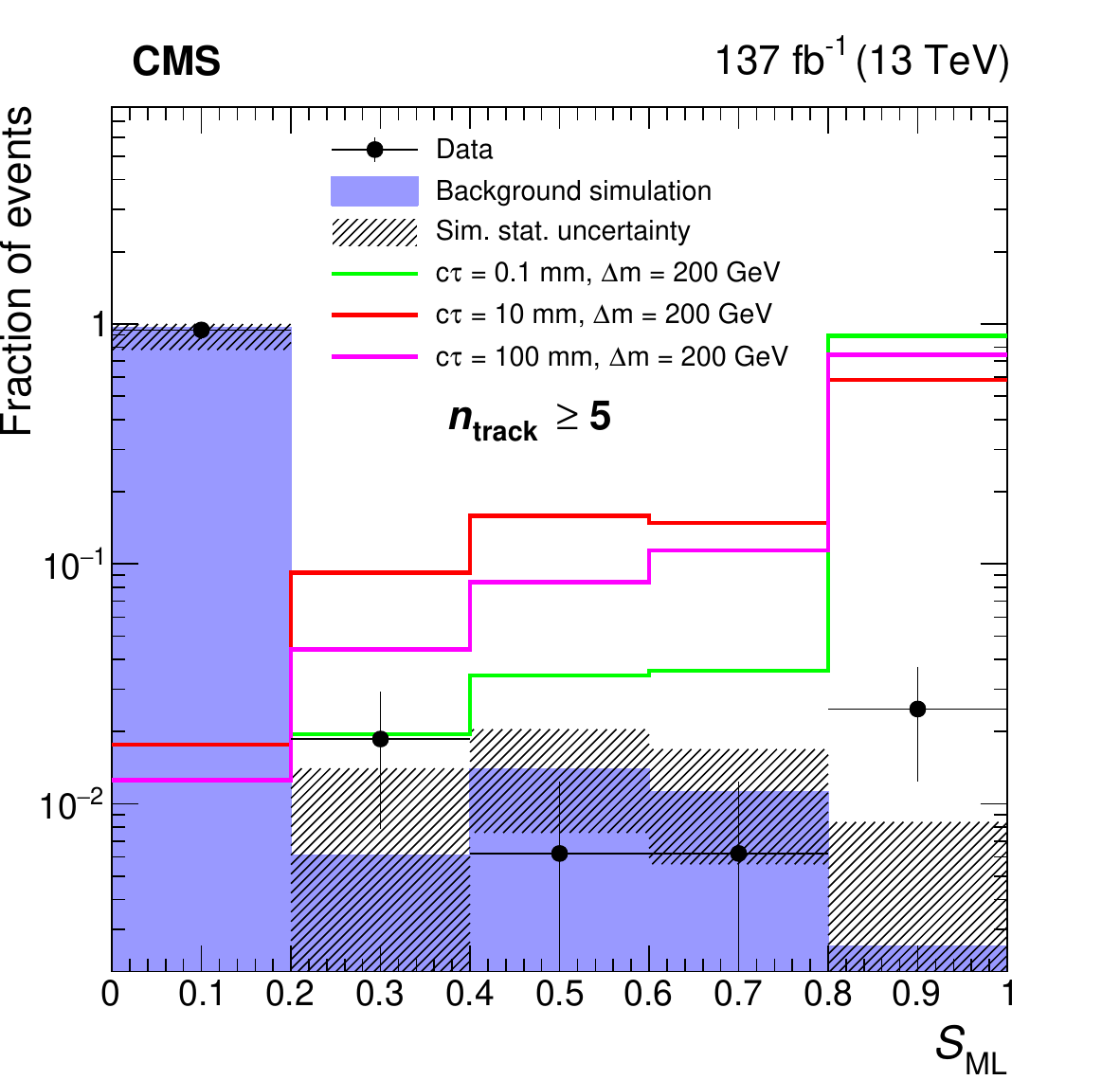}} 
\includegraphics[width=0.35\textwidth]{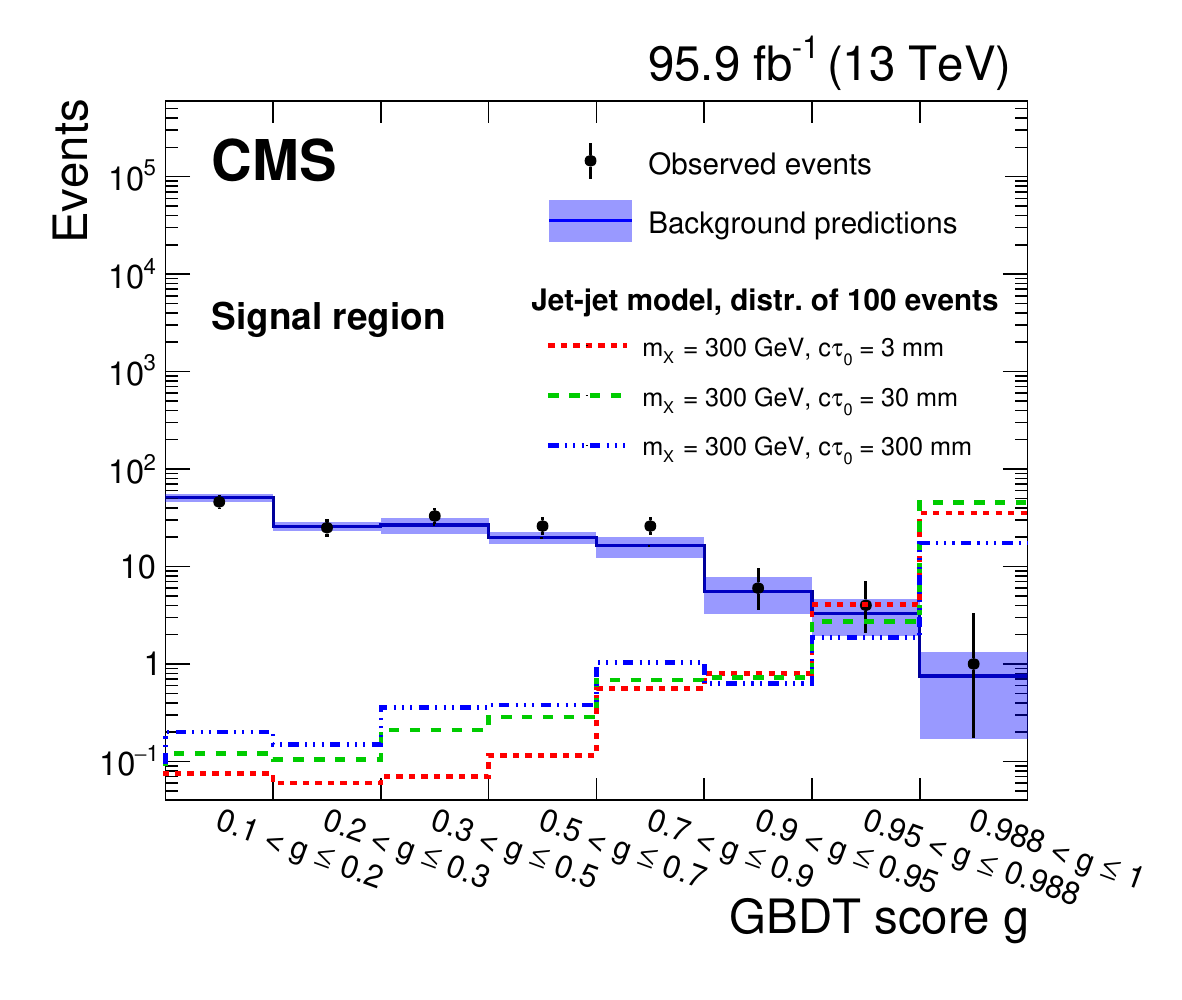} 
\caption{Highlights from searches with LLPs: Mass spectrum in the signal region from the HCSP search (top left); distribution of disappearing track mass obtained from dE/dx in the 1 long track baseline region from the 0- and 1-lepton disappearing track analysis (top-right); jet timing distribuion comparing data and background prediction from the delayed jet analysis (bottom-left); distribution of the interaction network score $S_{\text{ML}}$ for a selection with $n_{track} \ge 6$ from the later displaced vertex analysis (bottom-center); and distribution of the GBDT score comparing data with predicted backgrounds for the displaced jet analysis (bottom-right). }
\label{fig:highlights_llp}
\end{figure}

The first analysis in the LLP category searched for heavy stable charged particles (HSCPs) that would traverse the detector as slowly moving, highly ionizing tracks~\cite{CMS:2024nhn}. It targeted pair production of gluino R-hadrons, top squark R-hadrons, and long-lived staus, where R-hadrons are bound states of a heavy SUSY parton (such as a gluino or squark) with ordinary quarks or gluons, which can carry electric charge and travel measurable distances before decaying. The search exploited anomalously high ionization energy loss (dE/dx) in the silicon tracker, using two complementary approaches: an ionization-based selection that relies only on pixel or strip detector information, and a mass-based method combining dE/dx with the track momentum to calculate the HSCP candidate mass, which serves as the main discriminant. Figure~\ref{fig:highlights_llp} (top-left) shows the mass distribution in the signal region. Both approaches achieve similar sensitivity. The results set the most stringent constraints to date on several HSCP scenarios, excluding gluino R-hadrons up to 2.08 TeV, top squark R-hadrons up to 1.47 TeV, and staus up to 0.69 TeV.

Next, we have three searches for long-lived charginos, typically with $\Delta m(\cone, \none)$ within a few hundred MeV, that decay within the silicon tracker and produce the characteristic ``disappearing track" (DTk) signature: an isolated high-$p_{\mathrm{T}}$ track with missing hits in the outer tracker layers, little or no associated calorimeter energy, and no matching hits in the muon detectors. An early search targeted DTks with $p_T > 55$ GeV, selected using a cut-based approach based on track properties such as the number of hits in different subdetectors, impact parameters, and associated calorimeter energy~\cite{CMS:2020atg}.  The event selection is very inclusive, requiring at least one DTk, at least one jet, and $\met > 120$ GeV, and is partitioned into three generic signal regions with 4, 5, and $\geq 6$ missing outer hits on the DTk.  In this and the other two DTk searches, the dominant backgrounds arise from misreconstructed charged lepton tracks or spurious tracks from random alignments of hits in the tracker, and are estimated entirely from data. Results are interpreted in anomaly-mediated SUSY breaking scenarios with nearly mass-degenerate charginos and neutralinos. For a pure higgsino (wino) LSP, chargino masses are excluded up to 750 (884) GeV for a lifetime of 3 ns, with additional limits at shorter lifetimes down to 0.05--0.2 ns.

Another study incorporated DTks into the inclusive $M_{T2}$ search described in Section~\ref{sec:inclusive} to increase sensitivity to long-lived charginos produced either directly or in decays of heavier SUSY particles~\cite{CMS:2019ybf}. This search reoptimized the DTk identification, extending coverage to $p_T$ as low as 15 GeV and defining four track-length categories. After a preselection including $M_{T2} > 200$ GeV, events are divided into 68 search regions across data-taking years, binned by DTk length, DTk $p_T$, $n_{\mathrm{jets}}$, and $H_T$. In models where gluinos and squarks decay with equal probability to $\tilde{\chi}^0_1$, $\tilde{\chi}^+_1$, and $\tilde{\chi}^-_1$, gluino masses are excluded up to 2460 GeV and $\tilde{\chi}^0_1$ masses up to 2000 GeV, while light-flavor (top) squark masses are excluded up to 2090 (1660) GeV and $\tilde{\chi}^0_1$ masses up to 1650 (1210) GeV. Including the DTk component enhances sensitivity particularly in the compressed region, where mass difference between the parent particle and $\tilde{\chi}^0_1$ is small.

The third DTk search extended the reach of the $M_{T2}$ analysis by considering both hadronic and leptonic final states, with the latter using a low $\met$ requirement of 30 GeV~\cite{CMS:2023mny}.  DTk identification was refined through dedicated boosted decision trees trained separately for short, pixel-based tracks with $p_T > 25$ GeV and long, pixel-plus-strip tracks with $p_T > 40$ GeV. To maintain sensitivity to a broad range of chargino production modes, whether produced directly or in the decays of heavier SUSY particles, the analysis defined 49 nonoverlapping signal regions. These were categorized according to DTk type (short or long), the presence of electrons or muons, number of jets and $b$-tagged jets, $\met$, and, for the first time in a DTk search, the DTk ionization energy loss (dE/dx) in the inner tracker. Figure~\ref{fig:highlights_llp} (top-center) shows the distribution of DTk mass obtained from dE/dx in the 1 long track baseline region. 
In gluino, top squark, and bottom squark production scenarios, masses are excluded up to 2300, 1590, and 1540 GeV, respectively. For top and bottom squark production, charginos are excluded up to 850 (1210) GeV and 1050 (1400) GeV, respectively, for proper decay lengths of 10 (200) cm. In pure wino dark matter models, charginos are excluded up to 650 GeV, while in pure higgsino dark matter models the limit is 210 GeV.

We now turn to searches that exploit the time delay of objects from LLP decays. The first analysis to explore such signatures was the ``delayed photon" search, targeting long-lived $\none$ produced in gluino or squark decays and subsequently decaying to a photon and a weakly interacting gravitino, a typical GMSB signature~\cite{CMS:2019zxa}. The photon from $\none$ decay, originating at a displaced vertex, reaches the electromagnetic calorimeter at a non-normal impact angle and with a delayed arrival time on the order of nanoseconds. This distinctive combination was exploited using the timing capabilities of the ECAL, to identify delayed photons and strongly suppress backgrounds. In 2017, a dedicated single-photon trigger was introduced to select photons with non-normal entrance angles. Events are required to contain at least one delayed photon, at least three jets, and $\met > 70$~GeV. Signal extraction is performed in bins of photon timing and $\met$. For $\none$ proper decay lengths of 0.1, 1, 10, and 100 m, masses up to about 320, 525, 360, and 215 GeV are excluded, respectively.

A related search extended the use of ECAL timing to jets originating from LLP decays~\cite{CMS:2019qjk}. It targeted long-lived gluinos in a GMSB scenario, each decaying to a gluon and a gravitino. The gluon forms a jet whose energy deposits in the ECAL are both spatially displaced and delayed in time by several nanoseconds. This was the first application of ECAL timing to a search for displaced jets, that achieved a high background rejection while maintaining good signal efficiency. Events are required to have at least one delayed jet and significant $\met$. Signal extraction was done using jet timing, distribution of which is shown in Figure~\ref{fig:highlights_llp} (bottom-left).  For proper decay lengths of 0.3, 1, and 100 m, gluino masses up to 2.10, 2.50, and 1.90 TeV are excluded. 

Next, we have the searches reconstructing displaced vertices (DVs) from LLP decays.  An earlier DV search targeted pair-produced LLPs decaying into multijet or dijet final states, each producing a DV in the tracker~\cite{CMS:2021tkn}.  This analysis was particularly sensitive to mean proper decay lengths between 0.1 to 15 mm. Specifically, it looked for two vertices, each formed from the intersection of multiple charged-particle trajectories and displaced from the interaction region but within the radius of the beam pipe. DVs are reconstructed from charged particle tracks using a custom vertex reconstruction algorithm. The analysis selects events that contain at least two DVs each with five or more tracks and uses the distance between two vertices in the x-y plane as the signal discriminating variable. As pair-produced LLPs tend to be emitted back-to-back in the x-y plane, $d_{VV}$ is larger for the signal.  Pair produced RPV gluinos, top squarks, and neutralinos are excluded up to masses 2.29 TeV, 1.48 TeV, and 1.31 TeV for mean proper decay lengths between 0.1 and 100 mm. 

A more recent DV search was designed to be sensitive to LLPs with $c\tau \sim 1$--$1000$~mm, whose decay products produce at least one DV and $\met$~\cite{CMS:2024trg}. It improved upon~\cite{CMS:2021tkn} by including events with only one DV and enhances sensitivity to cases where only one LLP decays in the detector, or could be identified. It applied an ML technique by using an interaction network (IN), which eliminated the highly increased backgrounds an order of magnitude more effectively compared to cut-based methods. The IN output score $S_{\text{ML}}$ was used for signal extraction. 
Figure~\ref{fig:highlights_llp} (bottom-center) shows the distribution of $S_{\text{ML}}$ for a selection with $n_{track} \ge 6$.  For a split-SUSY scenario with $\tilde{g} \to q\bar{q}\none$, gluino masses up to 1.80 TeV are excluded for $c\tau = 1$--100 mm with $\Delta m = 100$ GeV, and up to 1.60 TeV for $c\tau = 1$--30 mm with $\Delta m > 50$ GeV. For a GMSB scenario with $\tilde{g} \to g\tilde{G}$, gluino masses up to 2.20 TeV are excluded for $c\tau = 0.3$--100 mm. This was the first CMS search to probe hadronically decaying LLPs with gluino-neutralino mass differences below 100 GeV, and it set the strongest limits to date for split-SUSY and for GMSB gluinos with $c\tau < 6$~mm. 

The most recent DV search extended the sensitivity to more compressed cases with $\Delta m$ as low as 25 GeV by targeting DVs with low momentum, high $\met$, greater than 400 GeV and an ISR jet with $p_T > 100$~GeV~\cite{CMS-PAS-EXO-24-033}.  Displaced vertices were reconstructed using a customized algorithm based on the so-called adaptive vertex fitter.  Signal regions are categorized using $\met$ and the variable $S_{xy}^{vtx} = L_{xy} / \delta L_{xy}$, where $L_{xy}$ is the transverse distance between the primary vertex and the DV, and $\delta L_{xy}$ its uncertainty. 
In a top squark coannihilation scenario, where the NLSP is a long-lived top squark and the LSP a bino-like neutralino, top squark masses between 400 and 1100 GeV are excluded. In a bino-wino coannihilation scenario, where the NLSPs are long-lived wino-like neutralino and prompt wino-like chargino with a bino-like LSP, wino-like neutralino masses between 220 and 550 GeV are excluded.

Displaced-jet searches provide a complementary strategy by targeting higher-momentum vertices associated with a dijet system. Two searches~\cite{CMS:2018qxv} and~\cite{CMS:2020iwv} based on 2016 data and 2017--2018 data, respectively,  looked for LLPs decaying into jets, with at least one decay vertex inside the tracker, but which is displaced from the production vertex by up to 550 mm in the plane transverse to the beam direction. They were designed to be as model independent as possible, given the wide range of BSM scenarios producing this signature. Displaced jets from LLP decays were clustered from calorimeter energy deposits, and identified by being associated to tracks displaced from the primary vertex, from which the decay vertex is reconstructed. In~\cite{CMS:2020iwv}, a special displaced jet trigger was also used, that recovered efficiency at high LLP masses. Events are required to have at least two jets and high $H_T$.  The properties of the tracks and the decay vertex are used to discriminate the signal, and in~\cite{CMS:2020iwv}, also through applying a gradient BDT algorithm, output score of which is shown in Figure~\ref{fig:highlights_llp} (bottom-right).. The approach probed a wide range of lifetimes and decay modes in the GMSB, RPV, and split SUSY frameworks, excluding gluinos up to 2.5 TeV and top squarks up to 1.6 TeV, depending on the model and lifetime.

Another complementary approach targeted LLP decays occurring in the outer regions of the tracker or in the calorimeters, which produce nearly trackless jets arriving out of time with respect to the primary collision. The analysis~\cite{CMS:2022wjc} requires events with large $\met$ and at least two jets tagged by a DNN discriminator trained on timing and tracking information. The tagger reduces SM backgrounds by over three orders of magnitude while maintaining a signal efficiency above 80\%. A dedicated background estimation extrapolates misidentification probability of the tagger from control regions with one or fewer tagged jets to the signal region. The search was interpreted in an electroweak chargino-neutralino production model with a long-lived neutralino decaying to a gravitino and a Higgs or $Z$ boson, excluding neutralino masses up to 1.18 TeV for $c\tau = 0.5$ m.

\vspace{10pt}	

\begin{figure}[H]
\centering
\includegraphics[width=0.90\textwidth]{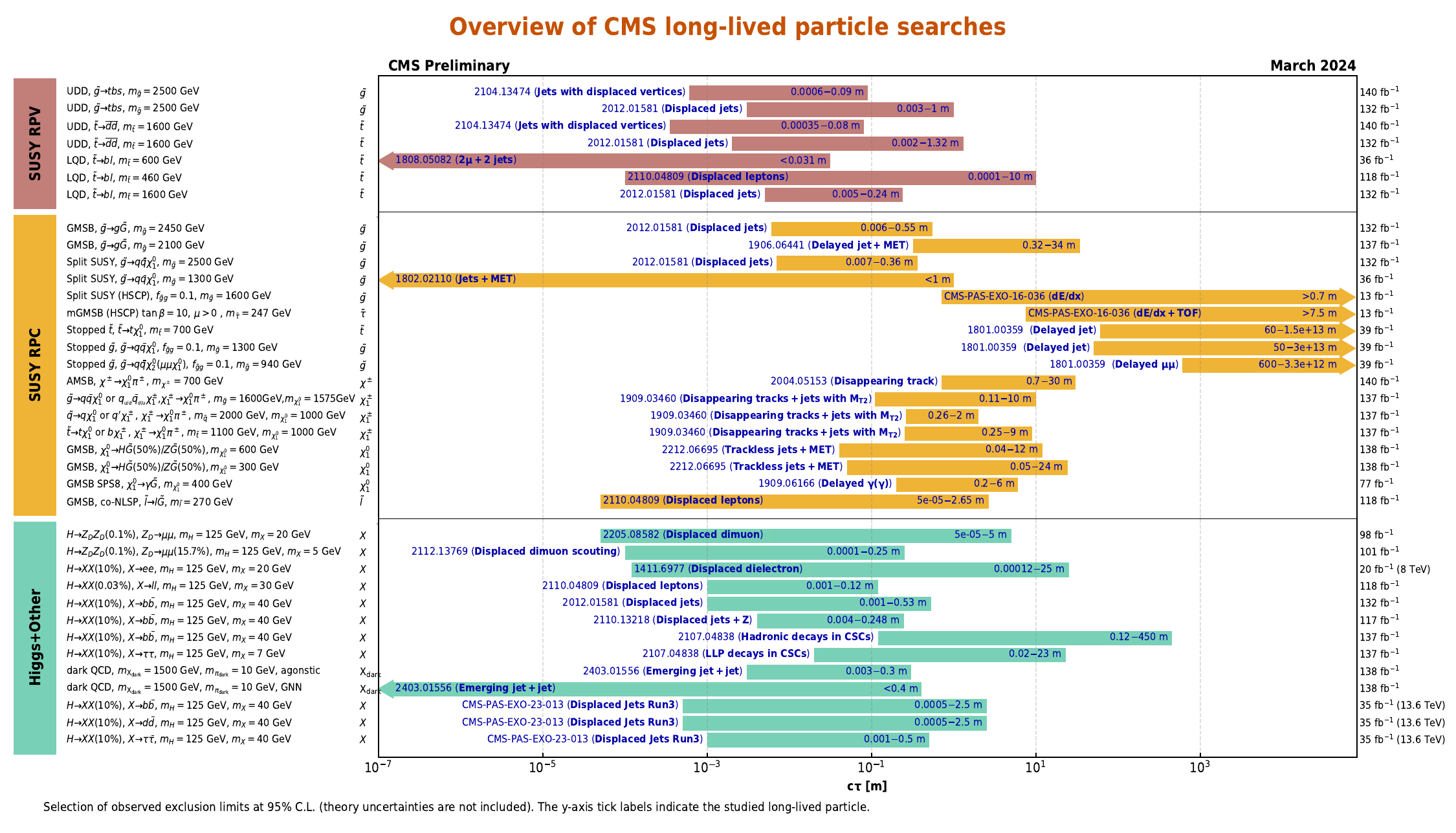} 
\caption{Summary of the proper lifetime reach of the LLP searches for particles and processes predicted by various SUSY and other new physics scenarios.  Interpretations were done for RPV SUSY, GMSB, split SUSY, and AMSB models. Gluino and squark mass exclusions reach up to about 2.5 TeV and 1.6 TeV, respectively, for proper lifetimes reaching over hundreds of meters. Chargino and neutralinos are probed up to masses of several hundred GeV and proper lifetimes of O(100 m).  Direct stau production has been constrained up to about 700 GeV. Together, these results cover proper lifetimes spanning more than ten orders of magnitude, from prompt-like decays down to hundreds of meters, significantly extending the SUSY coverage of Run 2.}
\label{fig:limits_llp}
\end{figure}

Apart from the dedicated analyses here, the hadronic $\tau$ plus \met analysis~\cite{CMS:2022syk} described in Section~\ref{sec:ewkslep}, and the isolated track plus \met analyis~~\cite{CMS-PAS-SUS-24-012} analysis described in Section~\ref{sec:compressed} also have long-lived components. Figure~\ref{fig:limits_llp} collectively presents the proper lifetime reach of the LLP searches for particles and processes predicted by various SUSY and other new physics scenarios.

\section{The SUSY picture after Run 2}

The diverse set of searches discussed above, spanning a broad range of signatures and strategies, provided sensitivity to different regions of the SUSY parameter space. Individual analyses probed complementary corners of this space, and taken together, they form a coherent picture that has been progressively refined throughout Run 2. To illustrate this more concretely, CMS performed three dedicated combination studies. The first two combined top squark searches and electroweakino/slepton searches, respectively. and interpreted them in the framework of simplified models. The third combined a variety of analyses with very different final states, and interpreted them in terms of the phenomenological MSSM. Together, these combinations provide the most global statement of CMS on SUSY after Run 2, and frame the discussion of what is excluded and what remains open.

\subsection{Collective reach in simplified models}
\label{sec:interp_sms}

\begin{figure}[H]
\centering
\includegraphics[width=0.32\textwidth]{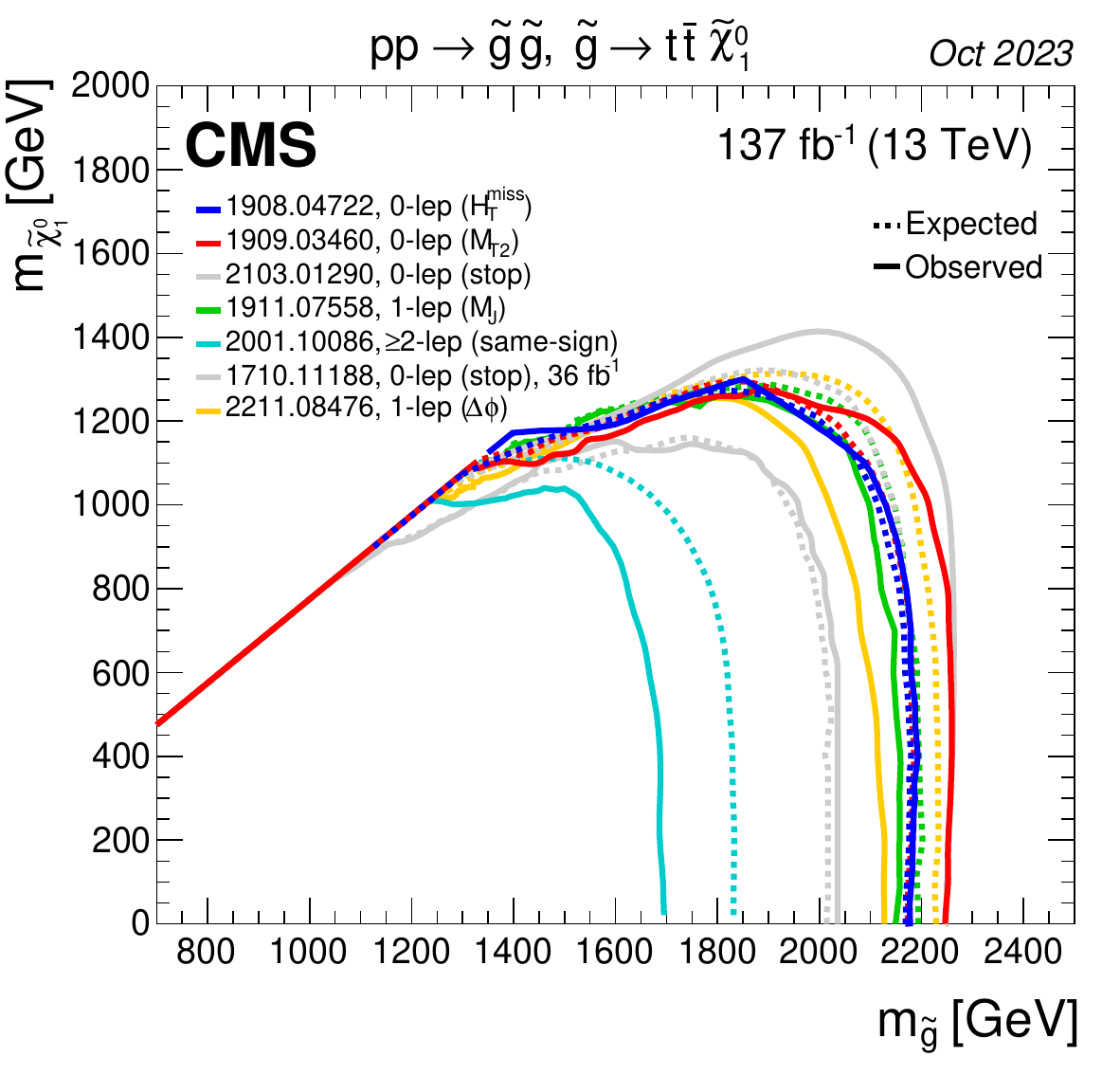} 
\includegraphics[width=0.32\textwidth]{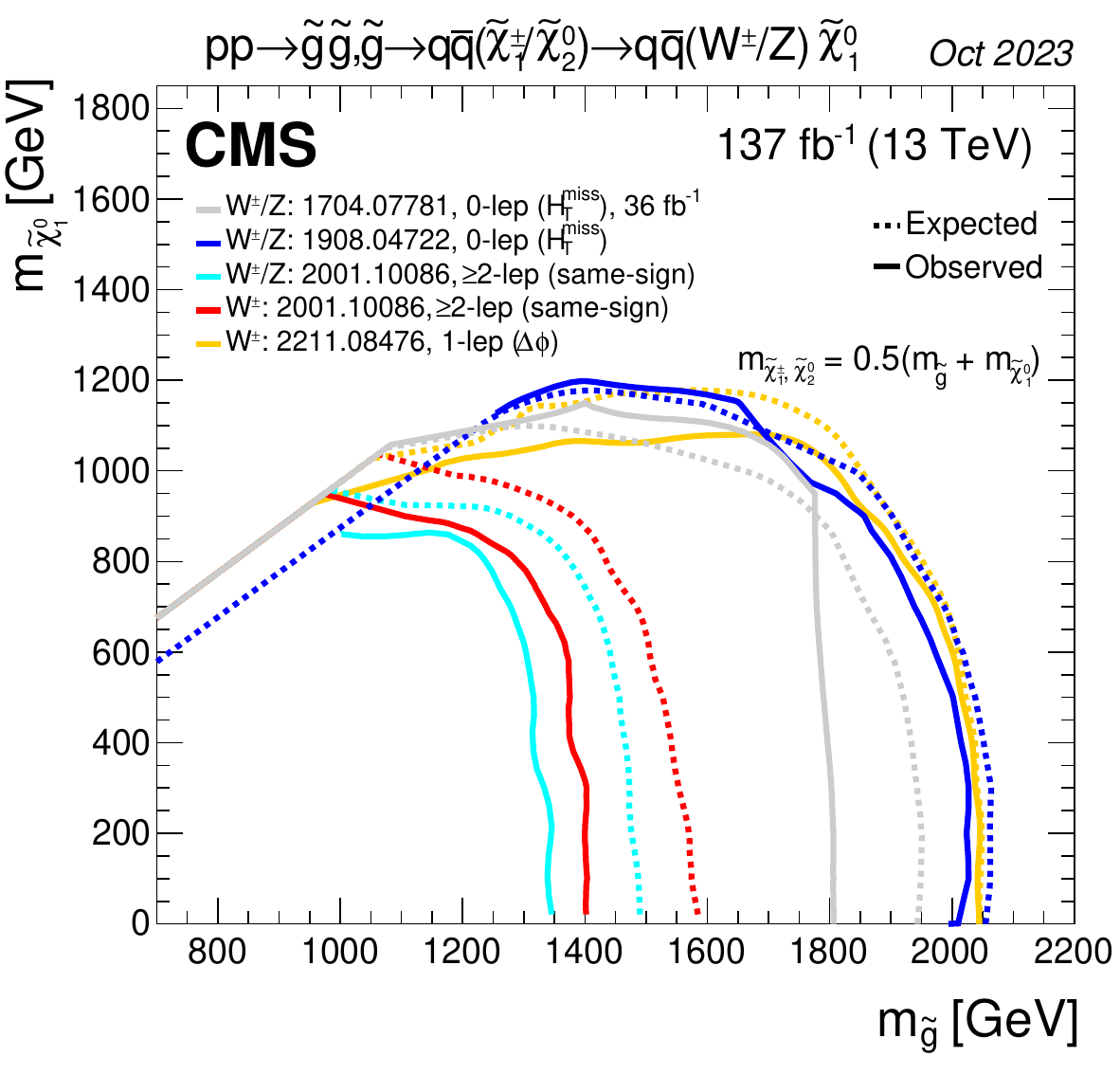} 
\includegraphics[width=0.32\textwidth]{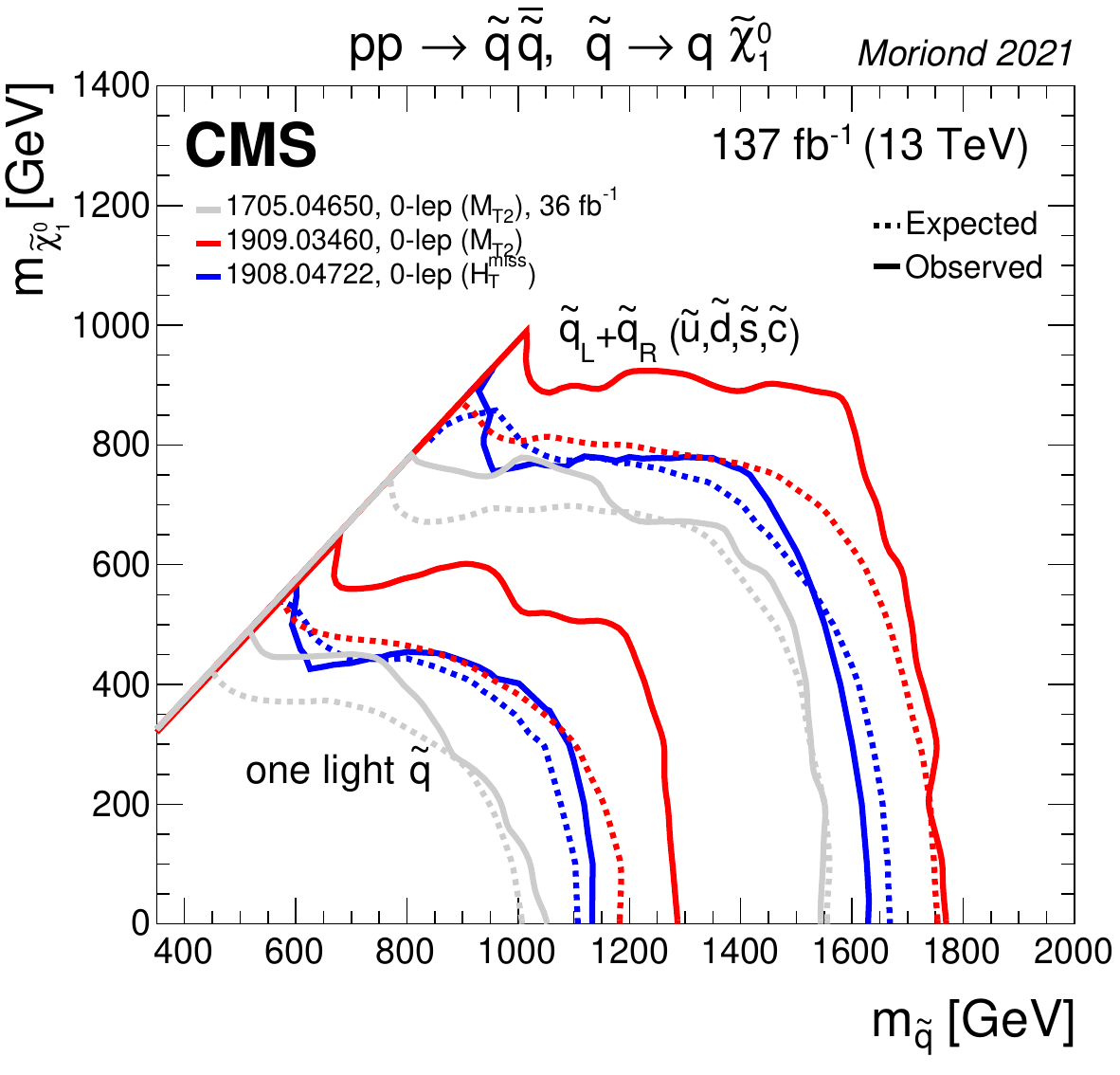} 
\caption{Mass exclusions for gluino and light squark versus $\none$ for three selected simplified models of direct production, from several analyses.}
\label{fig:limits_gluino}
\end{figure}

Figure~\ref{fig:limits_gluino} shows a representative summary of gluino and light-flavor squark searches in $R$-parity conserving simplified models, with branching fractions taken as unity. The exclusion contours are shown as a function of the gluino or squark mass versus the lightest neutralino mass. Expected and observed limits from several inclusive analyses are shown together comparatively. Although some contours appear to overlap, the underlying analyses rely on different final states and selections, and they probe distinct, and sometimes disjoint event samples.  This complementarity highlights the value of combining results, to extend sensitivity beyond the reach of any single search.

\begin{figure}[H]
\centering
\includegraphics[width=0.32\textwidth]{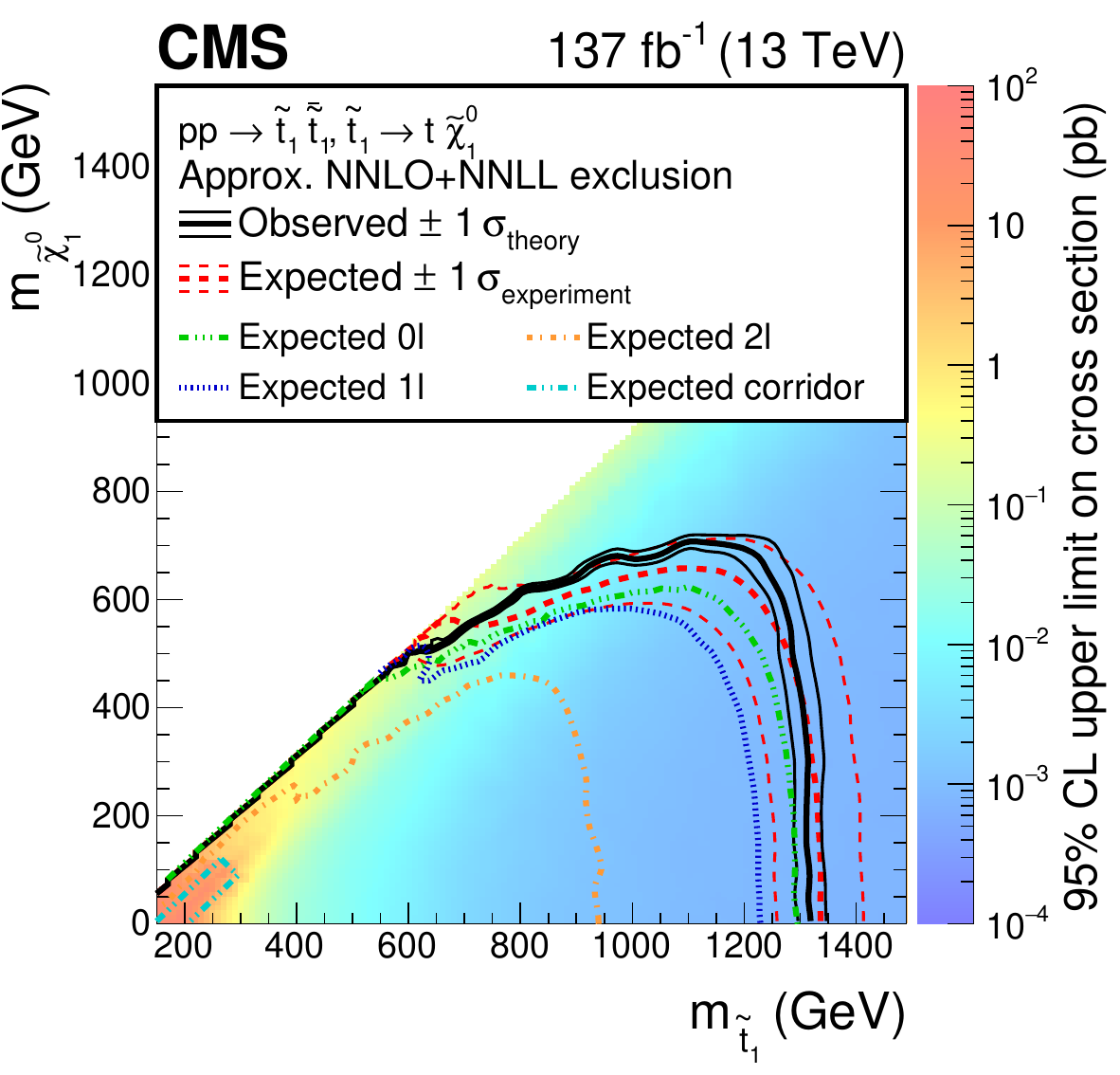} \quad
\includegraphics[width=0.32\textwidth]{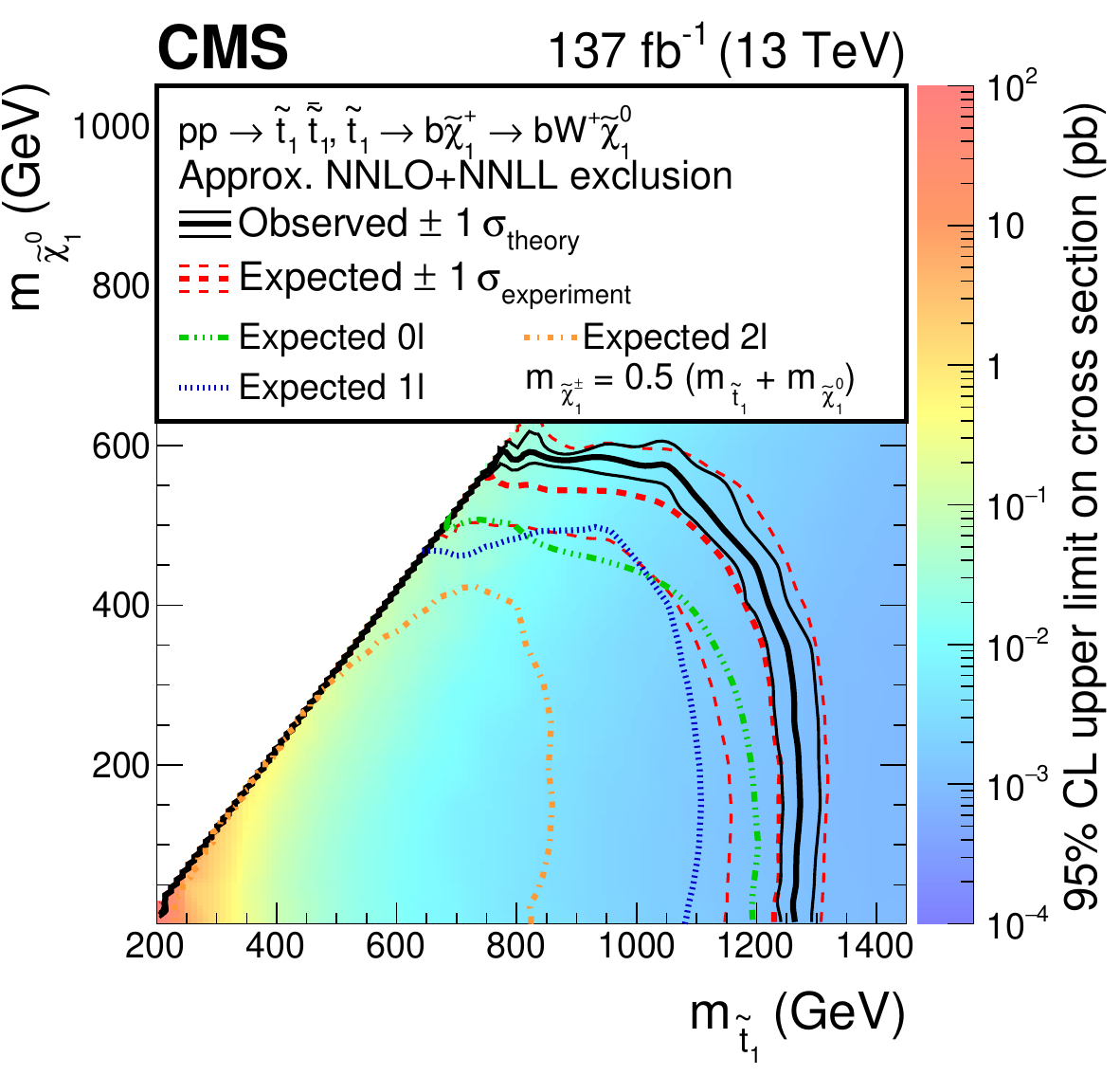}
\caption{Observed upper limits on the cross section, and expected and observed mass exclusions for $\tilde{t}$ versus $\none$ in $\tilde{t} \to t\none$ and $\tilde{t} \to b\cone \to bW^\pm\none$ models, from the combination of fully hadronic, single-lepton, dilepton, and corridor analyses. Expected limits from the individual analyses are also shown.}
\label{fig:limits_stopcomb}

\end{figure}

As mentioned in Section~\ref{sec:stop}, the fully hadronic, single-lepton, and dilepton top squark searches were combined in~\cite{CMS:2021eha}, together with the dedicated search targeting the stop corridor region. These searches were designed to be mutually exclusive from the start, making a statistical combination straightforwardly possible. Figure~\ref{fig:limits_stopcomb} shows the resulting exclusion limits on $\tilde{t}$ versus $\none$ masses for the $\tilde{t} \to t\none$ and $\tilde{t} \to b\cone \to bW^\pm\none$ models. As anticipated, the combination pushes the sensitivity beyond the reach of the individual analyses, especially for the $\tilde{t} \to b\cone \to bW^\pm\none$ case.  

Direct electroweakino production processes have cross sections about two to three orders of magnitude smaller than strong production of gluinos and squarks, while direct slepton production is over an order of magnitude lower compared to electroweakino production. Combinations are therefore particularly critical in this sector to increase sensitivity and achieve complementarity. An extensive study was performed to combine six searches targeting electroweakinos and sleptons~\cite{CMS:2024gyw}. The included analyses were: i) soft two opposite-charge or three-lepton plus $\met$ (the earlier search)~\cite{CMS:2021edw}, covering the compressed regions; ii) inclusive multilepton (same-charge dilepton and trilepton components)~\cite{CMS:2021cox}, probing moderate mass splittings; iii) opposite-charge same-flavor dileptons, both on-Z and off-Z~\cite{CMS:2020bfa}, sensitive to electroweakino decays via W/Z bosons or slepton decays; iv) single lepton plus $H\to b\bar{b}$ and $\met$~\cite{CMS:2023ktc}, targeting $\cone\ntwo \to W\none H\none$; v) $HH \to 4b$ plus $\met$~\cite{CMS:2022vpy}, targeting $\none$ or $\ntwo$ pair production decaying each to a Higgs boson and the LSP; and vi) boosted hadronic $WW/WZ/WH$ plus $\met$~\cite{CMS:2022sfi}, probing the largest mass splittings.  Unlike the top squark searches, these analyses were initially designed with only loose coordination. For the combination, modifications and synchronizations were introduced to ensure mutual exclusivity and enable a robust statistical combination. 

\begin{figure}[H]
\centering
\includegraphics[width=0.32\textwidth]{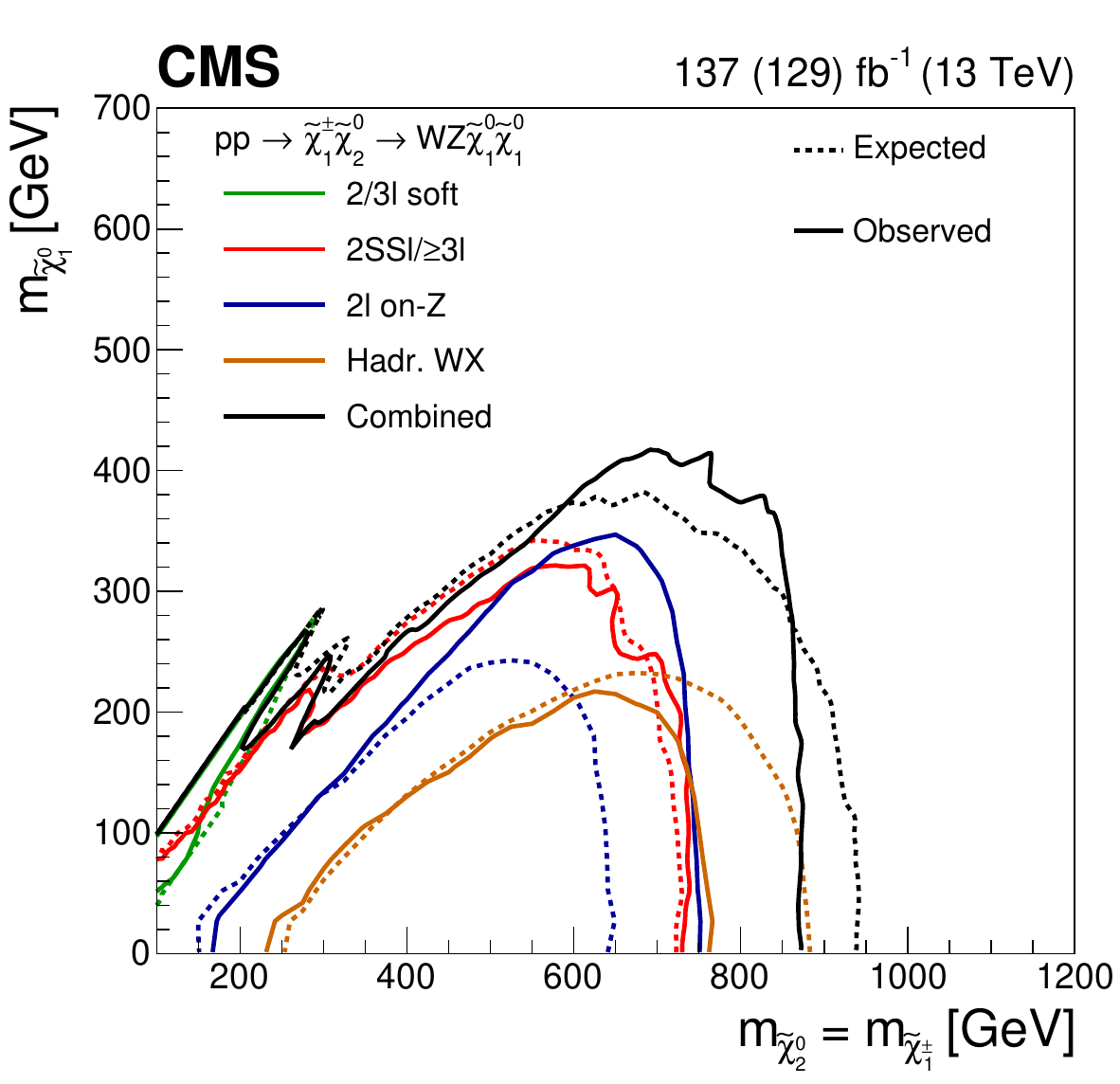} 
\includegraphics[width=0.32\textwidth]{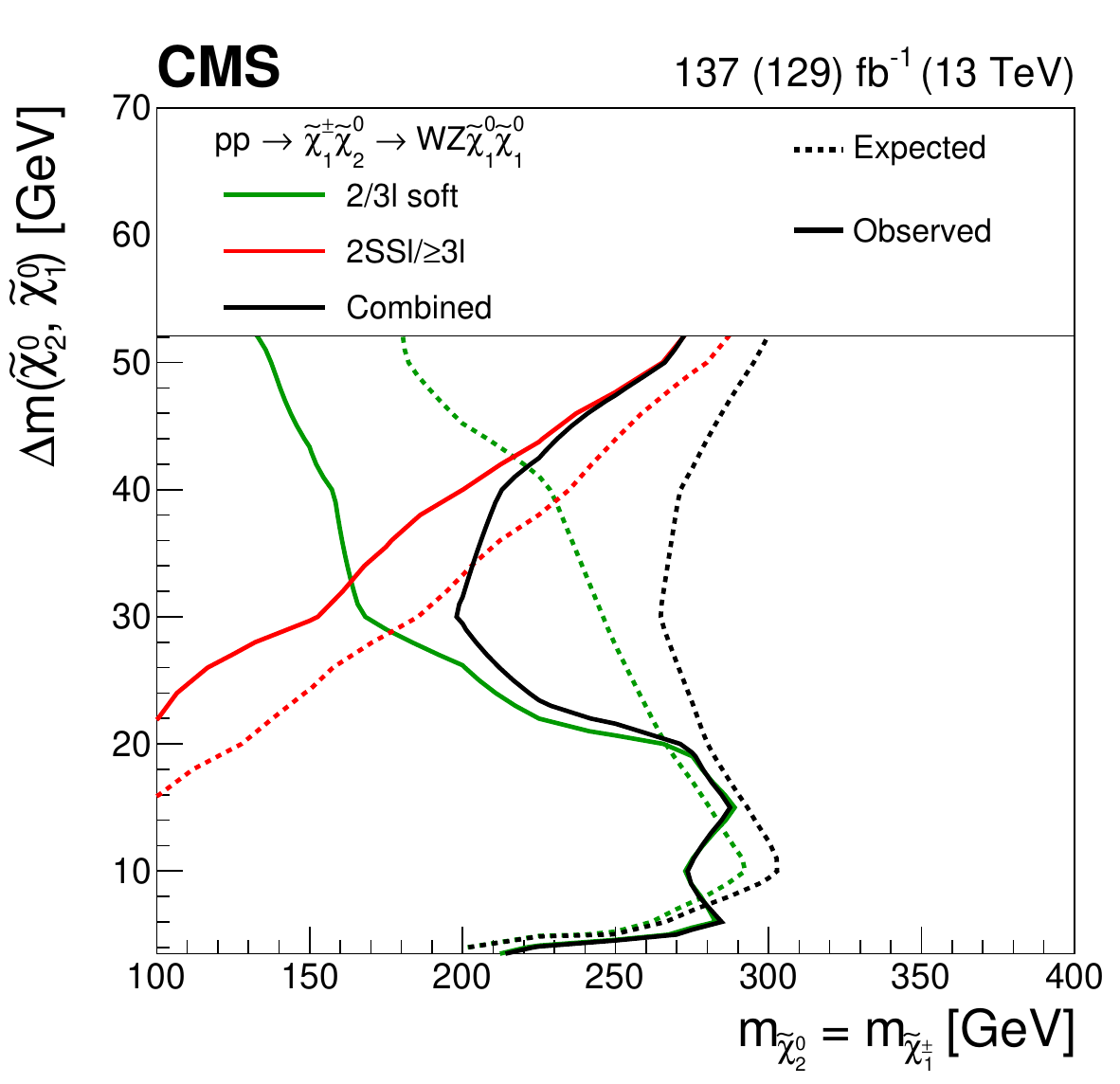} 
\includegraphics[width=0.32\textwidth]{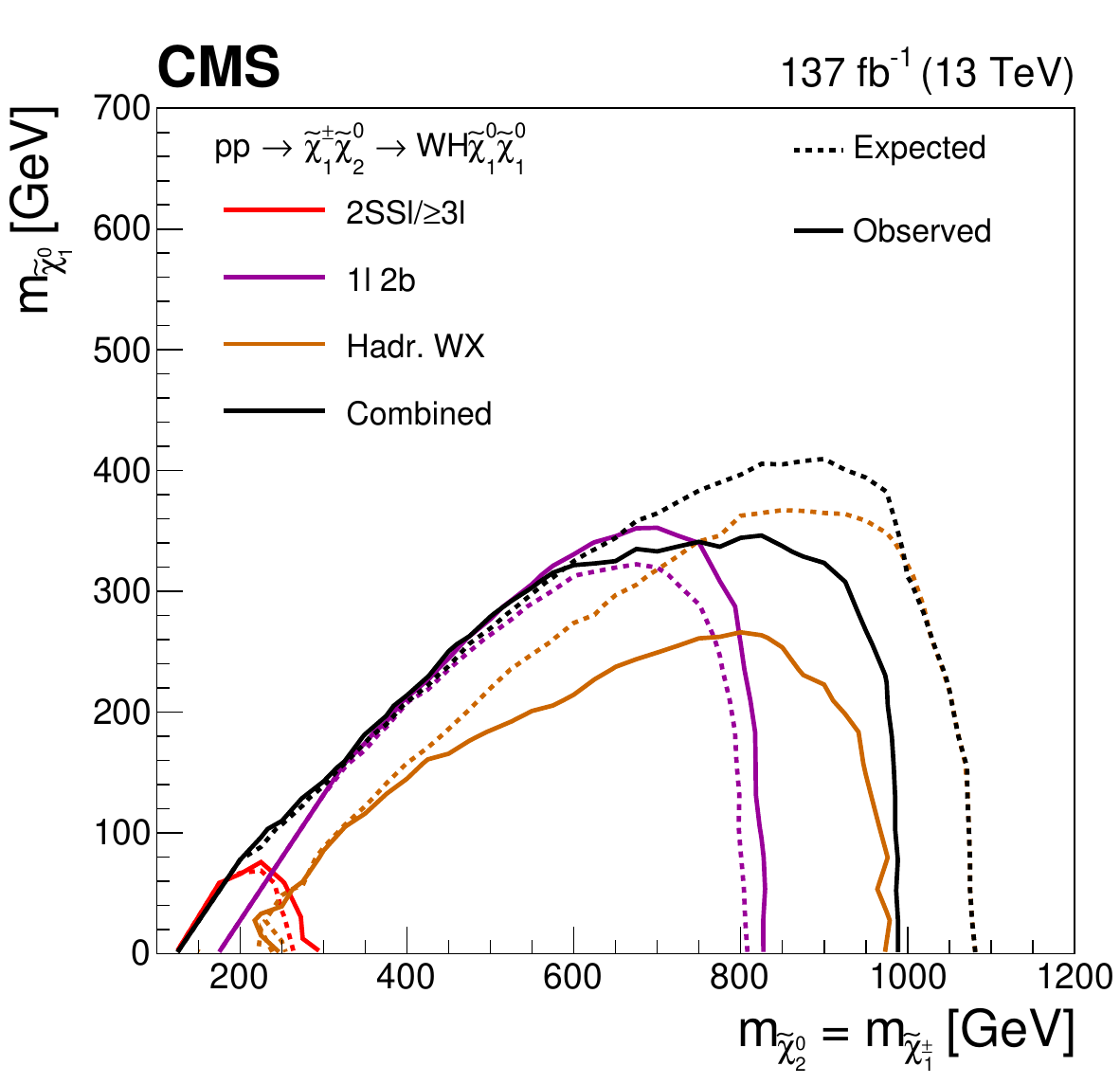} 
\includegraphics[width=0.32\textwidth]{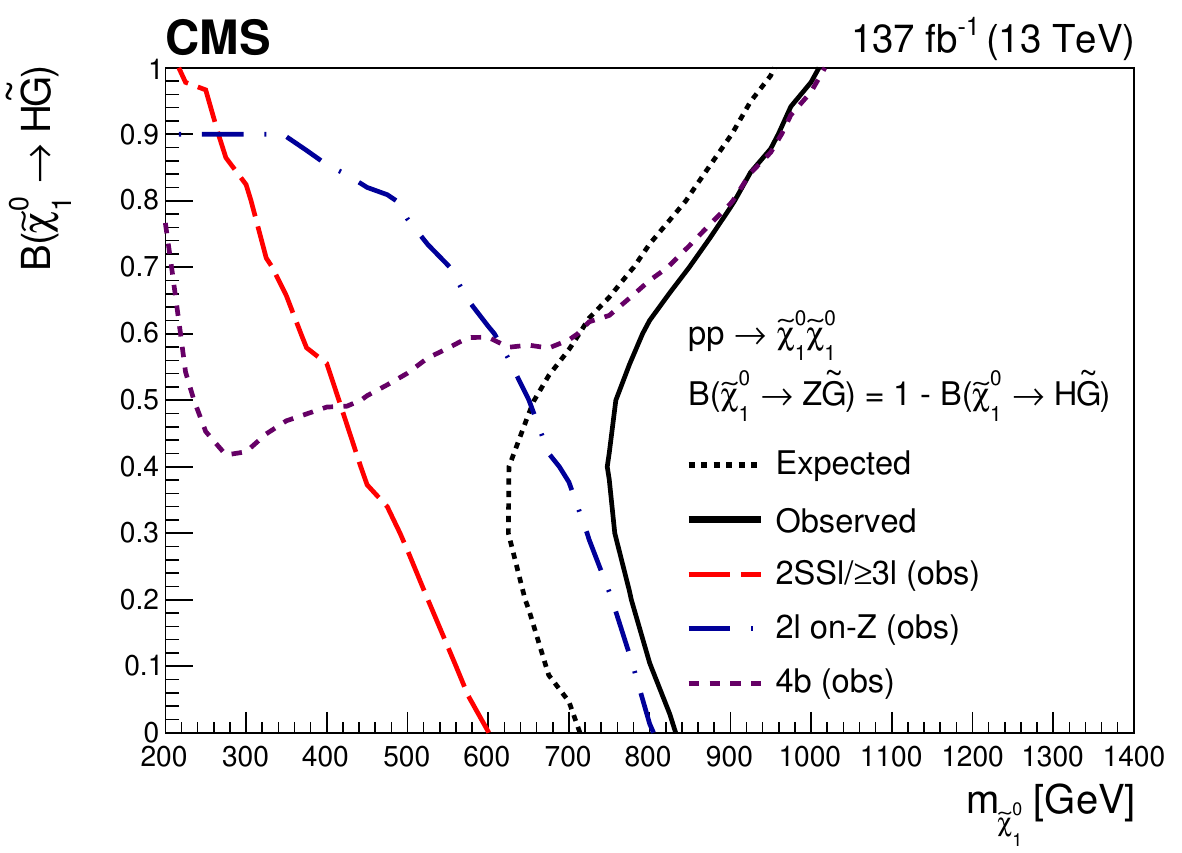} 
\includegraphics[width=0.32\textwidth]{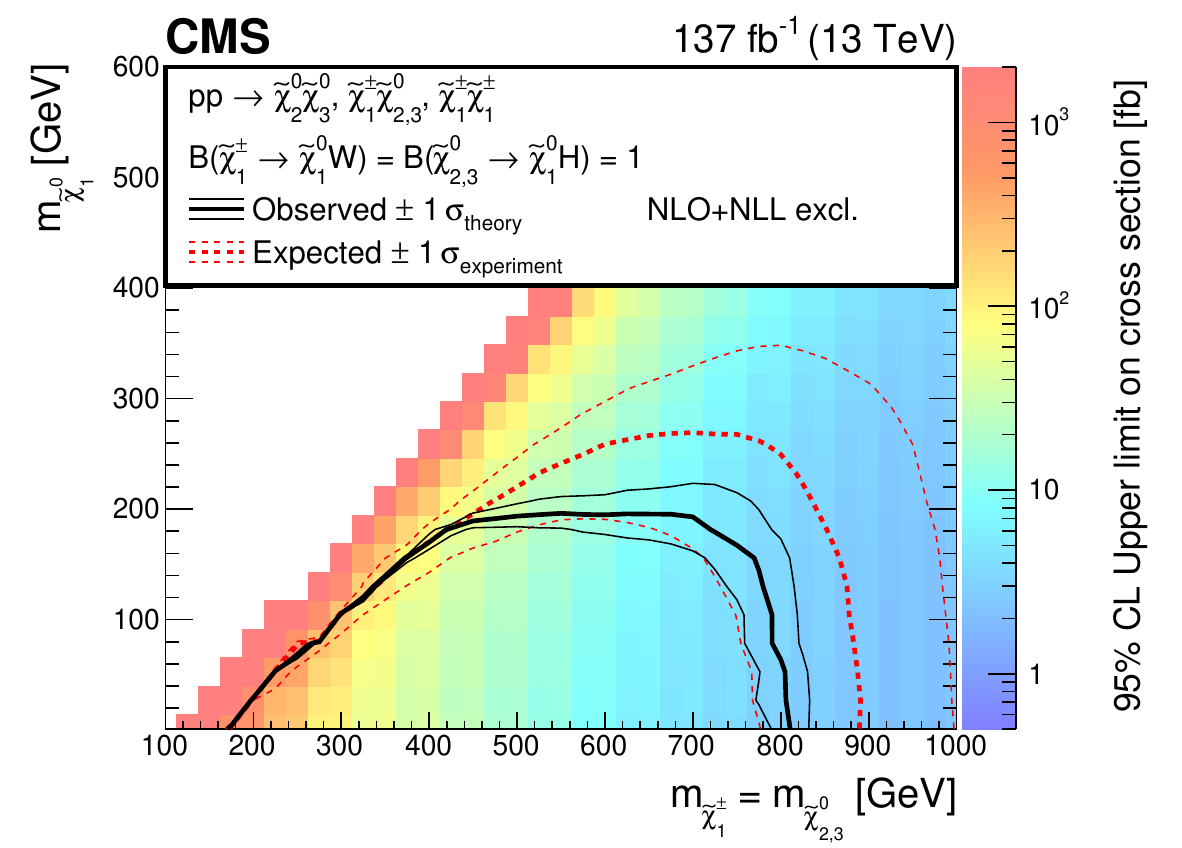} 
\includegraphics[width=0.32\textwidth]{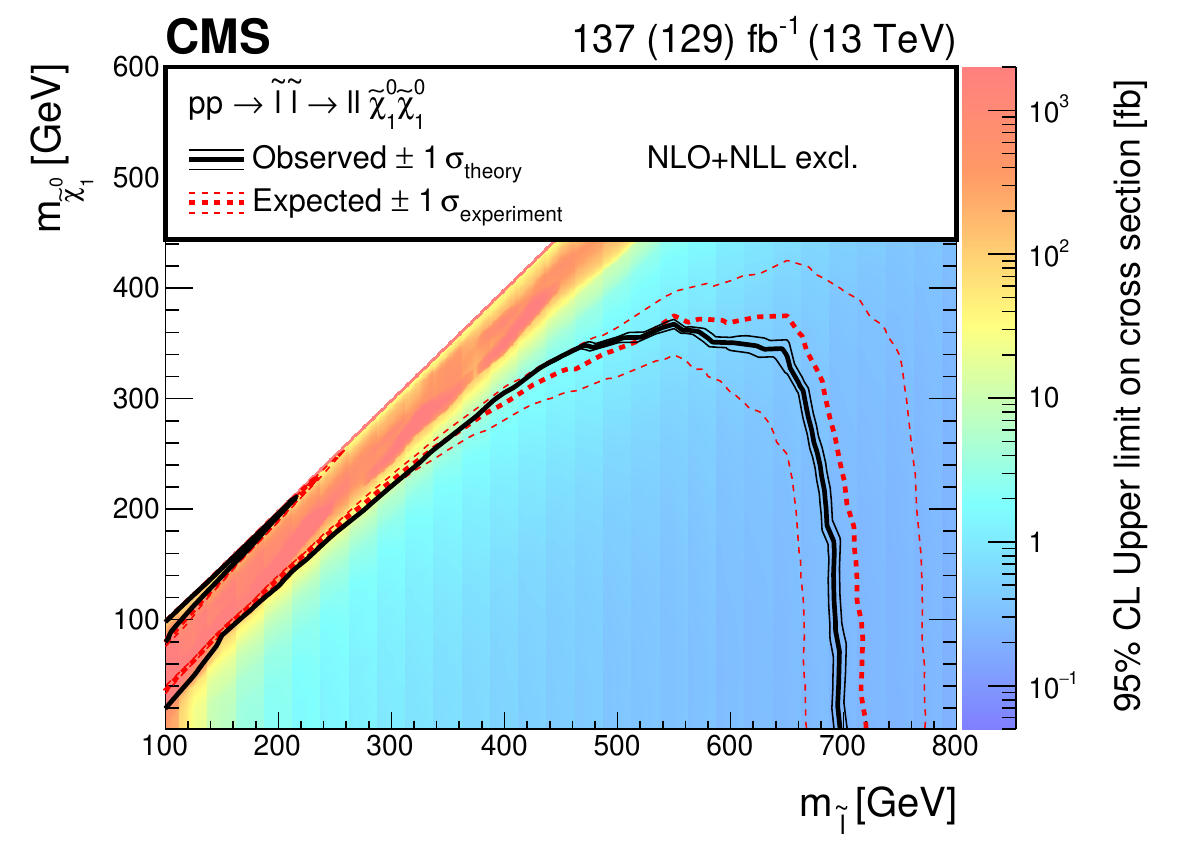} 
\caption{Reach of the electroweakino combination for four simplified scenarios: Wino-bino model in decays to WZ topologies for the noncompressed case (top-left), wino-bino model in WZ topologies for the compressed case (top-center), wino-bino model in WH topologies (top-right), GMSB-higgsino model (bottom-left), higgsino-bino model (bottom-center), and slepton-neutralino model (bottom-right).  Details are described in the text. }
\label{fig:limits_ewkslep}
\end{figure}

The combined results were interpreted in terms of four simplified models. The combination was performed through a simultaneous maximum likelihood fit to the signal and control regions of the searches described above, for each signal model. For different models, different subsets of signal regions were used, chosen to maximize sensitivity while avoiding overlaps. Figure~\ref{fig:limits_ewkslep} shows the combined limits for these models, together with the contributions from individual analyses in cases where multiple analyses entered. The first model considers the associated production of wino-like degenerate $\cone$ and $\ntwo$, decaying into a lighter bino-like $\none$ as $\cone \to W\none$ and $\ntwo \to Z/H\none$. The top-left and top-center plots show mass exclusion contours in the $\none$ versus $\cone/\ntwo$ plane for the WZ topology in the noncompressed region, and on the $\Delta m(\cone/\ntwo, \none)$ versus $\none$ plane for the compressed region, respectively, while the top-right plot shows the WH topology. In all cases, the combined result significantly exceeds the reach of the individual searches, and the complementarity of the contributing analyses is strikingly visible.

The second model is a GMSB-inspired scenario with pair production of higgsino $\none$ decaying to gravitinos as $\none \to H\tilde{G}$ or $\none \to Z\tilde{G}$.  The bottom-left plot shows the exclusion contours in the $\none$ mass versus $\none \to H\tilde{G}$ branching fraction, where once again, the complementarity between the search involving explicit Higgs identification and those without is explicitly visible.  The third model assumes nearly mass-degenerate $\ntwo$, $\nthree$, and $\cone$ that decay into a bino-like $\none$ and an SM boson.  The relevant production mechanisms in this higgsino-bino model are $\cone \ntwo$, $\cone \nthree$, $\cone \cone$, and $\ntwo \nthree$.  The bottom-center plot shows the upper limits on the cross section and the corresponding mass exclusion in this model in the degenerate Higgsino and $\none$ mass plane.  Finally, we have the slepton-neutralino model, with pair produced sleptons, each decaying as $\tilde{\ell} \to \ell \none$, for which the upper limits on the cross section and the mass exclusion are shown in the $\tilde{\ell}$ versus $\none$ mass plane in the bottom-right plot.

\begin{figure}[H]
\centering
\includegraphics[width=0.6\textwidth]{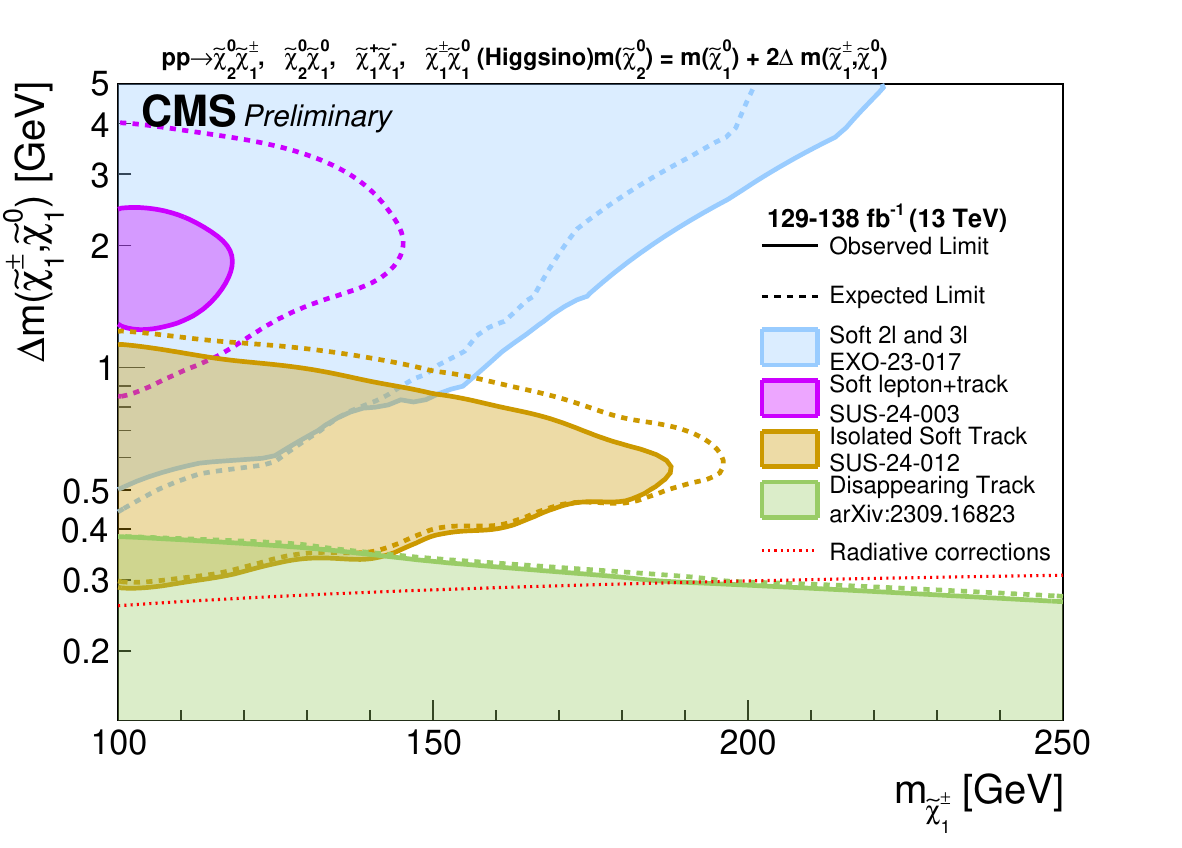} 
\caption{Summary of observed and expected mass exclusion on the simplified scenario with light, nearly mass-degenerate higgsinos based on four Run 2 searches.}
\label{fig:limits_higgsino}
\end{figure}

Several analyses were particularly targeting scenarios with light, nearly mass-degenerate higgsinos with masses related as $\Delta m^\pm(\cone, \none) = \frac{1}{2}\delta m^0(\ntwo, \none)$, which are particularly well motivated by arguments of  naturalness and by dark matter constraints.  These searches each targeted a different range of mass splitting, which also directly translates to proper lifetime of the chargino.  Figure~\ref{fig:limits_higgsino} shows the exclusion contours of these searches in the $\Delta m(\cone, \none)$ versus $\cone$ mass plane, and illustrates the vivid interplay between them.  For the largest mass splittings, above around 0.5 GeV the latest soft oppositely charged dilepton / trilepton search~\cite{CMS-PAS-EXO-23-017} provides the leading sensitivity.  In a subset of this range, for $\Delta m$ between 1 and 3 GeV, the soft lepton plus track search~\cite{CMS-PAS-SUS-24-003} contributes significantly.  Even though the exclusions overlap for both analyses, their selections are disjoint, thus allowing a future combination to improve cross section reach.  At smaller splittings, the isolated soft track search~\cite{CMS-PAS-SUS-24-012} extends coverage into a region that had long been difficult to probe and only became accessible with the analysis of this new final state. Finally, for $\Delta m$ below about 0.5 GeV, we enter the long-lived chargino regime, where the inclusive 0- and 1-lepton disappearing track search~\cite{CMS:2019zmd} is the most sensitive. 

\subsection{Interpretations in the pMSSM framework}

Simplified models are very useful for interpreting individual searches systematically and comparatively, but by design they focus on narrow slices of the SUSY landscape and only a few decay modes at a time. To obtain a more global view, we have also interpreted the CMS Run 2 results in the context of the phenomenological MSSM (pMSSM)~\cite{CMS-PAS-SUS-24-004}. The pMSSM is a realization of the MSSM with a reduced set of 19 free parameters defined at the SUSY scale. It assumes R-parity conservation, CP conservation, and minimal flavor violation, with the first two sfermion generations taken to be degenerate. Using the pMSSM allows a coherent treatment of a wide variety of SUSY spectra and decay chains, and provides a natural way to assess the collective impact of diverse analyses in a realistic full model.  

The study followed a Bayesian approach. The 19-dimensional pMSSM parameter space was scanned under the assumption that the lightest neutralino is the LSP. A Markov Chain Monte Carlo procedure was used to construct a prior, based on a likelihood that incorporates constraints from flavor physics, the Higgs boson mass, and LEP measurements. From this prior, about 500,000 parameter points were selected. For each point, events were generated and simulated, CMS analyses were applied, and the resulting signal yields were obtained. A statistical analysis was then performed to derive individual likelihoods, which were combined into a single CMS likelihood. The contributing analyses and signal regions were chosen to be disjoint or to have negligible overlaps. The set of searches included: the soft opposite-charge dilepton and trilepton search with $\met$~\cite{CMS:2021edw}, the soft lepton track search~\cite{CMS-PAS-SUS-24-003}, opposite-charge same-flavor dileptons (on-Z)~\cite{CMS:2020bfa}, the direct stau search~\cite{CMS:2022syk}, the single lepton $\Delta\phi$ search~\cite{CMS:2022idi}, the inclusive 0- and 1-lepton disappearing track search~\cite{CMS:2019zmd}, and the inclusive multijet + \met (RA2b) search~\cite{CMS:2023mny}.

\begin{figure}[H]
\centering
\includegraphics[width=0.28\textwidth]{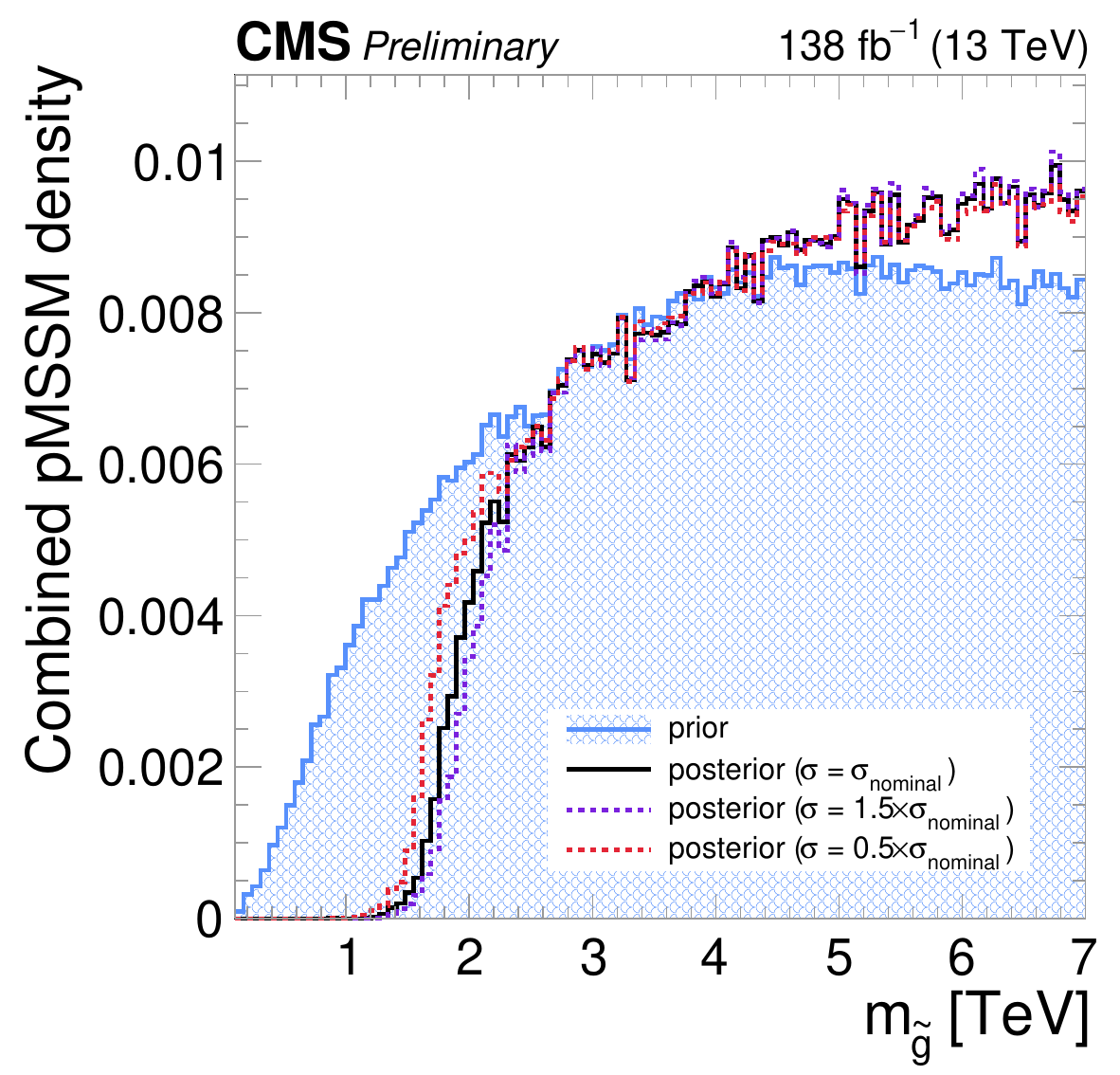} \quad
\includegraphics[width=0.28\textwidth]{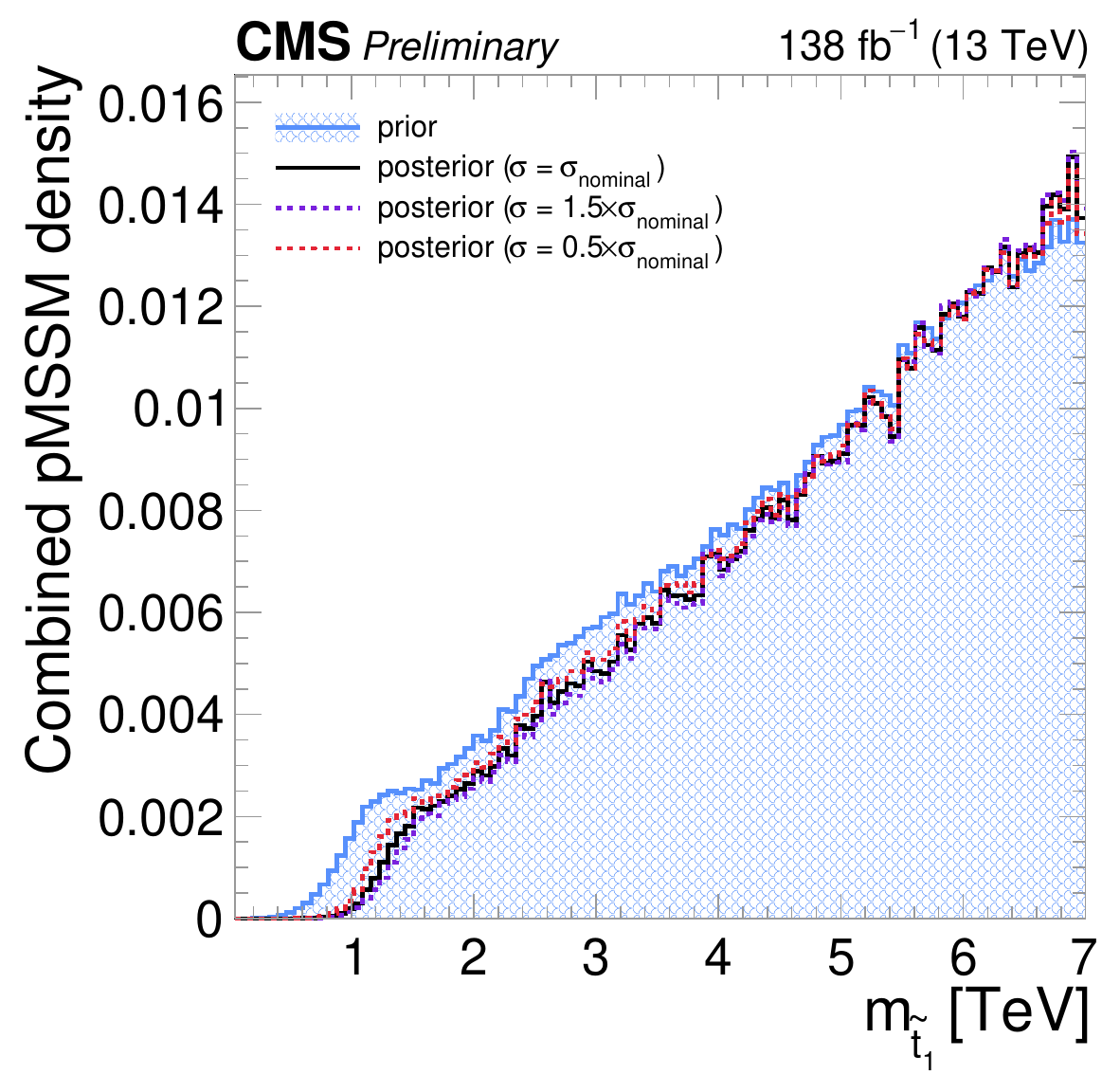} \quad
\includegraphics[width=0.28\textwidth]{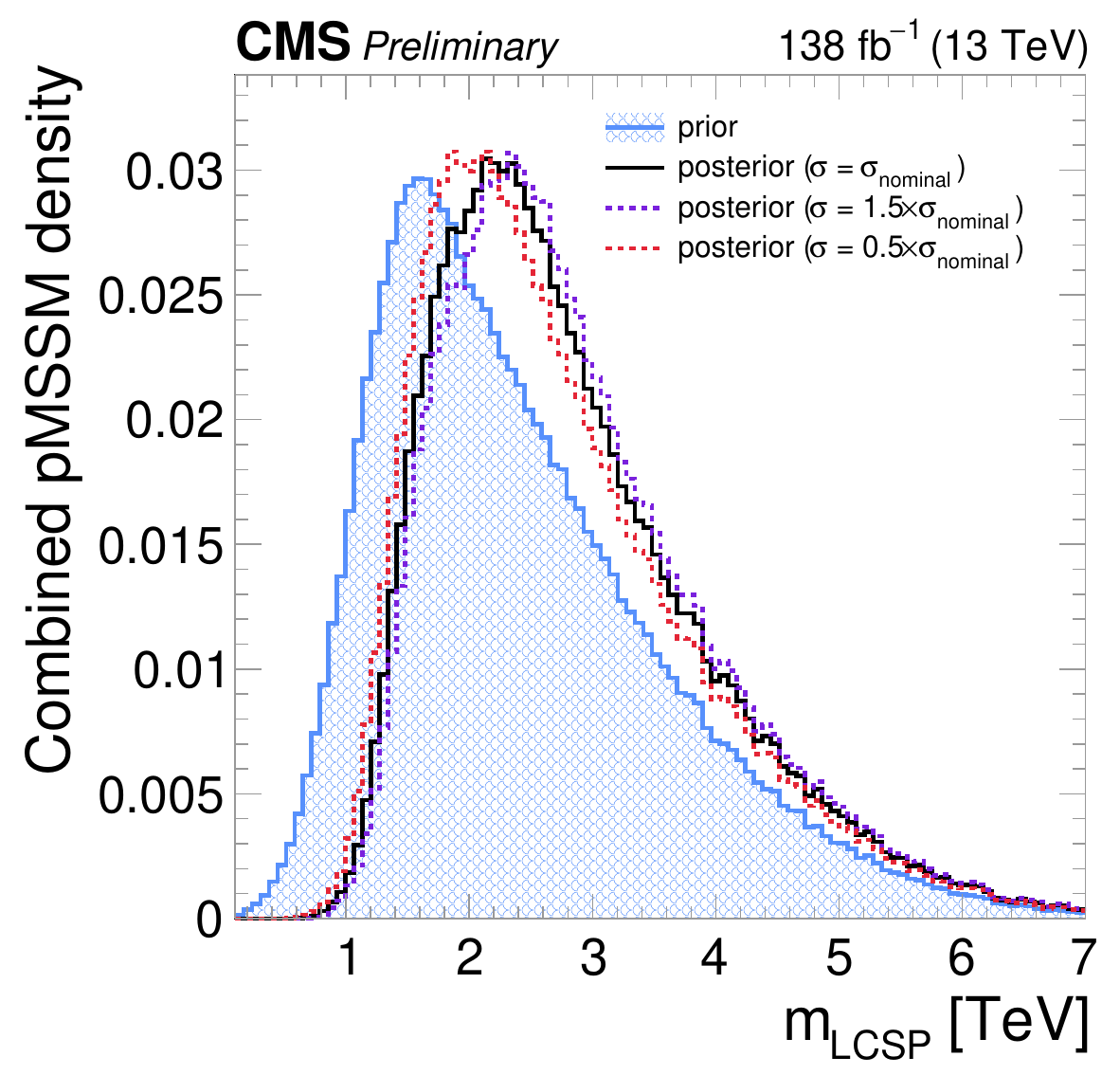} 
\includegraphics[width=0.28\textwidth]{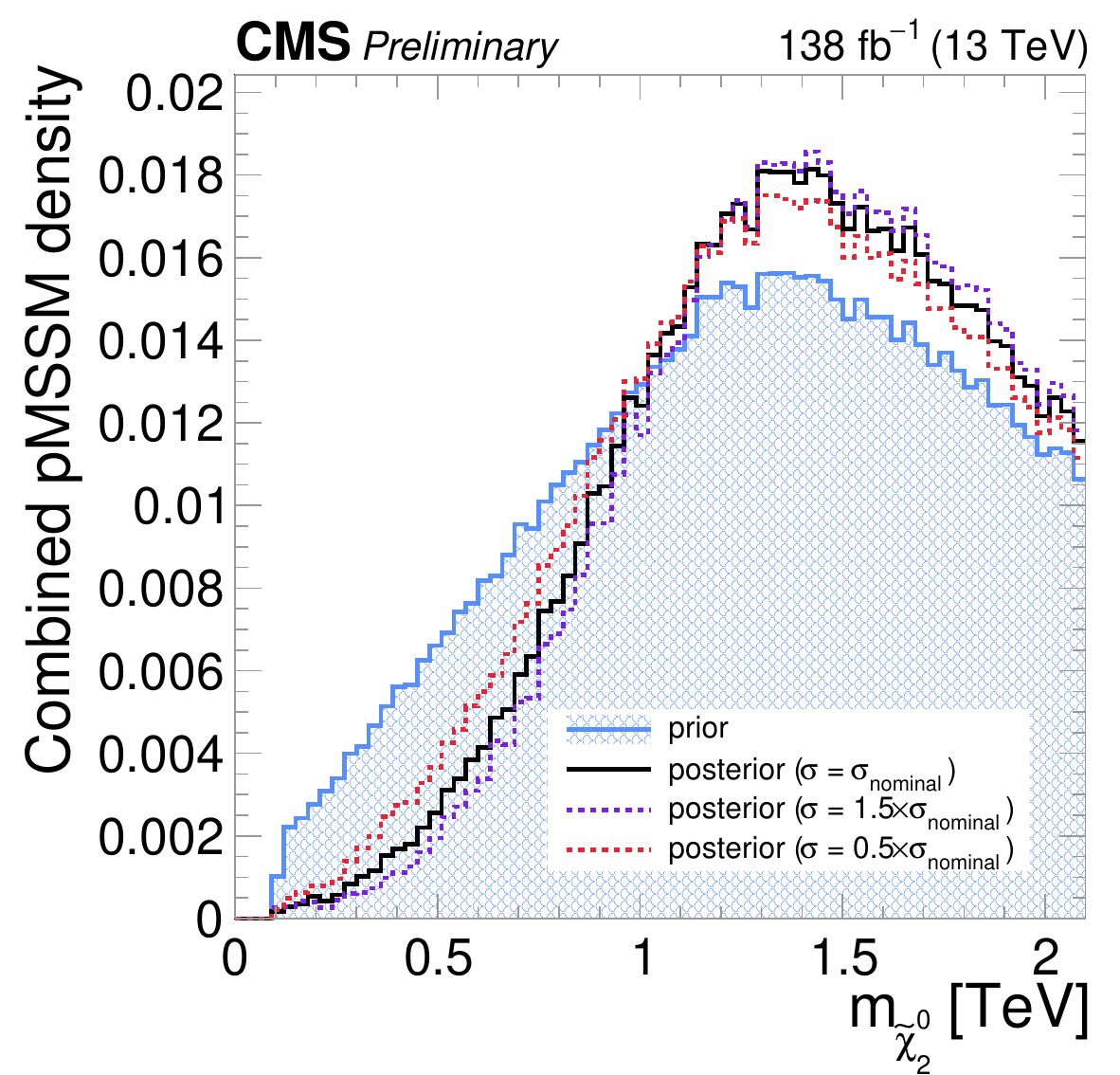} \quad
\includegraphics[width=0.28\textwidth]{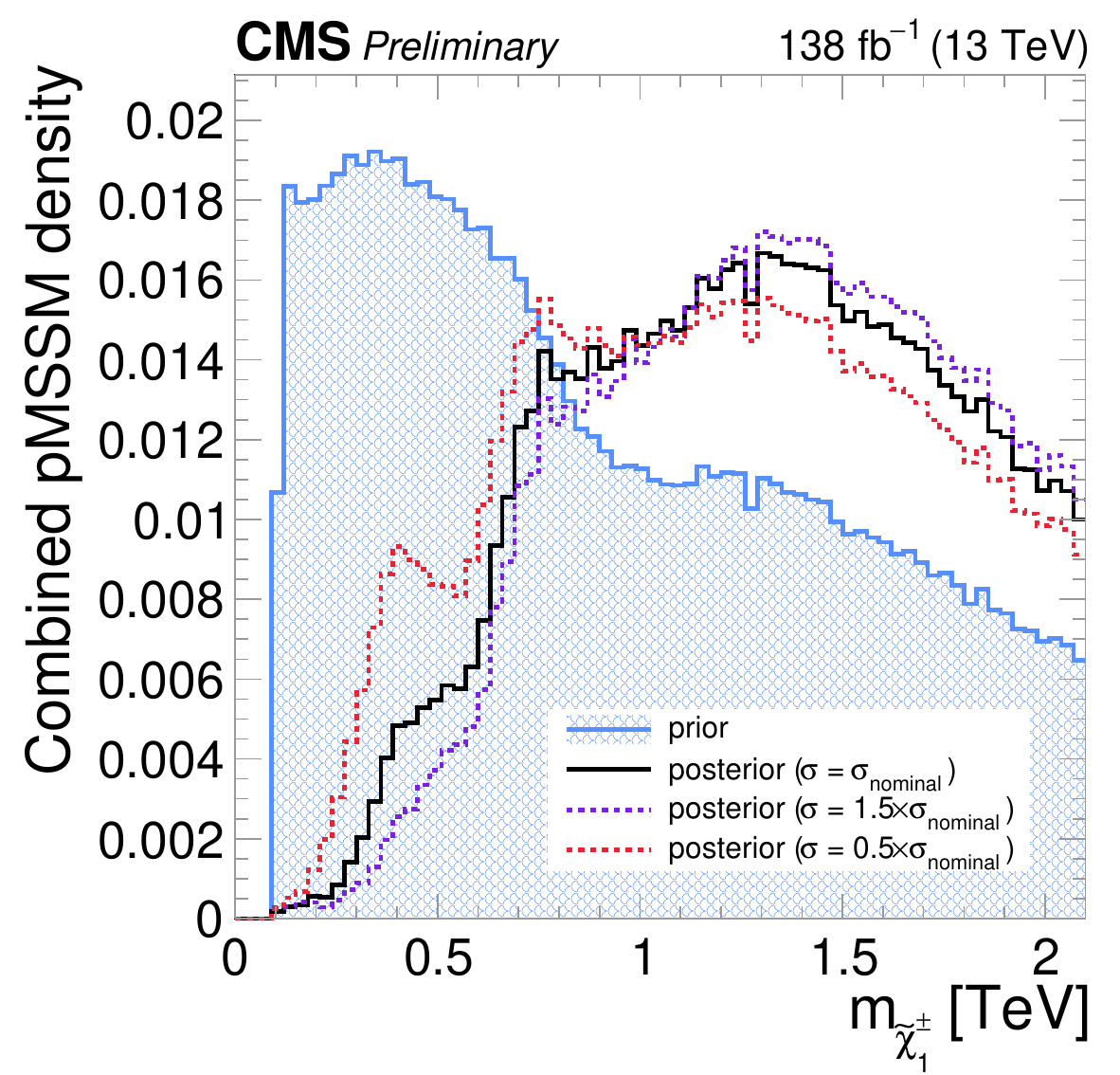} \quad
\includegraphics[width=0.28\textwidth]{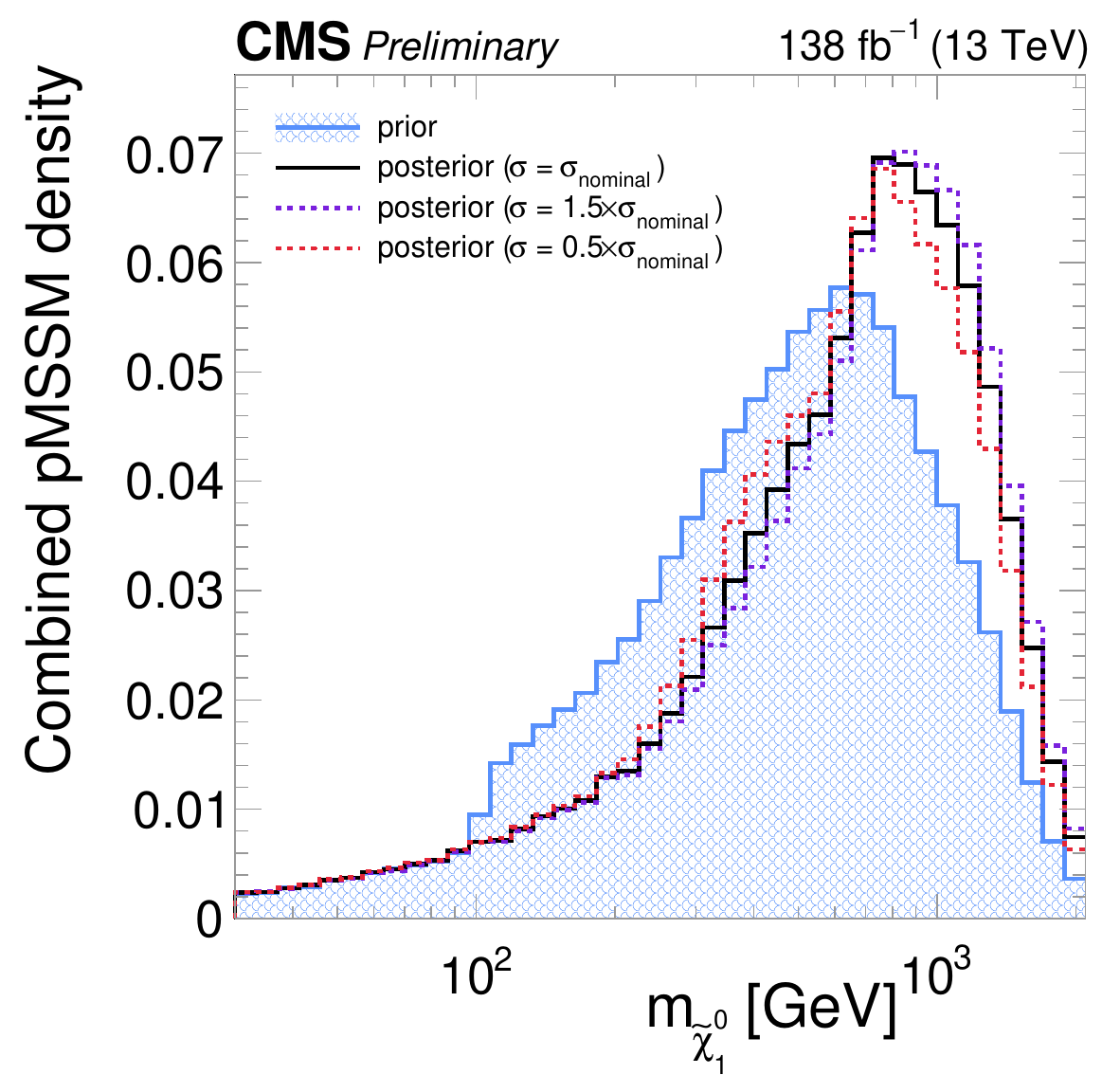} 
\includegraphics[width=0.28\textwidth]{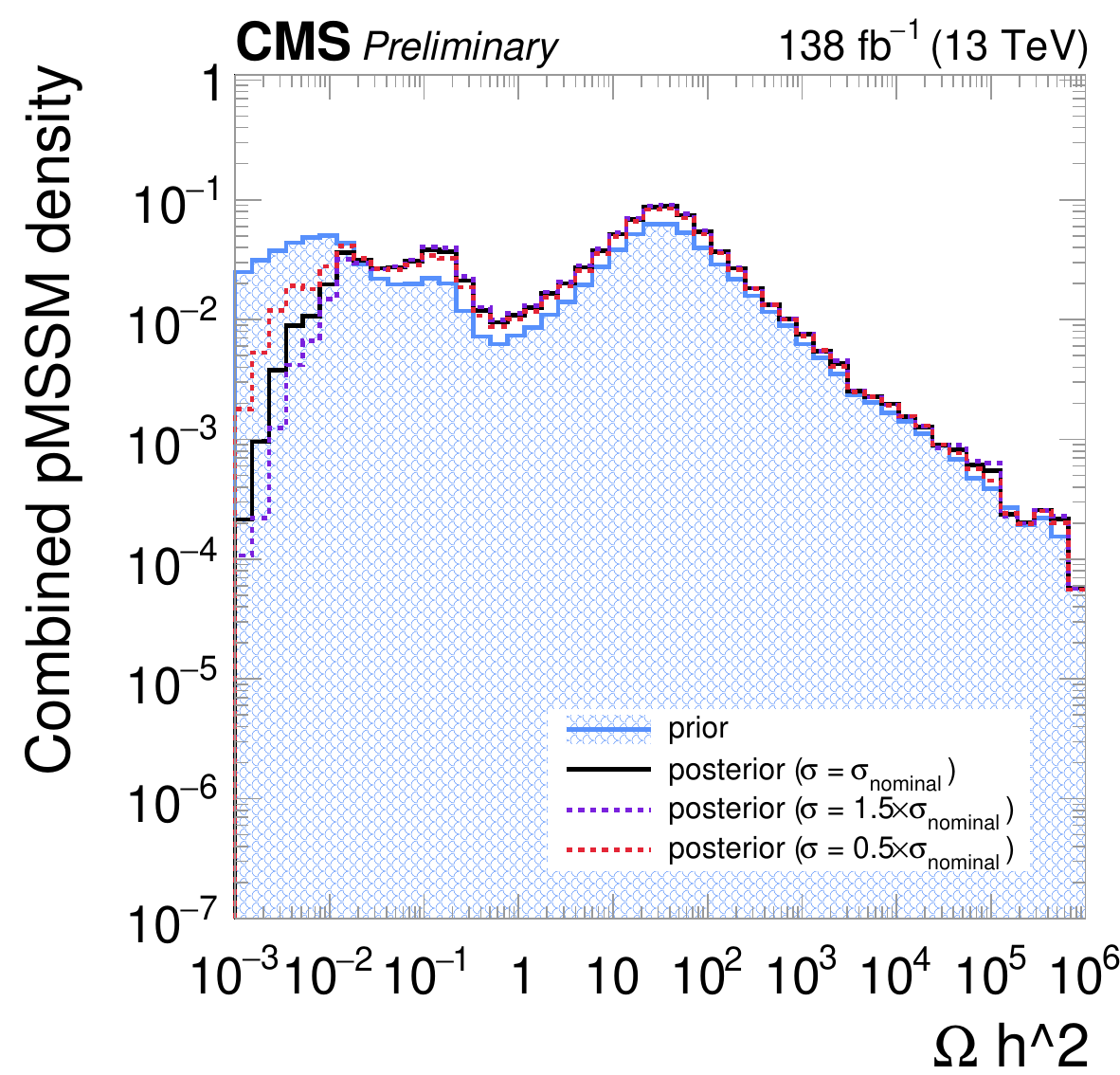} \quad
\includegraphics[width=0.28\textwidth]{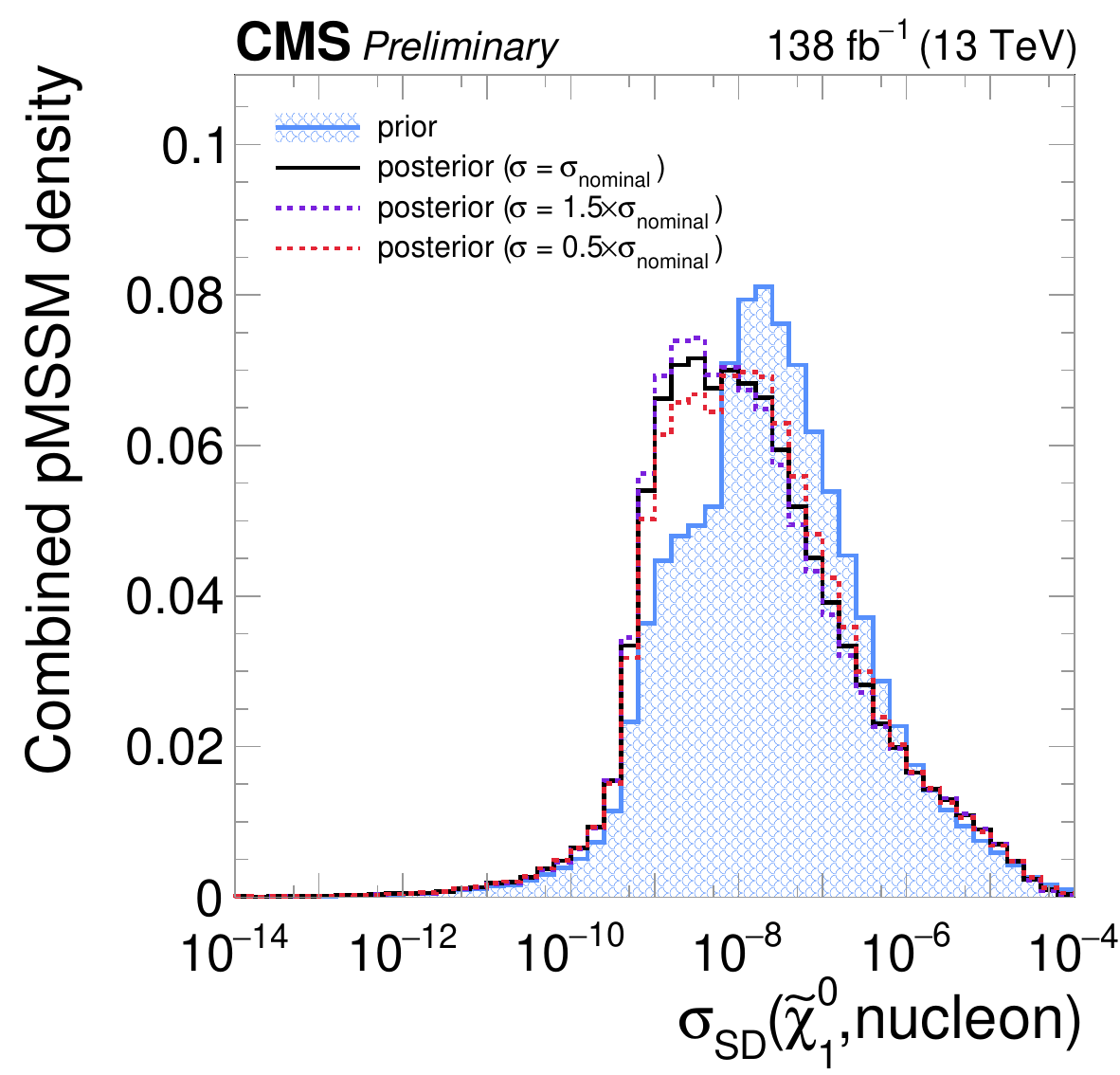} \quad
\includegraphics[width=0.28\textwidth]{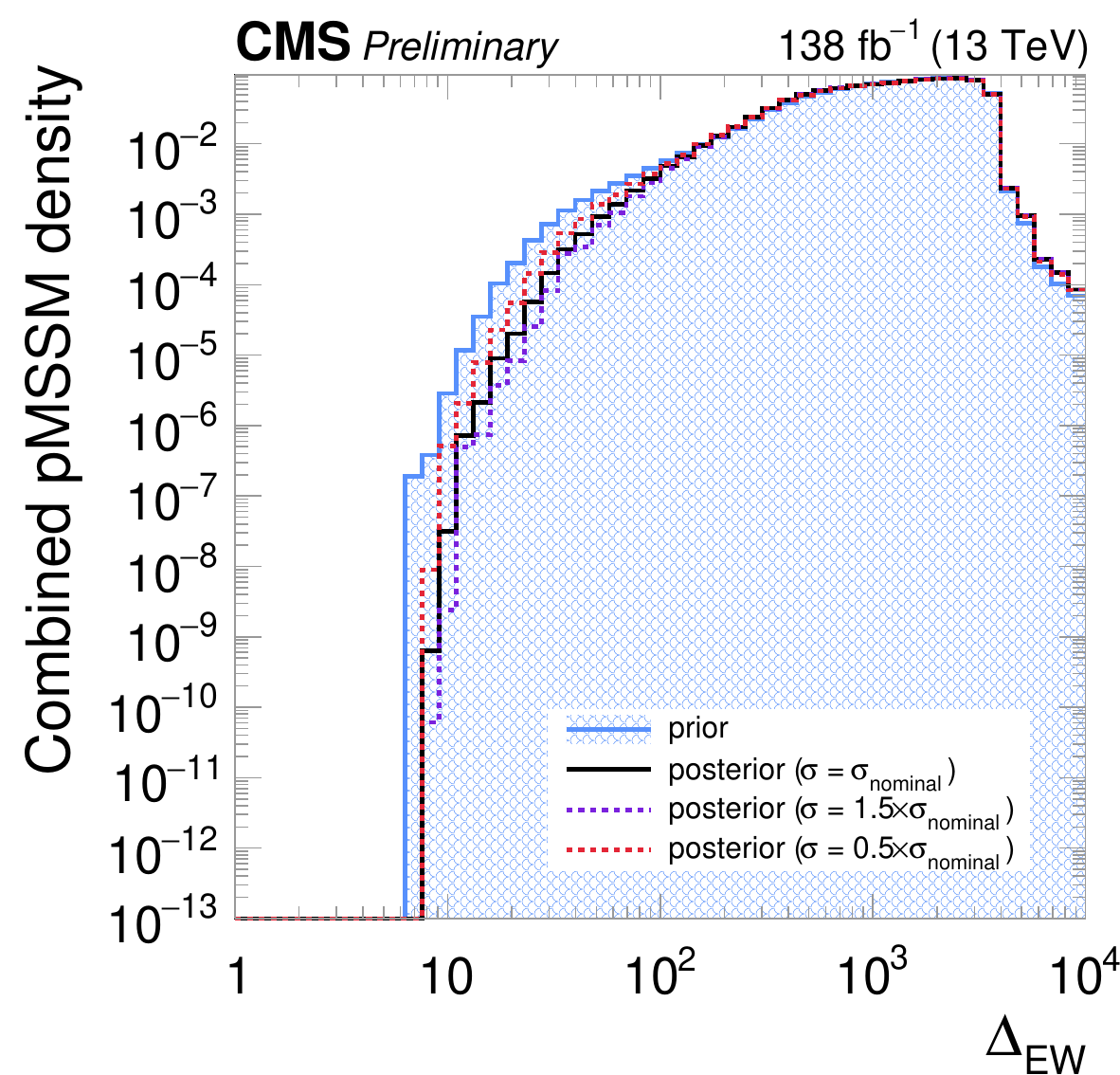} 
\caption{Marginalized prior and posterior distributions for gluino, top squark, lightest colored SUSY particle (LCSP), $\ntwo$, $\cone$, and $\none$ masses, as well as dark matter relic density $\Omega h^2$, spin-dependent $\none$--nucleon cross section $\sigma_{\text{SD}}(\none,\text{nucleon})$, and the electroweak fine-tuning measure $\Delta_{\text{EW}}$.  Also shown are the uncertainties obtained by varying cross sections by 1.5 and 0.5 times the nominal values.}
\label{fig:pmssm1dpost}
\end{figure}

Figure~\ref{fig:pmssm1dpost} shows the one-dimensional marginalized prior and posterior densities obtained from the combination of the analyses listed above.  The distributions are given for the gluino, top squark, lightest colored SUSY particle (LCSP), $\ntwo$, $\cone$, and $\none$ masses, as well as for the dark matter relic density $\Omega h^2$, the spin-dependent $\none$--nucleon cross section $\sigma_{\text{SD}}(\none, \text{nucleon})$, and a certain fine-tuning measure $\Delta_{\text{EW}}$. The shifts between the prior and posterior illustrate the impact of the CMS searches. Although these are one-dimensional projections, they effectively profile over the correlated effects of other sparticles, and carry the imprint of the full parameter correlations in the pMSSM

\begin{figure}[H]
\centering
\raisebox{\height}{\includegraphics[width=0.28\textwidth]{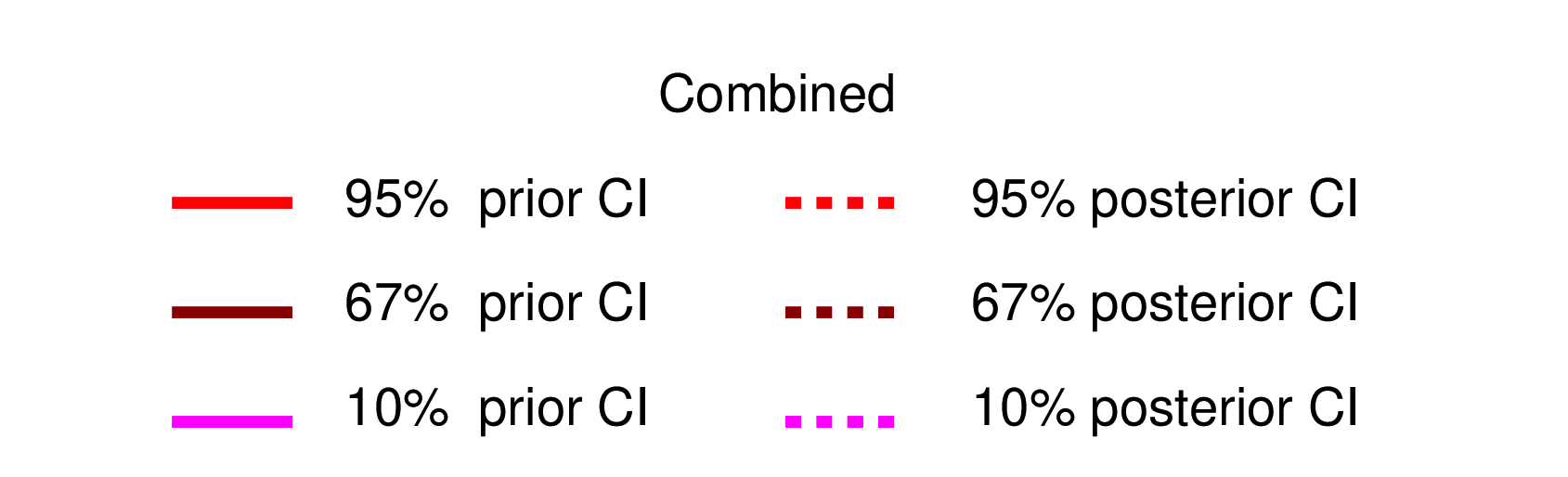}} \quad
\includegraphics[width=0.28\textwidth]{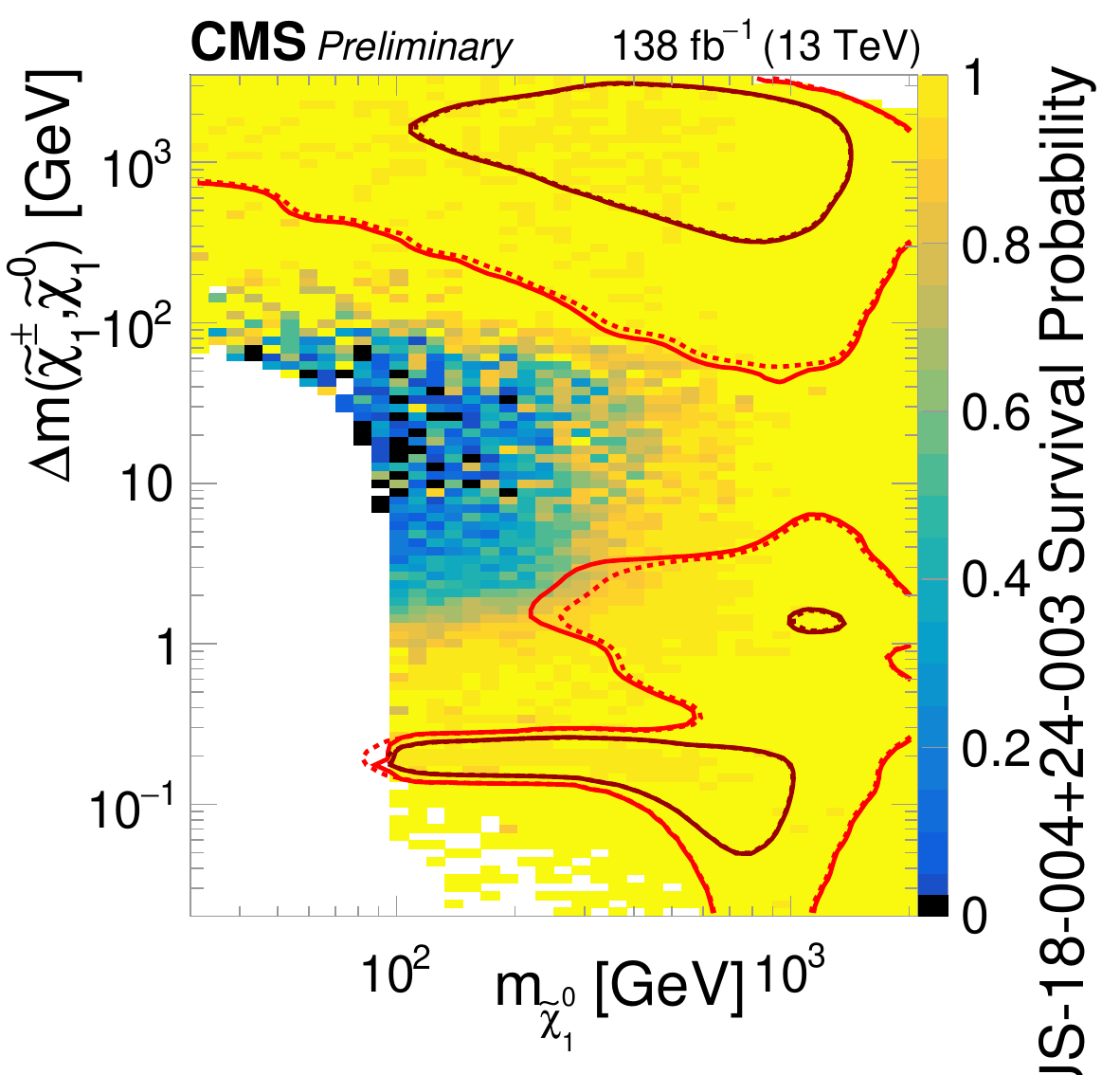} \quad
\includegraphics[width=0.28\textwidth]{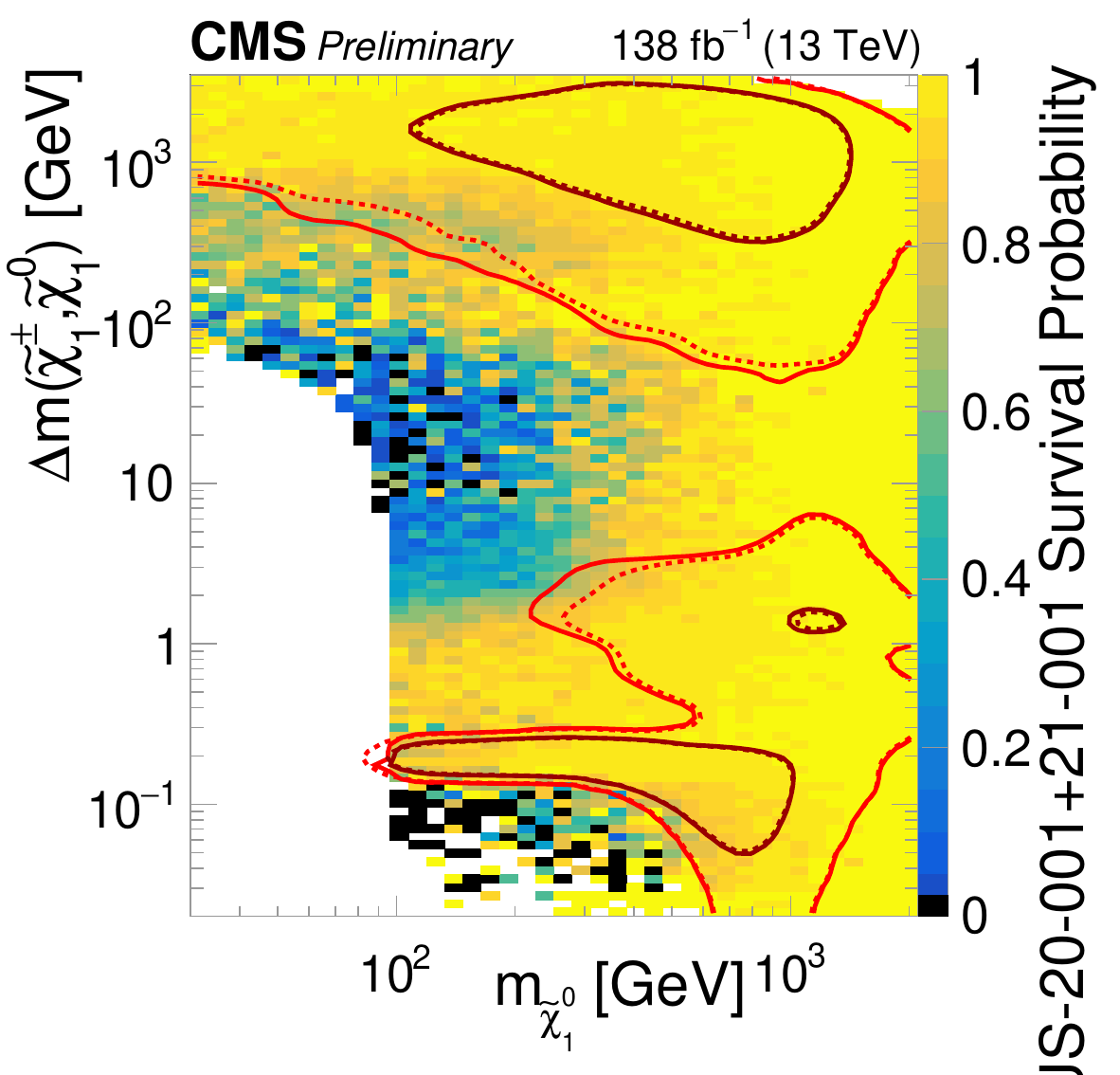} \\
\includegraphics[width=0.28\textwidth]{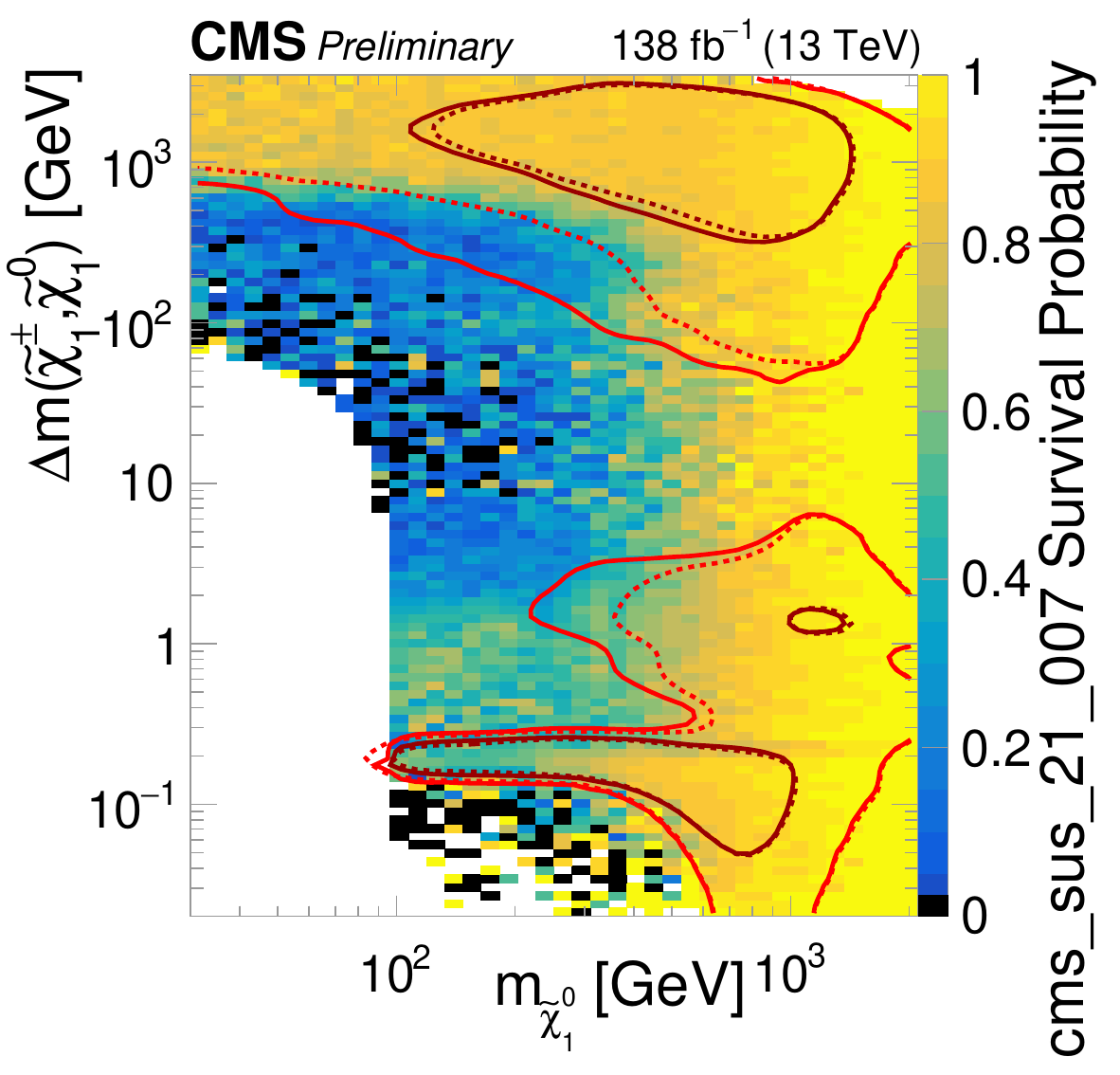} \quad
\includegraphics[width=0.28\textwidth]{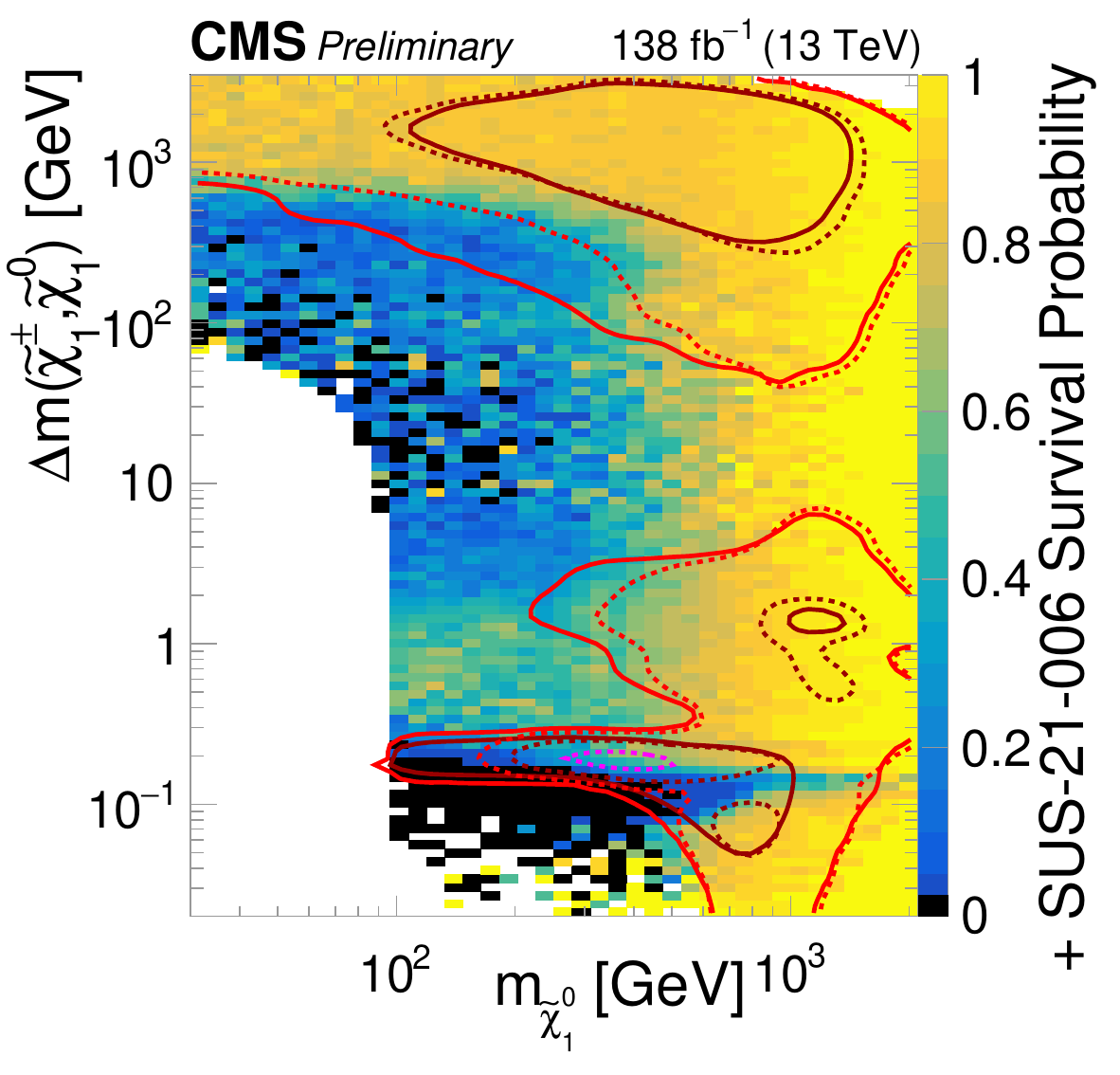} \quad
\includegraphics[width=0.28\textwidth]{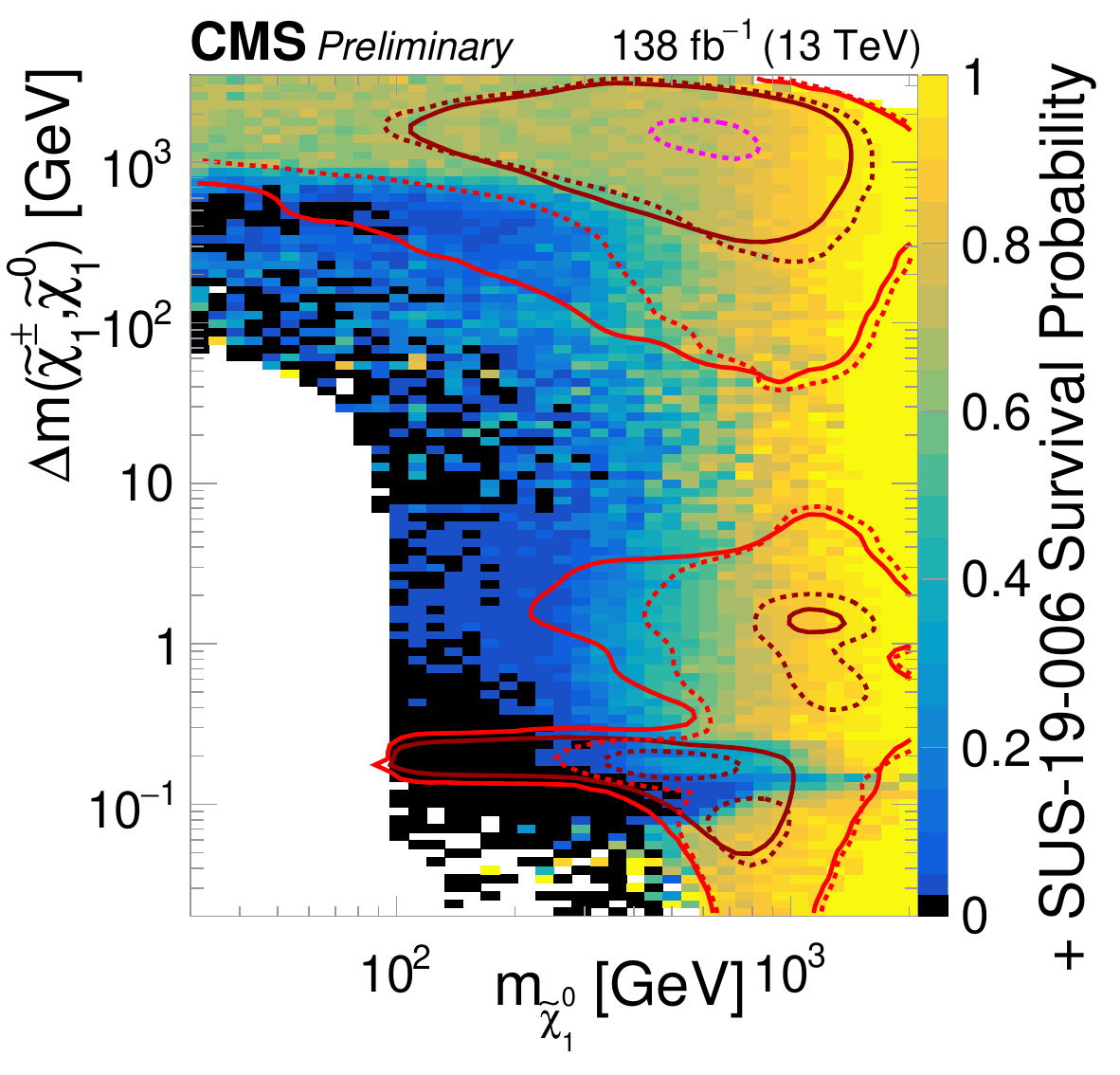} \\
\includegraphics[width=0.28\textwidth]{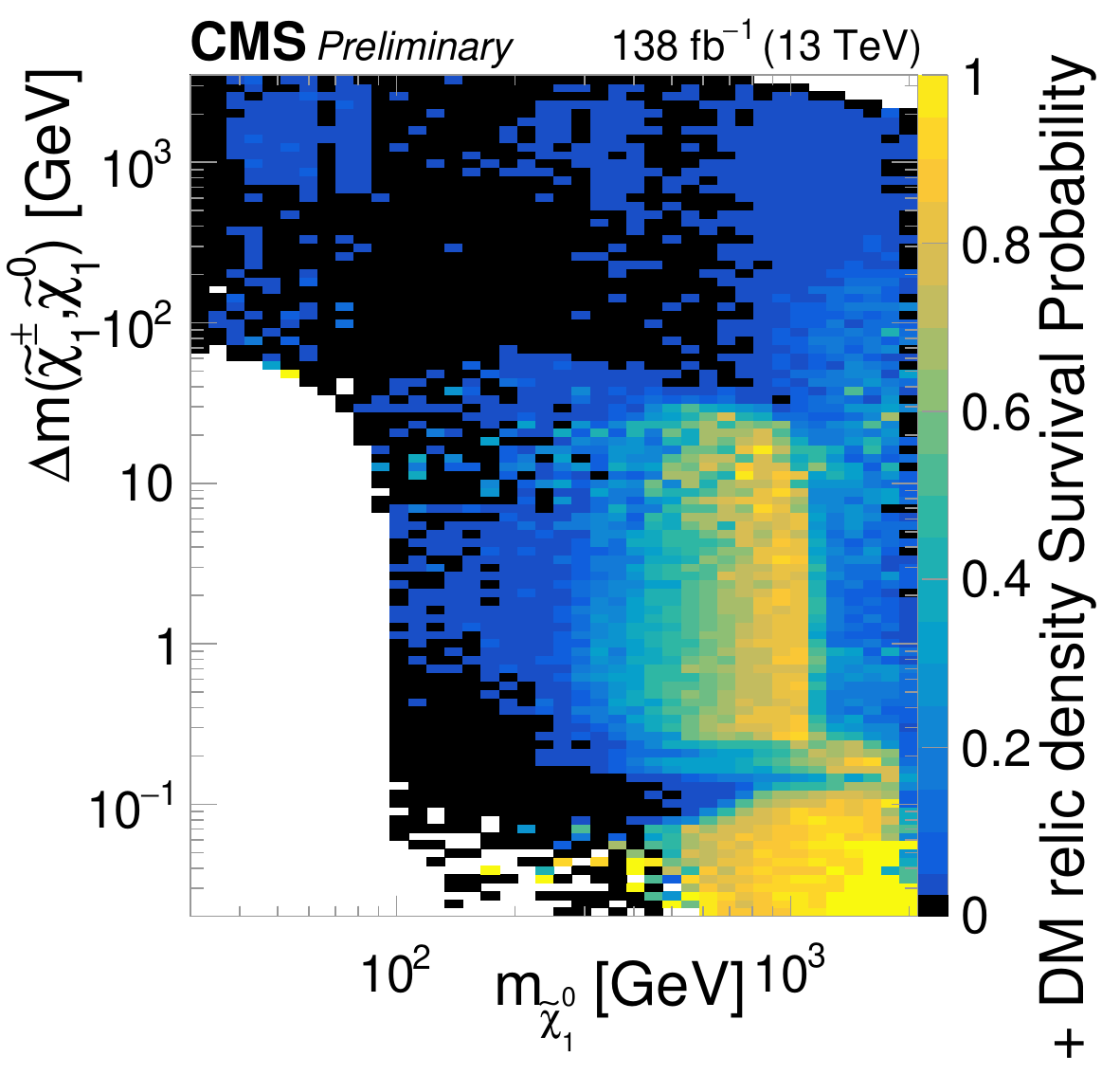} \quad
\includegraphics[width=0.28\textwidth]{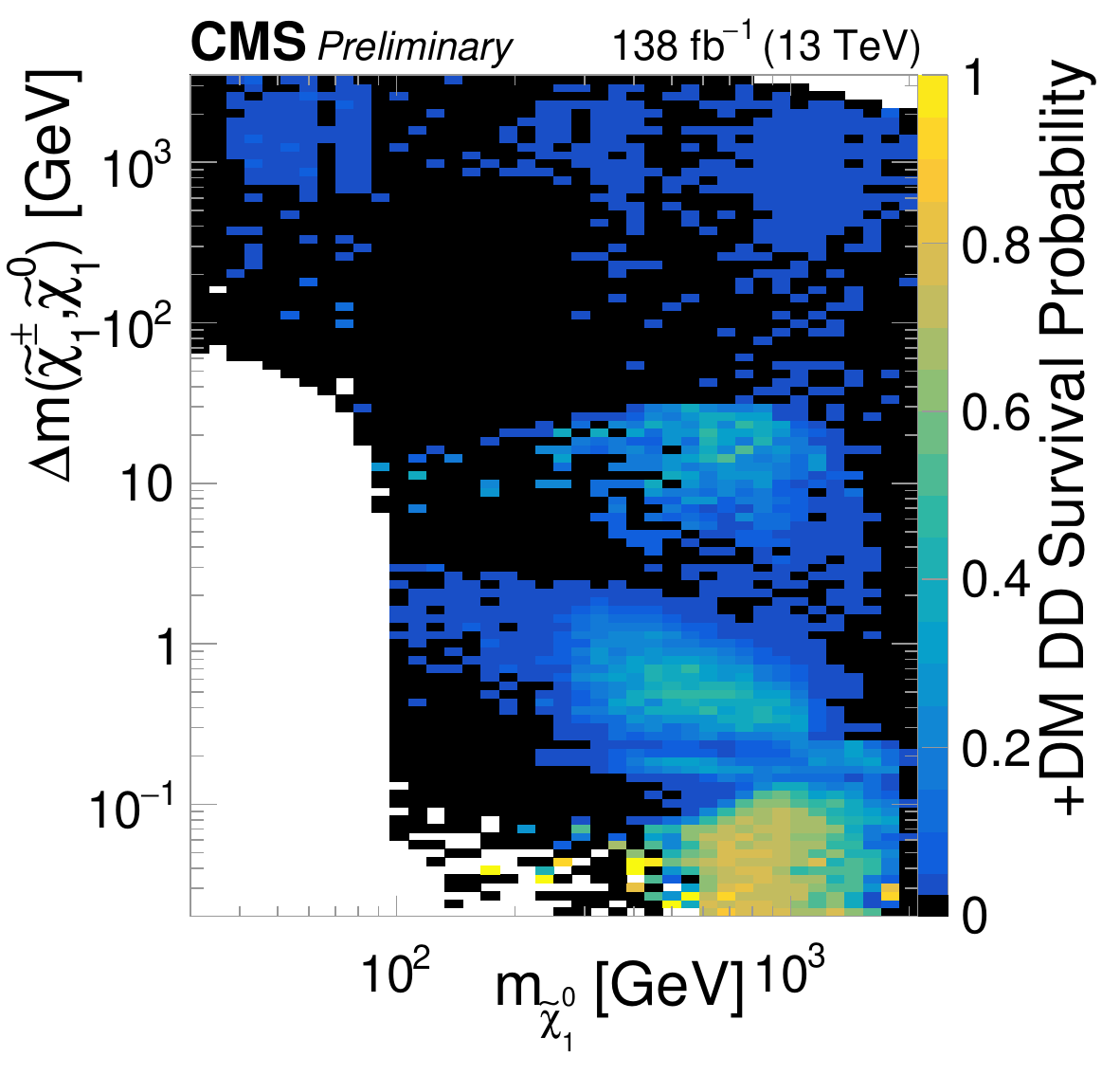} \quad
\includegraphics[width=0.28\textwidth]{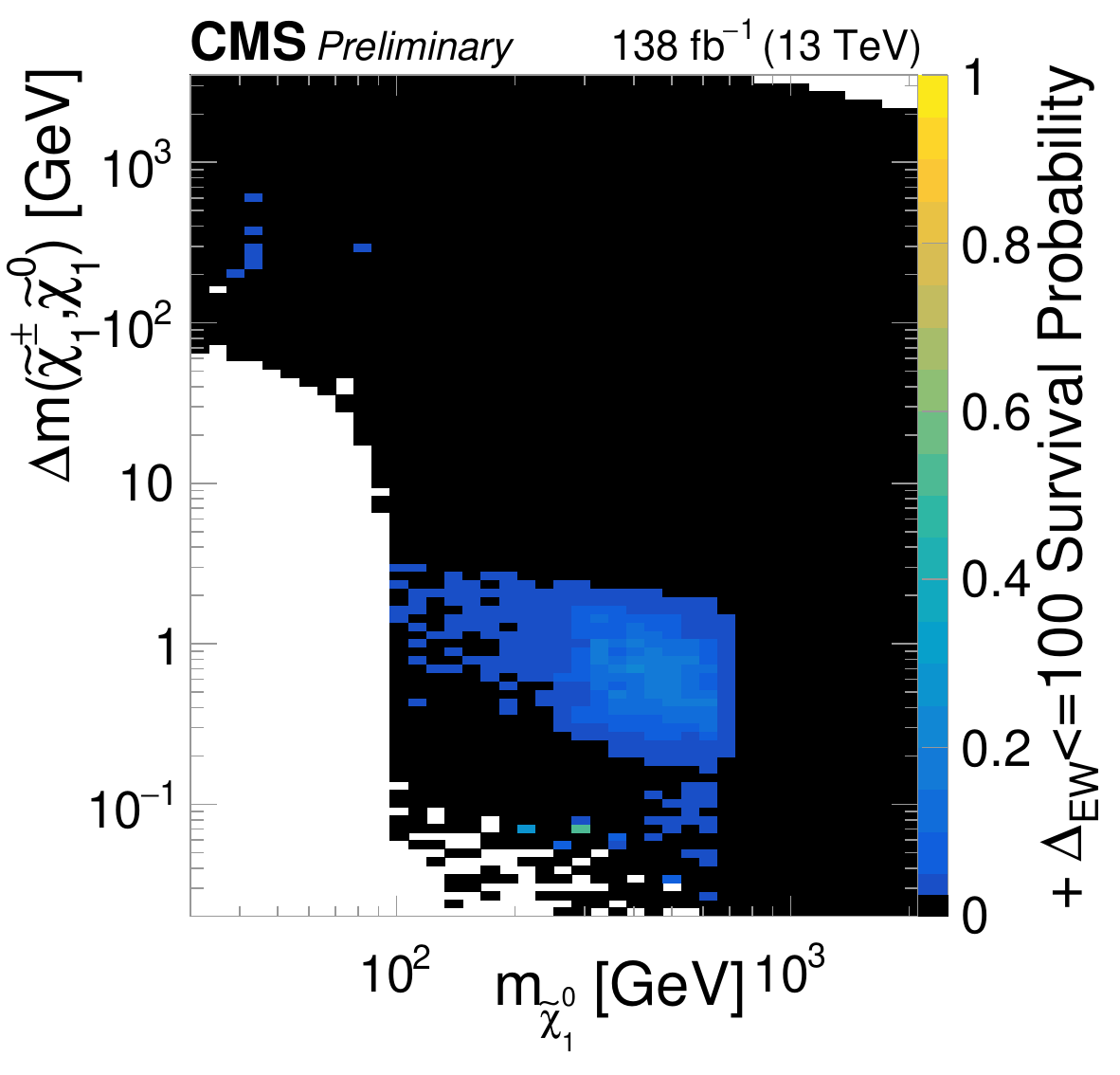} 
\caption{Survival probability distributions in two-dimensional projections in the $(m_{\none}, \Delta m(\cone,\none))$ plane, showing the progressive impact of successive CMS analyses and external constraints on the pMSSM prior. Black bins indicate where no pMSSM points survived the CMS analyses, and white indicate where no pMSSM points are present in the prior. Prior and posterior credibility interval contours are also shown.  Each panel corresponds to the addition of one or more constraints in the sequence described in the text.}
\label{fig:mn1dmflipbook}
\end{figure}

Another way to quantify the impact of any constraint is by the survival probability (SP), defined as the fraction of pMSSM points surviving the constraint, e.g., a 95\% exclusion by an analysis, out of the pre-constraint set.   Figure~\ref{fig:mn1dmflipbook} shows the progressive impact of a series of constraints on the pMSSM as a function of the $\none$ mass and $\Delta m(\cone, \none)$ with respect to the pre-CMS prior, expressed as SPs.  Also shown are prior and posterior credibility interval contours.  The order of applied constraints is soft 2/3$\ell$~\cite{CMS:2021edw} and soft lepton track~\cite{CMS-PAS-SUS-24-003} (top-center), opposite-charge same-flavor 2$\ell$ (on-Z)~\cite{CMS:2020bfa} and direct stau~\cite{CMS:2022syk} (top-right), single lepton $\Delta\phi$~\cite{CMS:2022idi} (middle-left), disappearing track~\cite{CMS:2019zmd} (middle-center), RA2b~\cite{CMS:2023mny} (middle-right), dark matter relic density (bottom-left), dark matter direct detection (bottom-center), and the fine tuning constraint $\Delta_{\text{EW}} < 200$.  We see that each CMS analysis  contributes to constraining the pMSSM.  We also see that dark matter and naturalness constraints are significant, and note that a very well-defined island survives for chargino masses lower than a TeV and mass differences around 1 GeV.  

\begin{figure}[H]
\centering
\includegraphics[width=0.28\textwidth]{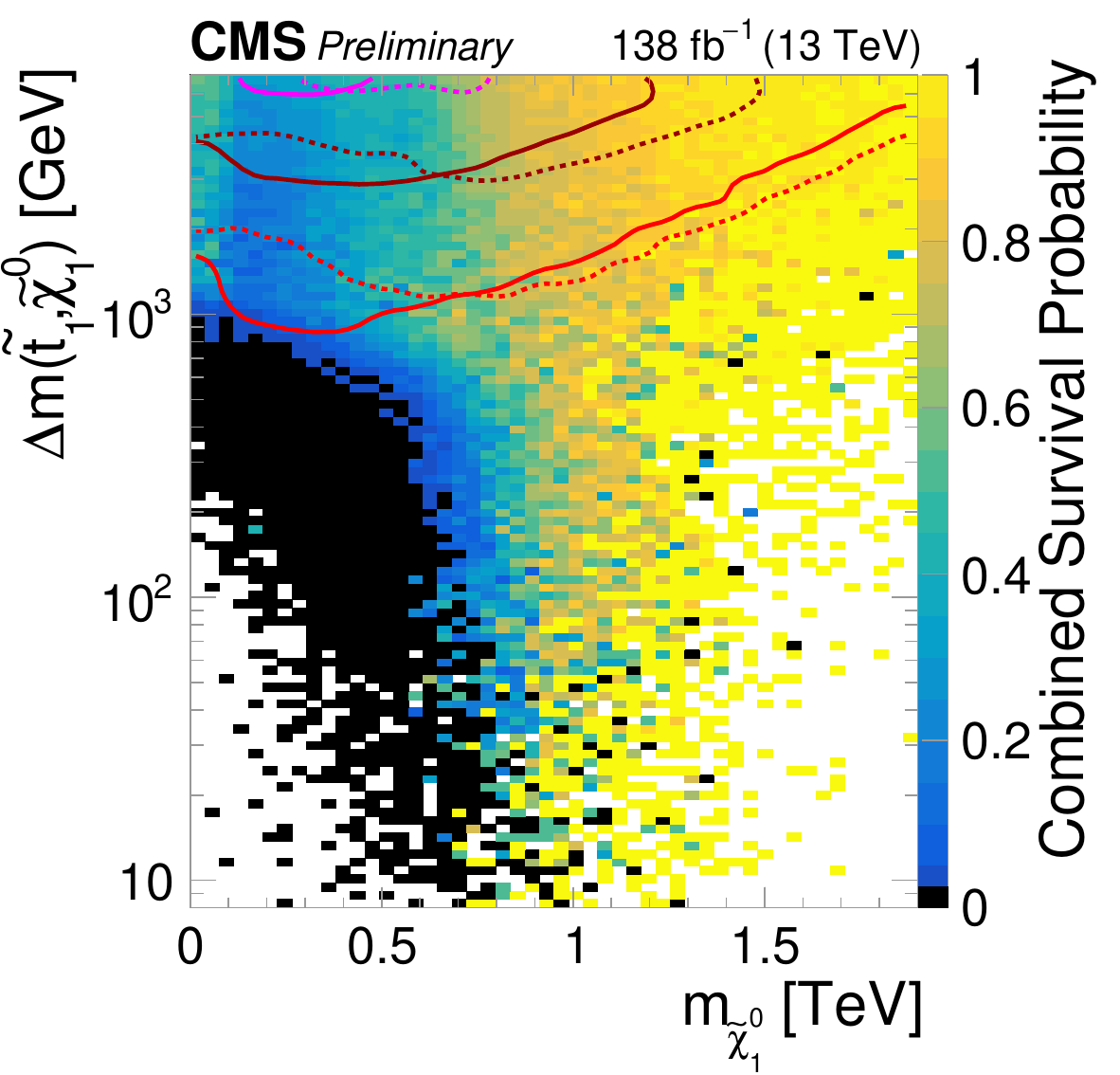} \quad
\includegraphics[width=0.28\textwidth]{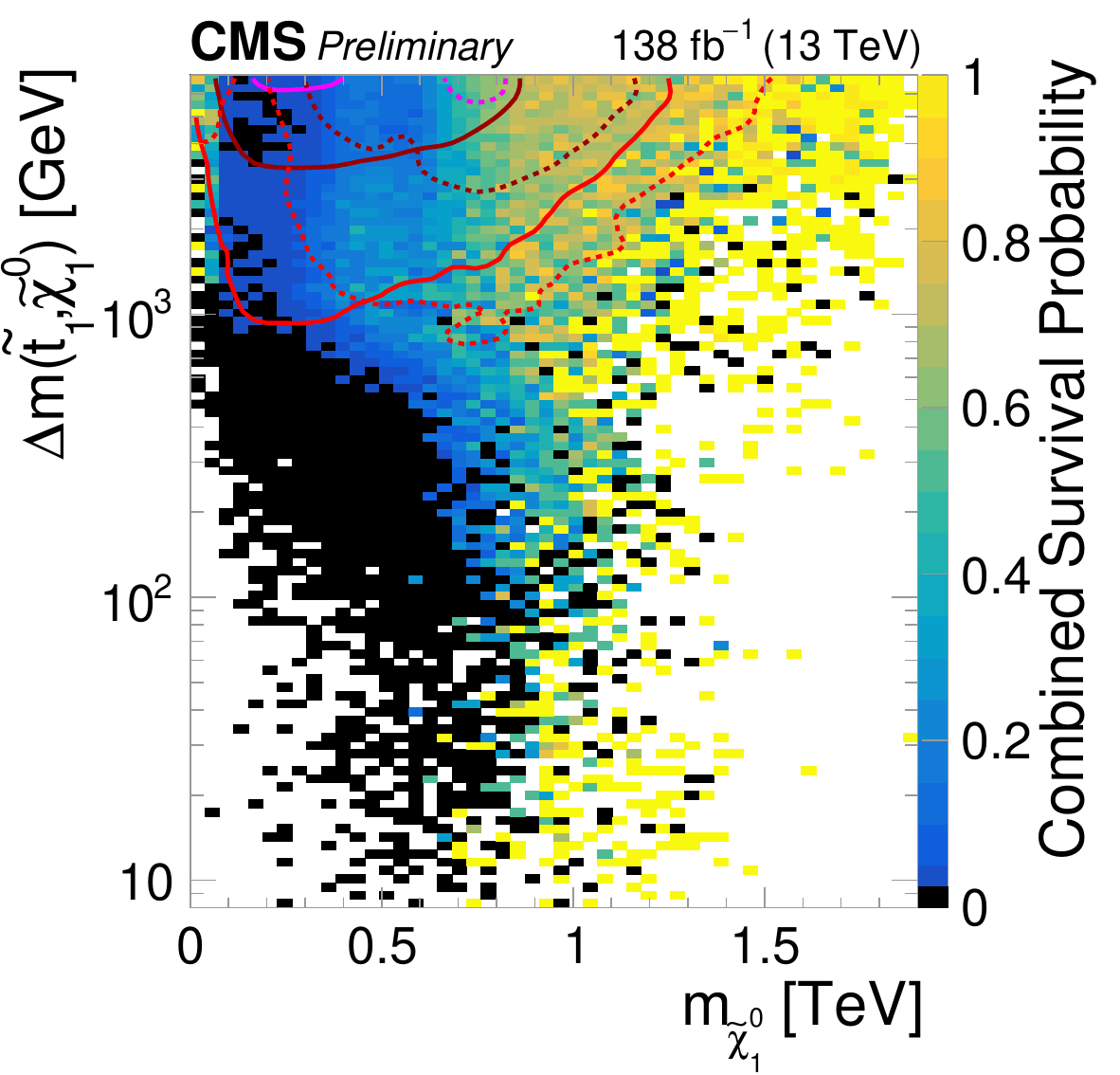} \quad
\includegraphics[width=0.28\textwidth]{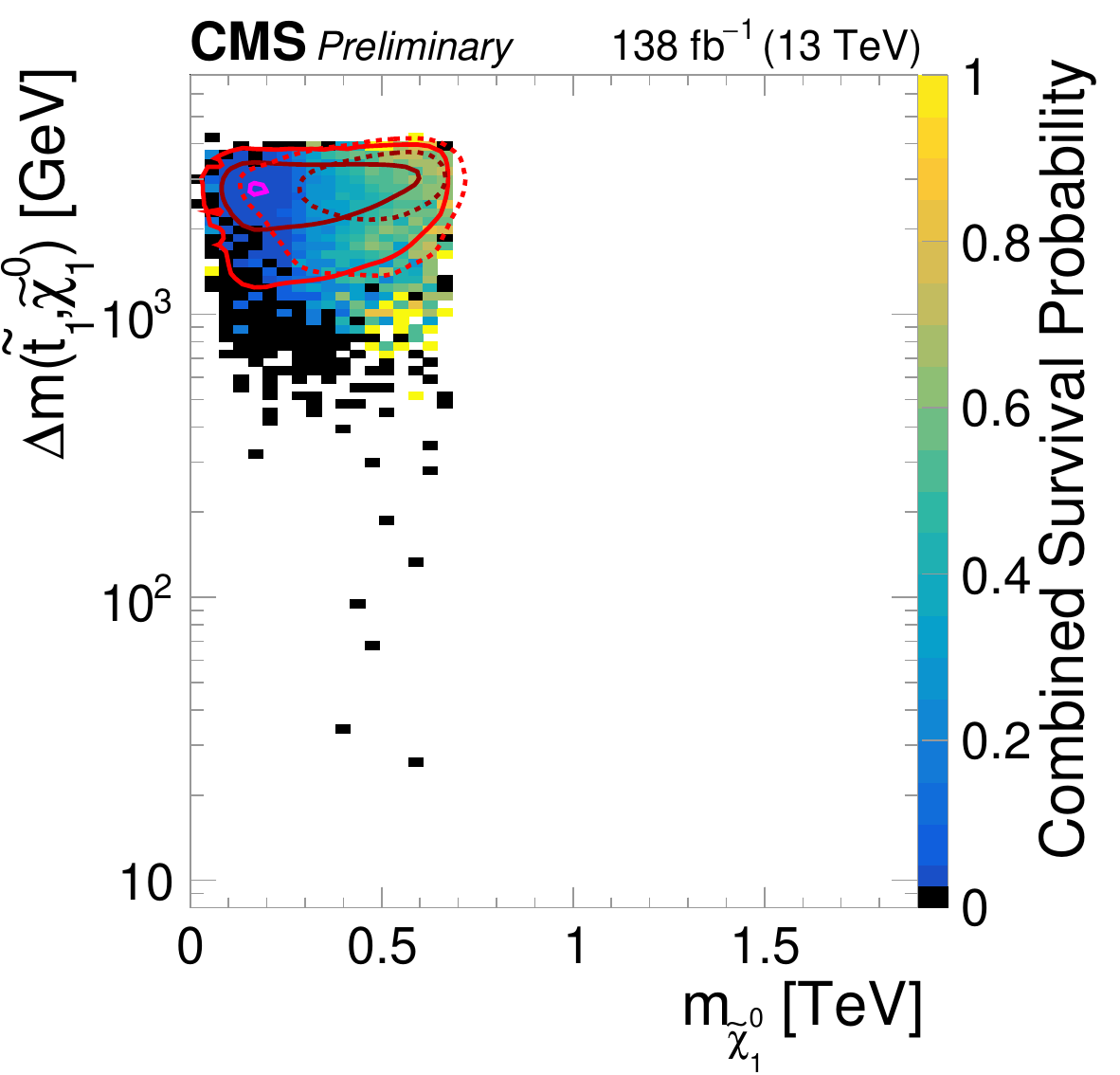} \\ 
\includegraphics[width=0.28\textwidth]{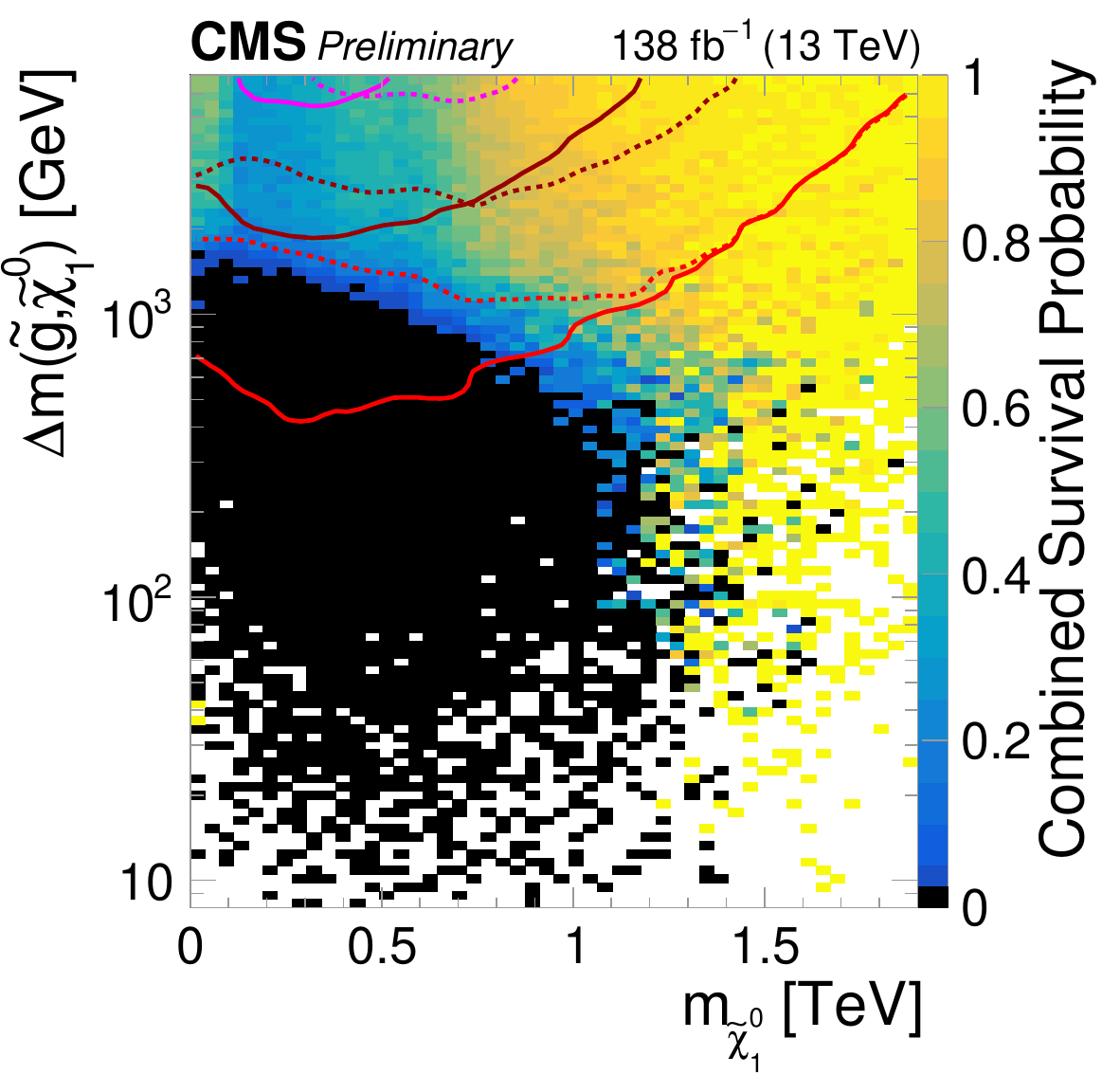} \quad
\includegraphics[width=0.28\textwidth]{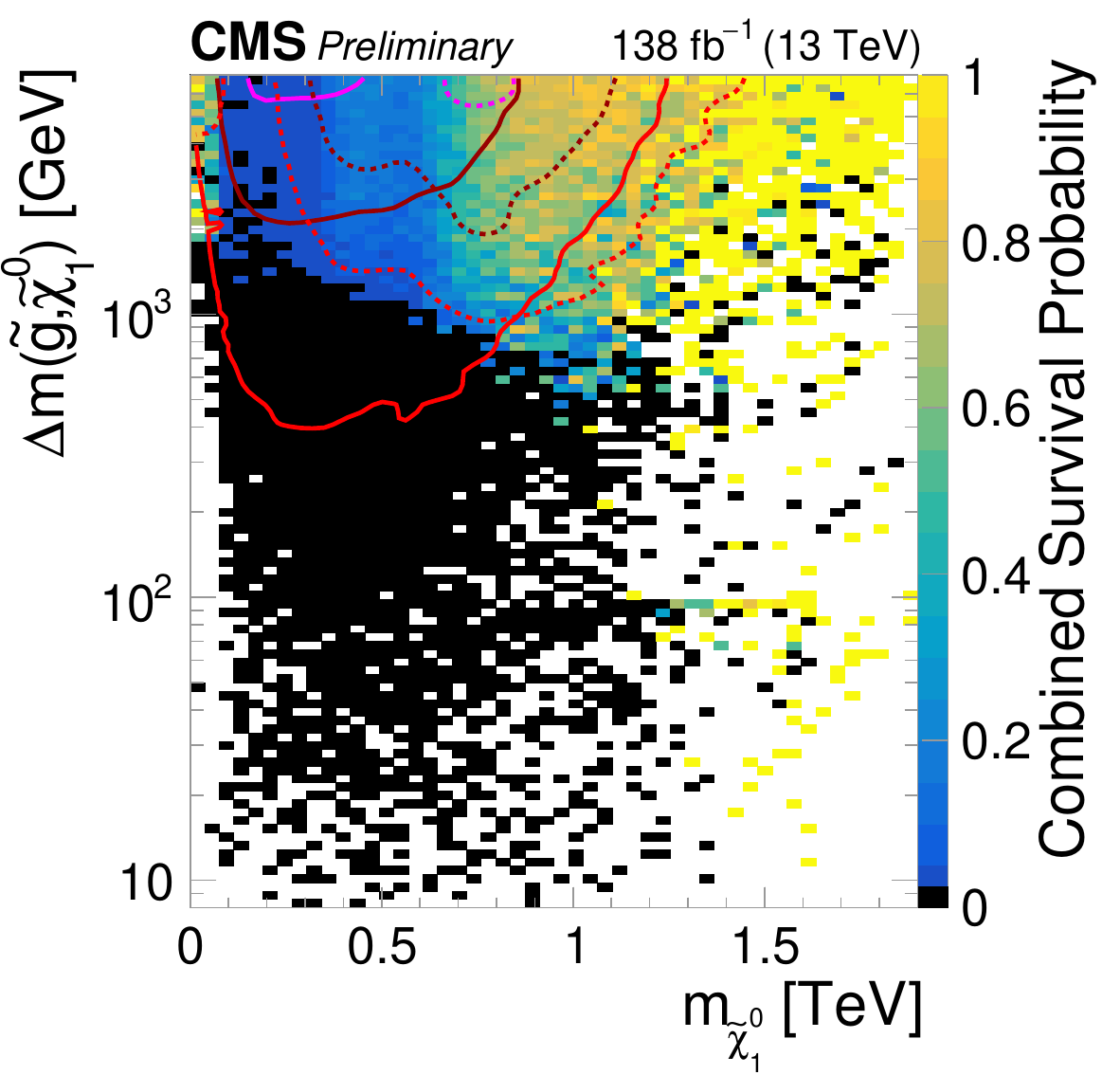} \quad
\includegraphics[width=0.28\textwidth]{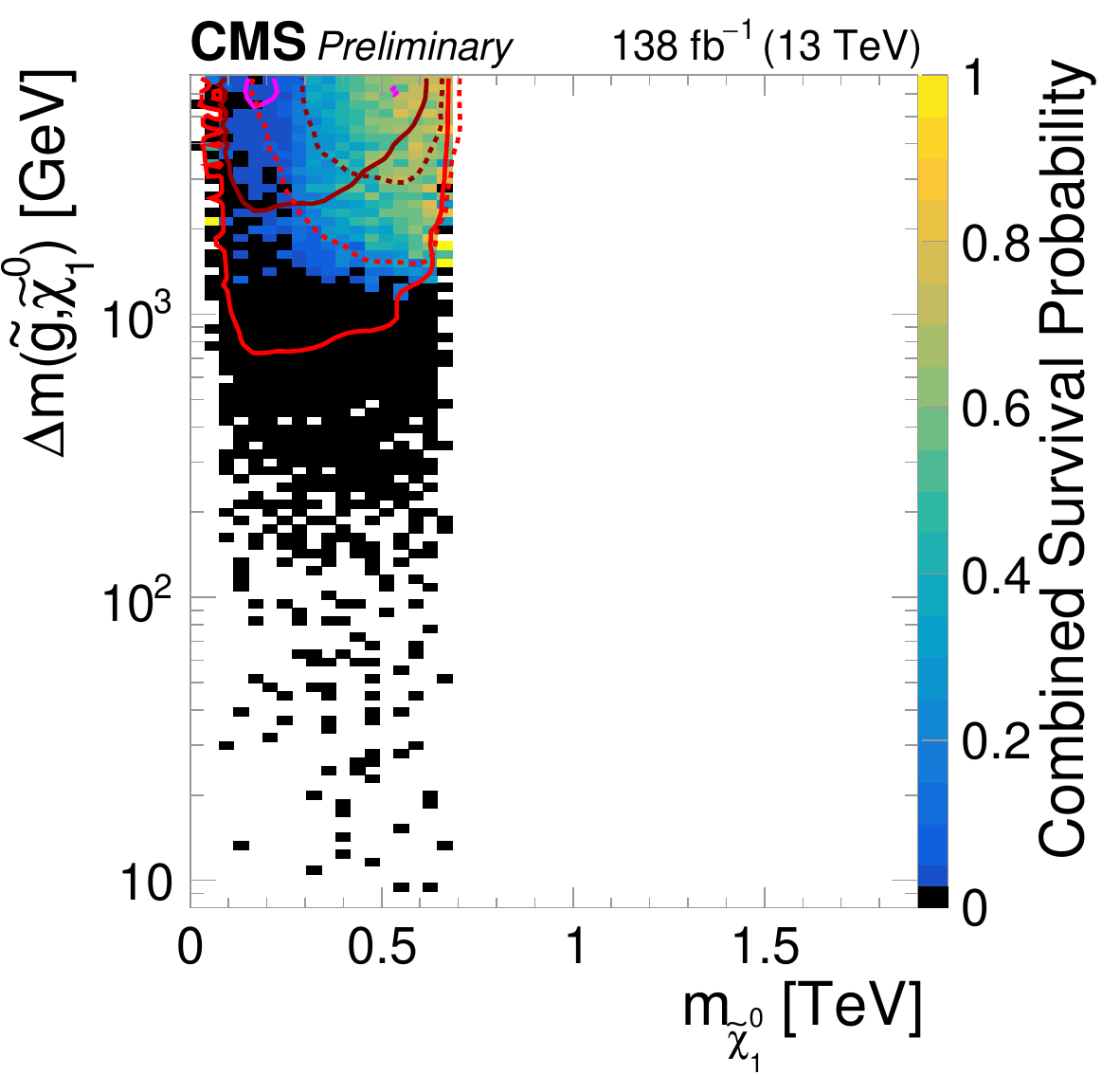} \\ 
\includegraphics[width=0.28\textwidth]{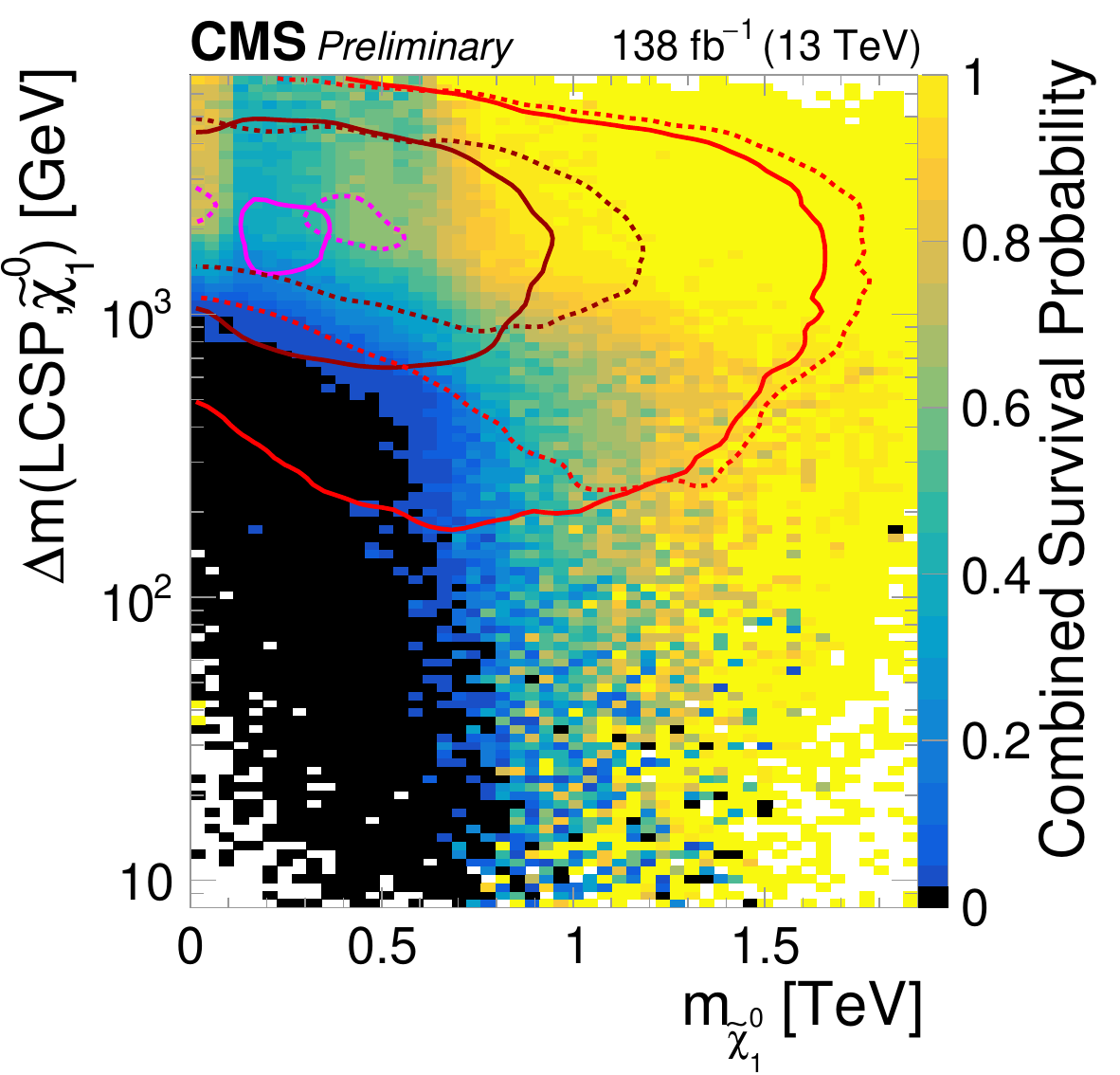} \quad
\includegraphics[width=0.28\textwidth]{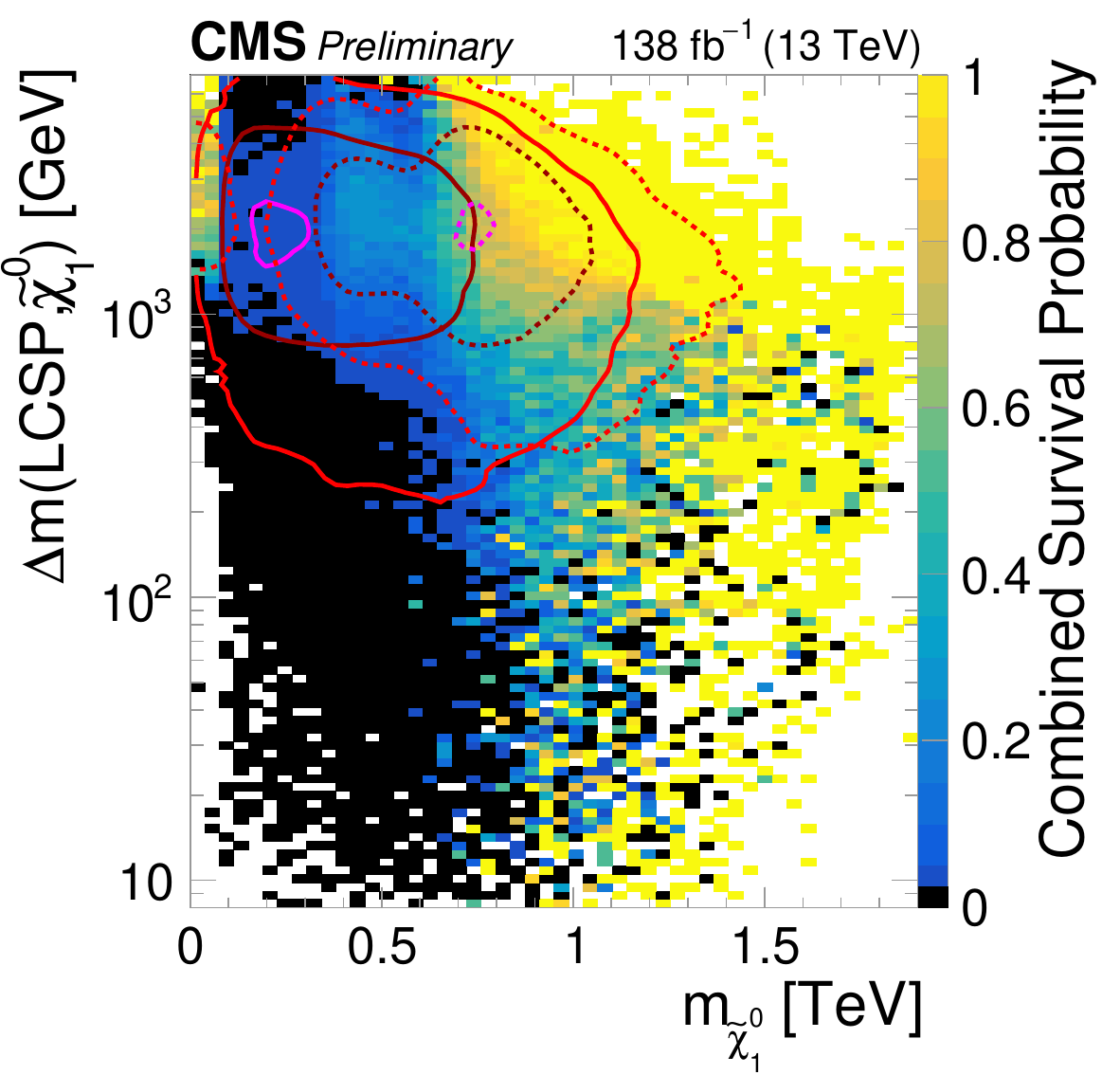} \quad
\includegraphics[width=0.28\textwidth]{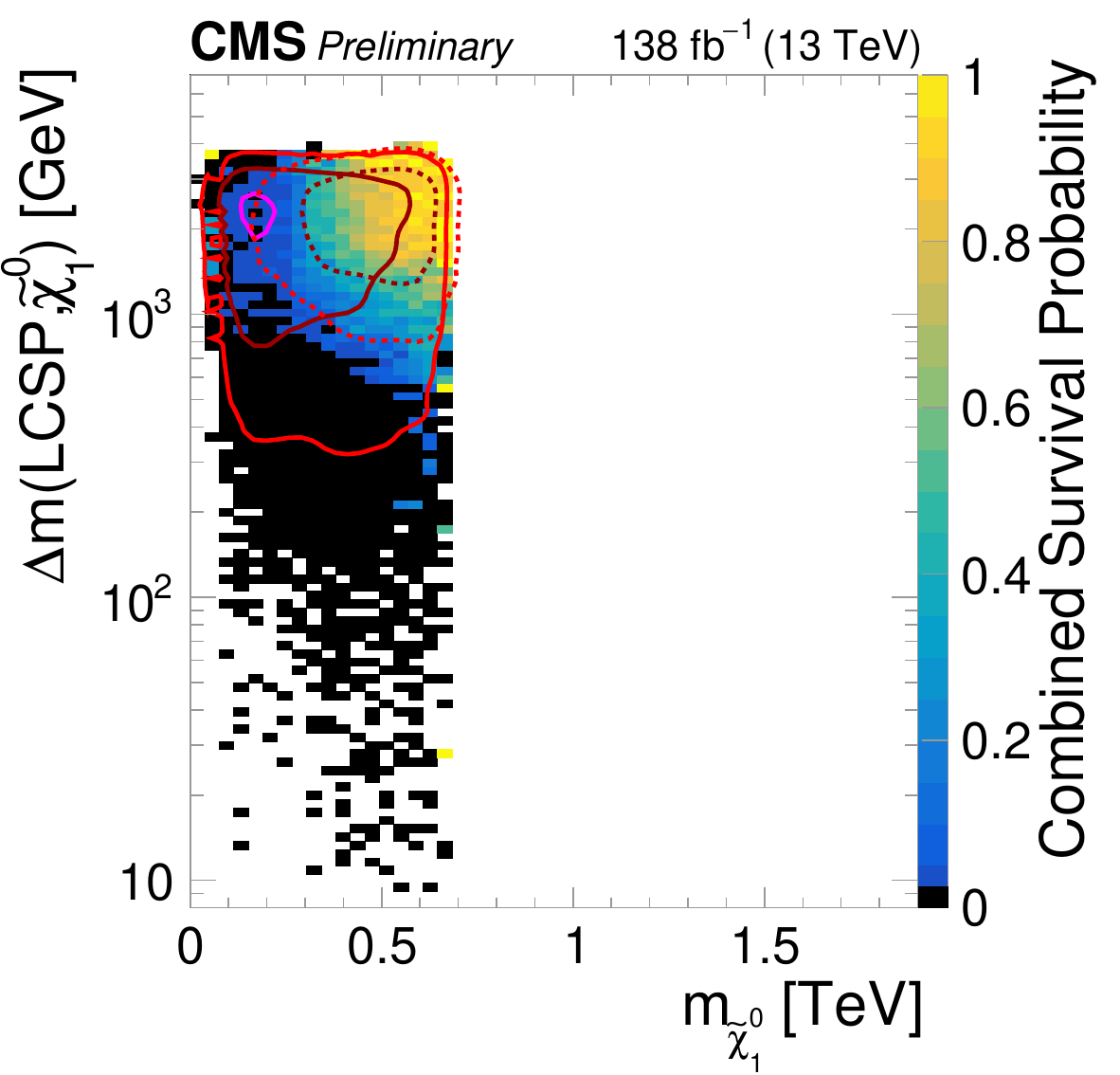} \\ 
\caption{Survival probabilities as a function of $\none$ mass and its mass differences with the lighter top and bottom squarks (top), gluino (middle), and LCSP (bottom). Results are shown for the nominal prior before CMS constraints (left), the prior including dark matter relic density and direct detection constraints (center), and the prior further constrained by $\Delta_{\text{EW}} < 200$ (right).}
\label{fig:pmssmSP}
\end{figure}

Figure~\ref{fig:pmssmSP} shows the survival probabilities as a function of the $\none$ mass and the mass differences between the $\none$ and the lighter top and bottom squarks (top), the gluino (middle), and the LCSP (bottom).  
For each projection, three priors are considered: the nominal prior before applying CMS constraints (left); the prior constrained by dark matter data, requiring the predicted relic density to be less than 110\% of the value measured by Planck and consistency with direct detection limits at 95\% CL (center); and the prior further constrained by naturalness, with $\Delta_{\text{EW}} < 200$ (right).  
Although the dark matter and naturalness requirements already impose strong constraints, CMS searches leave a clearly visible impact even within these restricted subsets of parameter space.  For all cases, we see an overall upward shift in the viable sparticle masses.  

The pMSSM interpretation illustrates how complementary analyses work together to constrain the model space and where important gaps in sensitivity remain. This provides concrete guidance for where future searches should focus.  It is important to note, however, that the present study includes only a subset of CMS analyses, and the ongoing inclusion of further results may modify the picture.

\section{To conclude}

During Run 2, CMS explored supersymmetry across nearly every accessible corner of phase space, pushing the use of detector output and analysis techniques to their limits. We continuously refined our methods, developed new strategies, and established broad coverage of the SUSY parameter space. No discovery has emerged so far, but we now have a sharper picture of what is excluded and where sensitivity is still lacking.  

Several Run 2 SUSY searches are still being finalized.  In parallel, we are working to ensure a lasting legacy by preserving analysis information for reuse and reinterpretation. CMS results are systematically published in HEPData with increasing amounts of auxiliary material, and recent analyses are working to release their statistical models. We are also encouraging preservation of analysis code and physics algorithms, both to ensure reproducibility, future combinations, and to transmit the accumulated knowledge of Run 2 to the next generation of analyses and analysts.  

With this foundation in place, SUSY searches are now moving forward with Run 3, with data taking underway since 2022 at 13.6 TeV and an integrated luminosity expected to exceed twice that of Run 2.  Developments in triggers, reconstruction, identification, and other analysis techniques are being further advanced.  Several analyses are already in preparation, building on these improvements with the goal of extending sensitivity and probing more challenging final states.
Particular attention is and will be directed to scenarios that remain open.  Light higgsinos are a prime example, where complementary decay chains involving jets, vector bosons, or Higgs bosons can be targeted.  Masses up to about a TeV remain both well motivated by theory and consistent with dark matter constraints.  Direct slepton and stau production is another important direction, especially in compressed mass regions where experimental sensitivity is limited.  More generally, limits on gluino and squark production will be pushed higher, through modern inclusive searches with broader  coverage, aided by improved boosted object identification.

Looking further ahead, the HL-LHC will bring a dramatic increase in dataset size, extended detector coverage, and new detector systems.  The new precision timing detector, in particular, will create new opportunities for long-lived particle searches. These next chapters will allow CMS to extend its sensitivity and probe increasingly challenging final states, reaching deeper into regions of parameter space that remain open after Run 2.
 
\section*{Acknowledgments}

I thank all colleagues in the CMS Collaboration who contributed to designing, building, and operating the experiment, and for providing the data, tools, and methods that enable the CMS SUSY program. I especially thank my CMS SUSY colleagues for their meticulous and creative efforts in coordinating, performing, and reviewing the analyses and studies presented in this review, and those still to come.

\noindent This work is supported by the Basic Science Research Program through the National Research Foundation of Korea (NRF) funded by the Ministry of Education under contracts NRF-2018R1A6A1A06024970 and RS-2008-NR007227.

\bibliographystyle{cms_unsrt}
\bibliography{HiHEPCMSSUSYReviewSekmen}  

\end{document}